\documentclass[iop]{emulateapj}
\usepackage[maxfloats=100]{morefloats}
\usepackage{graphicx}
\usepackage[varg]{txfonts}

\usepackage{natbib}
\usepackage{color,soul}

\bibpunct{(}{)}{;}{a}{}{,} 

\global\long\def\kmss{\,\mathrm{km\, s^{-1}}}
\global\long\def\kpc{\mathrm{\, kpc}}
\global\long\def\halpha{\mathrm{H}\alpha}
\global\long\def\her3{\mathrm{Her_3}}

\global\long\def\kmss{\,\mathrm{km\, s^{-1}}}
\global\long\def\msun{\,M_{\odot}}
\global\long\def\kpc{\mathrm{\, kpc}}

\global\long\def\msunpc{M\mathrm{_{\odot}\,pc^{-2}}}
\global\long\def\halpha{\mathrm{H}\alpha}
\global\long\def\sings{3.6\mu\mathrm{m}}
\global\long\def\ad2g{\mathrm{\alpha_{D2G}}}
\global\long\def\aad2g{\mathrm{\overline{\alpha}_{D2G}}}
\global\long\def\amw{\mathrm{\alpha_{MW}}}

\global\long\def\rchisq{\mathrm{\chi^2_r}}
\global\long\def\auni{M\mathrm{{_\odot}\,pc^{-2}\,(K\,km\,s^{-1})^{-1}}}
\global\long\def\aunif{\mathrm{\frac{M_{\odot}\,pc^{-2}}{K\,km\,s^{-1}}}}
\global\long\def\aunif{\frac{M_{\odot}\mathrm{\,pc^{-2}}}{\mathrm{K\,km\,s^{-1}}}}

\shorttitle{Molecular Gas in Mass Models of Galaxies}
\shortauthors{Frank et al.}

\begin{document}

\title{The Impact of Molecular Gas on Mass Models of Nearby Galaxies}

\author{B. S. Frank\altaffilmark{1,2},  W. J. G. de Blok\altaffilmark{1,2,3}, F. Walter\altaffilmark{4}, A. Leroy\altaffilmark{5} and C. Carignan\altaffilmark{2}}

\altaffiltext{1}{Netherlands Institute for Radio Astronomy (ASTRON), Postbus 2, 7990 AA Dwingeloo, The Netherlands}
\altaffiltext{2}{Department of Astronomy, University of Cape Town, Private Bag X3, Rondebosch 7701, South Africa}
\altaffiltext{3}{Kapteyn Astronomical Institute, University of Groningen, P.O. Box 800, 9700 AV Groningen, The Netherlands}
\altaffiltext{4}{Max-Planck Institute for Astronomy, K{\"o}nigstuhl 17, D-69117 Heidelberg, Germany}
\altaffiltext{5}{Department of Astronomy, The Ohio State University, 140 West 18th Avenue, Columbus, OH 43210, USA}
\newcommand{\myemail}{frank@astron.nl}

\begin{abstract}
We present CO velocity fields and rotation curves for a sample of nearby galaxies, based on data
from the HERACLES survey. We combine our data with literature THINGS, SINGS and KINGFISH results to
provide a comprehensive sample of mass models of disk galaxies inclusive of molecular gas.  We
compare the kinematics of the molecular (CO from HERACLES) and atomic (H\,{\sc i} from THINGS) gas
distributions to determine the extent to which CO may be used to probe the dynamics in the inner
part of galaxies. In general, we find good agreement between the CO and H\,{\sc i} kinematics with
small differences in the inner part of some galaxies. We add the contribution of the molecular gas
to the mass models in our galaxies by using two different conversion factors $\mathrm{\alpha_{CO}}$
to convert CO luminosity to molecular gas mass surface density - the constant Milky Way value and
the radially varying profiles determined in recent work based on THINGS, HERACLES and KINGFISH data.
We study the relative effect that the addition of the molecular gas has upon the halo rotation
curves for Navarro-Frenk-White (NFW) and the observationally motivated pseudo-isothermal halos. The
contribution of the molecular gas varies for galaxies in our sample --- for those galaxies where there
is a substantial molecular gas content, using different values of $\mathrm{\alpha_{CO}}$ can result
in significant differences to the relative contribution of the molecular gas and, hence, the shape
of the dark matter halo rotation curves in the central regions of galaxies.
\end{abstract}


\section{Introduction}

\label{cha:label-ch1}
Rotation curves of nearby galaxies provide strong evidence for the existence of dark matter
\citep[see][]{SOFUE:2001PD}. They were first obtained using optical spectroscopy
\citep{1970APJ...159..379R} and soon thereafter radio observations in the 21-cm line of neutral
hydrogen (H\,{\sc i}) revealed that rotation curves remain flat over many multiples of the optical
radius \citep{1973MNRAS.163..163W,1978A&AS...34..259V,1978PHDT.......195B,
1981A&A....93..106B,1981AJ.....86.1791B,1981AJ.....86.1825B}.

Studies of the rotation curves of the inner parts of galaxies are important in discerning competing
models of dark matter.  The cuspy dark matter density profiles predicted from $\Lambda$CDM
simulations require steeply rising rotation curves \citep{NFW_1997}, while pseudo-isothermal halos
representing a cored potential are seen observationally  \citep{BLOK:FK}. These halo models show
the largest differences in the centres of galaxies.  It is hence essential that we form a better
understanding of the kinematics in the inner regions of galaxies.  The central parts of spiral
galaxies, especially early-types, tend to have little or no H\,{\sc i}, making it difficult to
determine the central dynamics from observations of the H\,{\sc i} alone.  However, molecular gas is
typically concentrated in the centres of galaxies, and CO is an abundant tracer of the molecular gas
($\mathrm{H_{2}}$) content of galaxies. 

Compared to  H\,{\sc i}, CO is more easily observable at higher redshifts
\citep{2013ARA&A..51..105C}, which makes CO an alternative tracer of the dynamics in studies of the
Tully-Fisher relation (TFR).

The Five Colleges Radio Astronomy Observatory (FCRAO) survey \citep{FCRAO-I}, the Berkeley Illinois
Maryland Association Survey of Nearby Galaxies \citep[BIMA-SONG,][]{BIMA-SONG-II} and the recent
survey by \citet{2007PASJ...59..117K} are some of the most comprehensive surveys of the CO content
of spiral galaxies. More recently, the ATLAS-3D \citep{2011MNRAS.414..940Y} survey carried out
observations of early type galaxies, and the NUclei of GAlaxies collaboration \citep[NUGA,][for
example]{2003A&A...407..485G,2004A&A...414..857C,2005A&A...442..479K} studied the molecular
chemistry and dynamics of the inner parts of nearby active galaxies.  

Pioneering studies of the kinematics of spiral galaxies with CO observations were 
presented in \citet{1996APJ...458..120S,SOFUE_1997} and \citet{SOFUE_1999}.  More recently, concerted efforts
have been made to map a representative sample of nearby galaxies in H\,{\sc i} and CO with The H\,{\sc
i} Nearby Galaxy Survey \citep[THINGS,][]{FABIAN-WALTER:2008ZL} and The HERA CO Line Extragalactic
Survey \citep[HERACLES,][]{LEROY:2009FK}, respectively. These surveys, in combination with ancillary
data at other wavelengths, provide a comprehensive view of the gaseous ISM of nearby galaxies. 

In this paper, we study the kinematics of the CO in a sample of the THINGS galaxies using the
HERACLES data, and we extend the analysis of the dynamics performed in \citet{BLOK:2008UQ}
(hereafter dB08), by including the contribution of the molecular gas. We compare our results to the
analysis of the H\,{\sc i} kinematics, aiming to address the following questions: How similar are
the kinematics of the CO and the H\,{\sc i}?   How does the CO-TFR compare with the H\,{\sc i}-TFR
for the same sample of galaxies? What is the effect of adding molecular gas on the derived mass
model parameters?

This paper is organized as follows: in Section \ref{sec:the-data} we describe the CO and H\,{\sc i}
data; in Section \ref{sec:velfields} we describe how we compute the velocity fields. The rotation
curve derivations are described in Section \ref{sec:rotcur-derivation}. In Section \ref{sec:rotcurs}
we present rotation curves. In Section \ref{sec:tfr} we present the TFR. In Section
\ref{sec:massmodels} we outline the motivation and the method used in constructing mass models, and
we present the results of the mass modelling in Section \ref{sec:massmodelresults}. Finally, we
summarize our study of the dynamics of the galaxies in our sample in Section
\ref{sec:discussion-summary}.
\section{The Data}
\label{sec:the-data}
\begin{table*}
	\caption{The THINGS-Model geometrical parameters\label{tab:The-THINGS-Model-Tilted-Ring}.}
	\centering
		\begin{tabular}{c c c c c c r r r}
		\hline \hline
		\multicolumn{1}{c}{Name} & Morphology & $\alpha(2000)$ & $\delta(2000)$ & $D$ &
		$\mathrm{log}(\frac{D_{25}}{0\farcm1}$) & \multicolumn{1}{c}{$V_{\mathrm{sys}}$} & \multicolumn{1}{c}{$\langle i \rangle$} & \multicolumn{1}{c}{$\langle pa \rangle$}\\
		&  & $(\mathrm{^h\,^m\,^s})$ & $(\mathrm{^{^\circ}\,^{'}\,^{''}})$  & ($\mathrm{Mpc}$) & & \multicolumn{1}{c}{($\kmss$)} & \multicolumn{1}{c}{($^\circ$)} & \multicolumn{1}{c}{($^\circ$)} \\
		(1) & (2) & (3)  & (4) & (5) & (6) & \multicolumn{1}{c}{(7)} & \multicolumn{1}{c}{(8)}& \multicolumn{1}{c}{(9)} \\
		\hline 
		NGC 0925 & SABd 	& $02\,27\,16.5$	& $+33\,34\,43.5$		& $9.2$ 	& $2.0$ & $546.3$ & $66.0$ & $286.6$ \\ 
		NGC 2403 & SABcd 	& $07\,36\,51.1$	& $+65\,36\,02.9$ 		& $3.2$ 	& $2.2$ & $132.8$ & $62.9$ & $123.7$ \\ 
		NGC 2841 & SAb 		& $09\,22\,02.6$	& $+50\,58\,35.4$ 		& $14.1$ 	& $1.8$ & $633.7$ & $73.7$ & $152.6$ \\ 
		NGC 2903 & SABbc 	& $09\,32\,10.1$	& $+21\,30\,04.3$ 		& $8.9$ 	& $2.0$ & $555.6$ & $65.2$ & $204.3$ \\ 
		NGC 2976 & SAc pec 	& $09\,47\,15.3$	& $+67\,55\,00.0$ 		& $3.6$ 	& $1.8$ & $1.1$ & $64.5$ & $334.5$ \\ 
		NGC 3198 & SBc 		& $10\,19\,55.0$	& $+45\,32\,58.9$ 		& $13.8$ 	& $1.8$ & $660.7$ & $71.5$ & $215.0$ \\ 
		NGC 3521 & SABbc 	& $11\,05\,48.6$	& $-00\,02\,09.2$ 		& $10.7$ 	& $1.9$ & $803.5$ & $72.7$ & $339.8$ \\ 
		NGC 3627 & SABb 	& $11\,20\,14.9$	& $+12\,59\,29.5$ 		& $9.3$		& $2.0$ & $708.2$ & $61.8$ & $173.0$ \\
		NGC 4736 & SAab 	& $12\,50\,53.0$	& $+41\,07\,13.2$ 		& $4.7$ 	& $1.8$ & $306.7$ & $41.4$ & $296.1$ \\ 
		NGC 5055 & SAbc 	& $13\,15\,49.2$	& $+42\,01\,45.3$ 		& $10.1$ 	& $2.0$ & $496.8$ & $59.0$ & $101.8$ \\ 
		NGC 6946 & SABcd 	& $20\,34\,52.2$	& $+60\,09\,14.4$ 		& $5.9$ 	& $2.0$ & $43.7$ & $32.6$ & $242.7$ \\ 
		NGC 7331 & SAb 		& $23\,57\,49.7$	& $-32\,35\,27.9$  		& $3.9$ 	& $1.9$ & $818.3$ & $75.8$ & $167.7$ \\ 
		\hline 
	\end{tabular}
	\tablecomments{(1) Name of galaxy; (2) Morphology from \protect\url{ned.ipac.caltech.edu}
	(3) right ascension (J2000); (4) declination (J2000), centre positions
	from \citet{TRACHTERNACH:2008FK}; (5) distance as given in \citet{FABIAN-WALTER:2008ZL};
	(6) size as given in \citet{FABIAN-WALTER:2008ZL}; (7) adopted systemic
	velocity; (8) average value of the inclination; (9) average value
	of the position-angle of the receding side, measured from north to
	east and in the plane of the sky; (7), (8) and (9) as presented in
	\citet{BLOK:2008UQ}.}
\end{table*}
In this work we use data from the THINGS and HERACLES surveys. Both surveys
have similar spatial and velocity resolution (HERACLES: $13''$ and
$2.6\kmss$, THINGS: $\sim10''$ and a velocity resolution 
$\leq\,5.2\kmss$).

The observational details of the HERACLES\footnote{http://www.mpia.de/HERACLES/} survey are
described in \citet{LEROY:2009FK}. The survey used the IRAM 30-m telescope to map the CO
$J=2\rightarrow1$ transition (rest frequency $\sim230.538$ GHz) in a sample of nearby galaxies.

We focus on a subset of the HERACLES galaxies for our analysis.  These galaxies were selected as
follows: 34 galaxies were observed as part of the THINGS survey. dB08 performed an analysis of the
dynamics on 19 of the galaxies with intermediate inclinations (i.e., neither face- nor edge-on).  Of
the 19 dB08 galaxies, the 12 that have molecular gas detected in them form our sample.  The general
properties of these galaxies are summarized in Table \ref{tab:The-THINGS-Model-Tilted-Ring}.  These
properties are taken from the THINGS papers: general properties follow from
\citet{FABIAN-WALTER:2008ZL}, centre positions follow from \citet{TRACHTERNACH:2008FK} and
kinematical data (e.g., inclination) follow from dB08.

We use the THINGS\footnote{http://www.mpia.de/THINGS/} date cubes derived using natural weighting and the
associated data products. 

\begin{figure*}
	\centering
	\includegraphics[width=17cm]{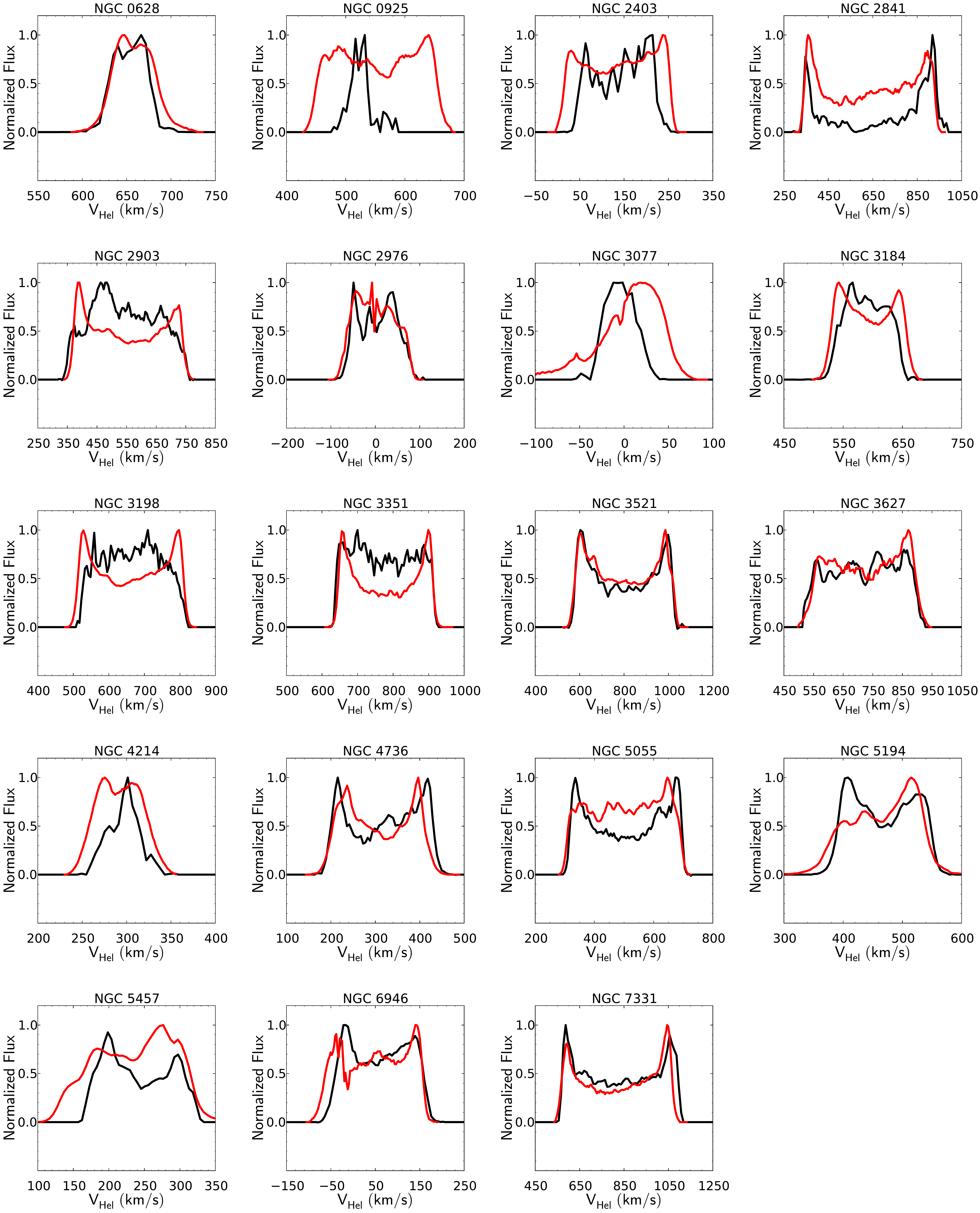}
	\caption{CO (black) and H\,{\sc i} (red) emission line global profiles 
	from the HERACLES and THINGS surveys respectively. The profiles are 
	normalized by their peak fluxes.}
	\label{fig:hico-profiles}
\end{figure*}

\begin{figure*}
	\centering
	\includegraphics[width=17cm]{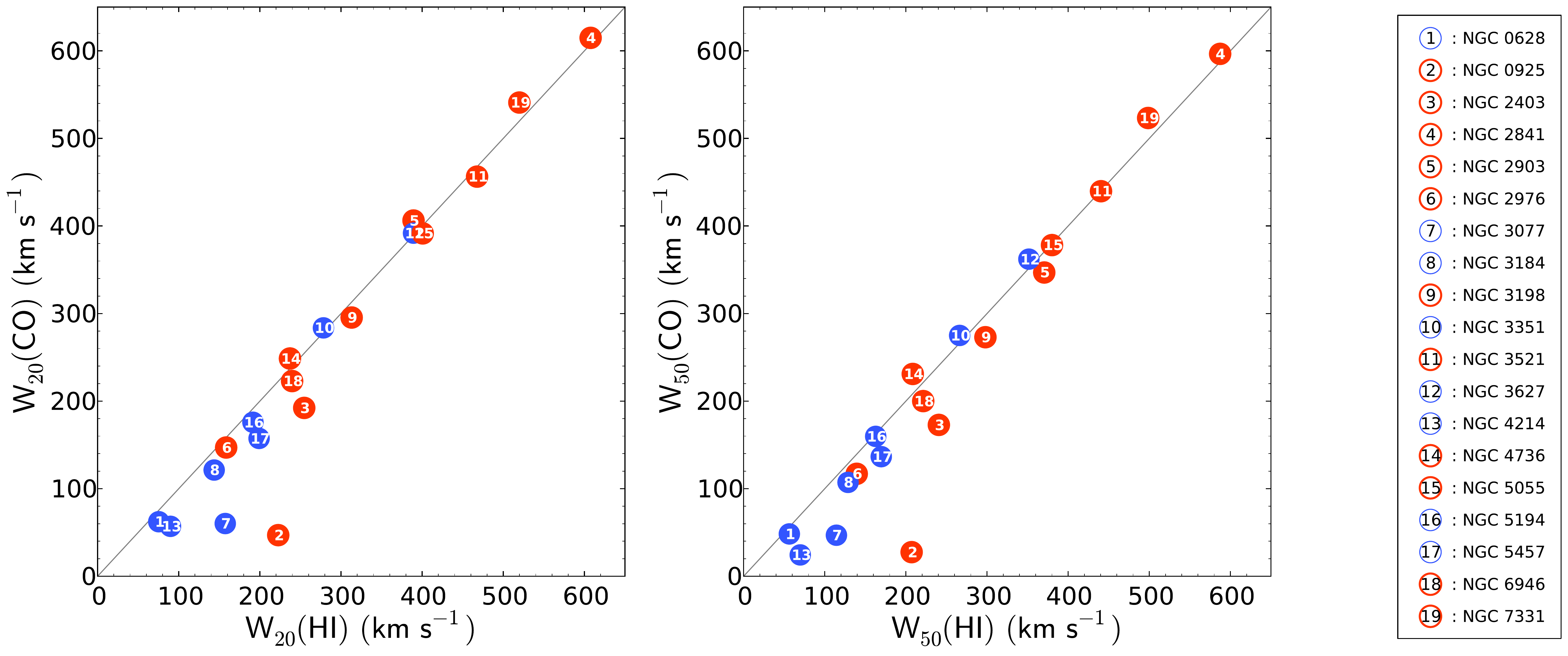}
	\caption{\label{fig:hico-w20w50}A comparison of the CO and H\,{\sc i} linewidths at
	$20\%$ and $50\%$ of maximum intensity, respectively. The left subplot
	shows the $W_{20}$ comparison and the right subplot shows the $W_{50}$
	comparison. The line indicates a unity ratio. The number in each circle corresponds to a galaxy in our
	sample, as shown in the legend on the right. The colour of each circle
	corresponds to whether or not the galaxy is part of the sub-sample for which we will perform a
	kinematical analysis in this paper --- galaxies in our sample are coloured in red symbols; the
	others are plotted blue. The typical uncertainties in the linewidths are approximately half the
	channel width of $2.6\kmss$. The corresponding error bars, if plotted on this scale, will be
	smaller than the markers. }
\end{figure*}

The velocity resolution of HERACLES was $2.6\kmss$ and Hanning smoothing was used to increase the
signal-to-noise for various galaxies. This resulted in effective velocity resolutions of $5.2\kmss$
for NGC 925, NGC 2403, NGC 2903, NGC 2976, NGC 3198, NGC 3627, NGC 4736, NGC 5055, NGC 6946 and
$10.4\kmss$ for NGC 2841, NGC 3521 and NGC 7331. In all cases the velocity smoothing makes a
negligible difference to the emission line profile and the data products derived from the cubes. The
spatial resolution in these cubes is $13''$.  We only consider channels within the velocity range
bound by the H\,{\sc i} emission as estimated using the H\,{\sc i} position-velocity \textit{pV}
diagram. We then masked the cubes using the masks determined in \citet{LEROY:2009FK} and
recalculated the average noise-per-channel in regions which did not contain any CO emission.  These
values are slightly different from those derived in \citet{LEROY:2009FK}, since the noise is a
function of frequency and therefore the value depends slightly on the channels chosen. In Table
\ref{tab:noise} we present the noise values used in this work, compared with those from
\citet{LEROY:2009FK}.  The masking left a few anomalous pixels corresponding to noise-peaks, which
we carefully removed with a round of manual masking. We used these cubes to derive the CO global
profiles which we compare with the  H\,{\sc i} profiles as published in
\citet{FABIAN-WALTER:2008ZL}. In Figure \ref{fig:hico-profiles} we plot the H\,{\sc i} and CO global
profiles for all the THINGS galaxies detected by HERACLES, i.e., also including the galaxies not in
our sample. In Figure \ref{fig:hico-w20w50} we plot a comparison of the CO and H\,{\sc i} linewidths
at the $20\%$ and $50\%$ levels respectively. Some profile shapes lead to ambiguous definitions of
$W_{50}$, e.g., NGC 2903 and NGC 2976. For these galaxies we choose the largest value. Figure
\ref{fig:hico-w20w50} shows that the CO linewidths are lower than the H\,{\sc i} linewidths, on
average.  This is because the distribution of the CO does not extend as far as the H\,{\sc i} and
does not trace the full intrinsic velocity-width of a galaxy, \citep[see, e.g.,][]{BLOK:2014AA}. 

\begin{table}
	\caption{\label{tab:noise}Noise in H\,{\sc i} and CO data cubes.}
	\centering
	\begin{tabular}{c c c c}
		\hline \hline
		Galaxy & $\sigma_{\mathrm{CO}}$ & $\sigma_{{\mathrm {CO'}}}$  & $\sigma_{{\rm {H\sc{I}}}} $ \\
		   &  (${\rm {mK}}$) &  (${\rm {mK}}$) & ${\mathrm {(mJy\,beam^{-1}})}$ \\
		   (1) & (2) & (3) & (4) \\
		\hline
		NGC 0925 & 22 & 20 & 0.57 \\
		NGC 2403 & $\ldots$ & 24 & 0.38 \\
		NGC 2841 & 44 & 26 & 0.65 \\
		NGC 2903 & 23 & 24 & 0.41 \\
		NGC 2976 & 21 & 24 & 0.36 \\
		NGC 3198 & 17 & 20 & 0.33 \\
		NGC 3521 & 22 & 25 & 0.40 \\
		NGC 3627 & $\ldots$ & 27 & 0.43 \\
		NGC 4763 & 23 & 23 & 0.33 \\
		NGC 5055 & 24 & 34 & 0.36 \\
		NGC 6946 & 25 & 29 & 0.55 \\
		NGC 7331 & 20 & 21 & 0.44 \\
		\hline
	\end{tabular}
	\tablecomments{(1) Name of Galaxy; (2) noise in CO cube as in
	\citet{LEROY:2009FK}, except for NGC 2403, which was not included in that paper;
	(3) Noise in CO cube as measured in present work; (4) Noise in H\,{\sc i} natural weighted 
	cube as in \citet{BLOK:2008UQ}.}
\end{table}
\section{Velocity Fields}
\label{sec:velfields}
Returning to the rotation curve sample listed in Table \ref{tab:The-THINGS-Model-Tilted-Ring} --- we
use the Hanning-smoothed HERACLES cubes to derive velocity fields for the CO emission.  There are
two commonly used methods to compute velocity fields from image cubes - calculating the Intensity
Weighted Mean (IWM) of profile values and fitting functions (e.g., Gaussians) to the profiles in an
image cube.  For asymmetric profiles the IWM of a profile can be affected by the presence of tails
to higher and lower velocities, and hence does not always provide an accurate representation of the
gas velocity. We therefore fit Gauss-Hermite polynomials of order 3 to the profiles along each pixel
in the image cubes, using the prescription described in \citet{VANDERMAREL_FRANX_1993}, and as in
dB08.  This allows us to account for asymmetry in the profiles, hence determining a more accurate
estimate of the gas bulk velocity. We denote these Gauss-Hermite velocity fields as ``$\her3$
velocity fields''. We use the \texttt{GIPSY} \citep[Groningen Image Processing
SYstem,][]{1992ASPC...25..131V} task \texttt{XGAUFIT} to compute the $\her3$ velocity fields from
the masked data as described above.

We also compute the IWM velocity fields from the masked HERACLES cubes using the \texttt{GIPSY} task
\texttt{MOMENTS}. In the derivation of both the IWM and $\her3$ velocity fields we reject values
less than $3\,\sigma_{\mathrm{CO'}}$ where $\sigma_{\mathrm{CO'}}$ is the average noise-per-channel
indicated in Table \ref{tab:noise}.

In the appendix we plot the IWM and $\her3$ velocity fields for each galaxy in our sample, as well
as the H\,{\sc i} velocity fields calculated in dB08.
\section{Rotation Curve Derivation}
\label{sec:rotcur-derivation}
We use the $\her3$ velocity fields to calculate the rotation curves using a tilted-ring model
\citep{BEGEMAN:1989QF}. In the tilted-ring analysis, a two-dimensional velocity field of a galaxy is
decomposed into a set of rings, each with an associated set of  parameters: the systemic velocity
$V_{\mathrm{sys}}$, the centre position $(x_{0},y_{0})$ on the sky, the inclination angle $i$
defined as the angle between the normal to the plane of the galaxy and the line-of-sight, the
position angle $pa$ of the major-axis on the sky, and the circular velocity $V_{c}$. A tilted-ring
model is thus characterized by a set of parameters
$(x_{o},y_{o},V_{\mathrm{sys}},i,pa,V_{\mathrm{c}})$ for each ring and can therefore vary with
radius.

Assuming the gas moves in purely circular orbits, the observed line-of-sight velocity $V$ at an
arbitrary position $(x,y)$ on a ring of radius $r$ can be expressed as:
\begin{equation}
	V(x,y)=V_{\mathrm{sys}}+V_{\mathrm{c}}(r)\sin(i)\cos(\theta)\label{eq:ROTCUR}
\end{equation}
where $\theta$ is the azimuthal or position angle with respect to
the receding major-axis, measured in the plane of the galaxy. It is
related to the position angle of the major-axis in the plane of the
sky by the following relations:
\begin{equation}
	\cos(\theta)=\frac{-(x-x_{0})\sin(pa)+(y-y_{0})\cos(pa)}{r}
\end{equation}
\begin{equation}
	\sin(\theta)=\frac{-(x-x_{0})\cos(pa)-(y-y_{0})\sin(pa)}{r\cos(i)}
\end{equation}
For each ring, the parameters are solved using a least squares algorithm
to obtain an optimal fit.

We use the \texttt{GIPSY} task \texttt{ROTCUR} to calculate rotation curves from the $\her3$
velocity fields, solving for kinematic parameters along rings which are spaced by half-a-beamwidth,
i.e., Nyquist sampled.  We use the filling-factor for each ring to determine whether to include it in
the tilted-ring fit. We define the filling factor as the ratio of the area with significant
signal and the total area of each ring. A filling-factor of $5\%$ is used as a cutoff, since solving
for parameters on rings with lower filling-factors does not lead to useful results.

We calculate two different types of rotation curves. Firstly, we use the tilted-ring model
parameters as determined in dB08: we fix them and simply apply them to the CO velocity field solving
for $V_{\mathrm{c}}$ only. We refer to these models as the THINGS models (TM). For galaxies where
there is a central depression in the H\,{\sc i} distribution and therefore no corresponding values
(e.g., NGC 7331), we extrapolate the tilted-ring parameters from the innermost point by assuming
that these parameters remain constant. 

Secondly, we determine the tilted-ring model for a subset of the galaxies in our sample for which
the CO coverage is sufficiently good to constrain a separate tilted-ring model. We refer to these
models as the HERACLES models (HM). For the HM we use an iterative method, where we use TM as a
starting point. We then alternate between solving for $(V_{\mathrm{sys}},x,y)$ and $(pa,i)$, each
time holding the other set of parameters fixed.  We repeat this until the model parameters converge.
We assume that the centre and the systemic velocity of the galaxy do not vary from ring to ring.  We
smooth radially varying position-angle and inclination values using a five-point boxcar smoothing
algorithm to suppress small scale variations. After the parameters have converged, we fix them and
solve for $V_{\mathrm{c}}$ alone.

To estimate the uncertainties along each ring we use the same prescription as dB08. The errors are
defined as the quadratic sum of the dispersion in velocity values along each tilted-ring and the
uncertainty due to the approaching and receding side of the galaxy, which is defined as a quarter of
the difference. The rotation curves from the approaching and receding sides are calculated by
running \texttt{ROTCUR} on the corresponding half of the galaxy only. In general, we find the
dispersion in velocities to be the larger contributor to the total error.

For the galaxies in the dB08 sample HMs could be derived for NGC 2403, NGC 2976, NGC 3198, NGC 3521,
NGC 5055 and NGC 7331. For the other galaxies there is either too little detected CO emission to
constrain the models (for NGC 925 and NGC 2841), or the low inclinations prohibits the independent
constraint of the kinematic parameters (for NGC 4736 and NGC 6946). For NGC 2903 the effect of the
strong bar makes it difficult to solve for a tilted-ring model since the CO velocities are heavily
influenced by bar streaming motions. NGC 3627 is part of the Leo Triplet and the observations show
signs of tidal interaction with the neighbouring galaxies. For these six galaxies we therefore only
calculate a TM. 

\subsection{Beam Smearing}
The high resolution of HERACLES allows us to study the distribution of the CO in great detail.
However, despite the relatively high spatial resolution, the finite beam size can still lead to beam
smearing effects near the central parts of the galaxies. This would affect the derived rotation curve and
becomes more significant for galaxies at a high inclination. To quantify the possible effect that
beam smearing may have on the derived rotation curves, we did a simple study using model galaxies.
We used the \texttt{GIPSY} task \texttt{GALMOD} to construct model galaxies with a constant
inclination of $70^{\circ}$, a Gaussian scale height of $100\,\mathrm{pc}$, a constant 
column density across the disk of $2\times 10^{21}\,\mathrm{cm}^{-2}$ and a velocity dispersion of
$8\kmss$. This model is very similar to that presented in dB08.

The rotation curve used as input to \texttt{GALMOD} rises linearly to a maximum of $250\kmss$, and
stays constant at this maximum for the rest of the disk. We used four different rotation curves to
quantify the effect of beam smearing. We varied the radius at which the rotation curve reached the
maximum velocity, using radii of $6.5''$, $13''$, $26''$ and $60''$ respectively. With these input
rotation curves we can study how the effect of beam smearing changes when the rotation curve reaches
a maximum within a resolution element, to when the rotation curve reaches its maximum at larger
radii. 

\begin{figure}
	\centering
	\resizebox{\hsize}{!}{\includegraphics{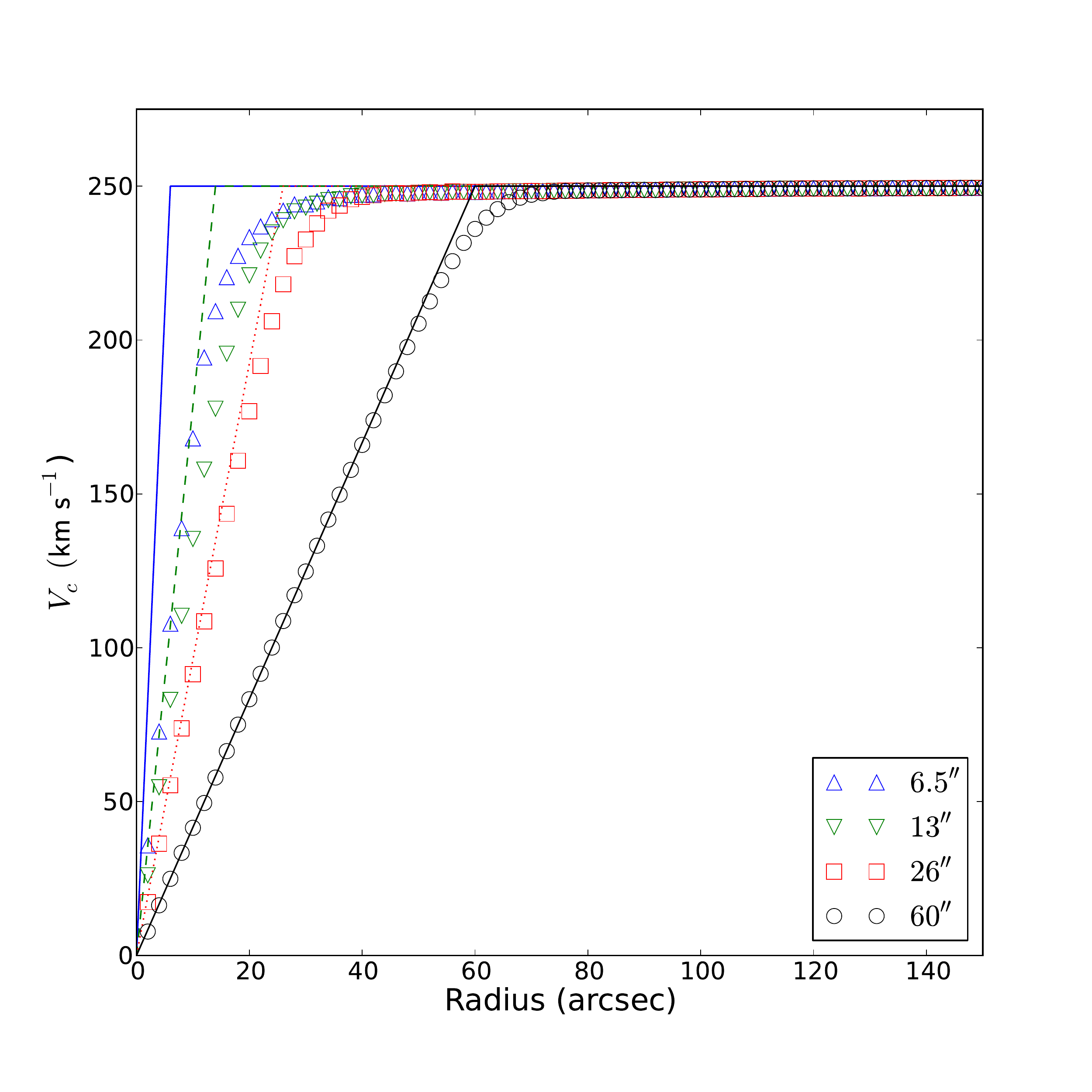}}
	\caption{\label{fig:beam-smearing}Comparison of model rotation curves with resultant
	observed rotation curves. The blue-solid, green-dashed, red-dotted and black-solid lines
	represent input rotation curves with a linear rise to a maximum velocity of $250\kmss$
	within $6.5''$, $13''$, $26''$ and $60''$ respectively. The blue upward-triangles, green
	downward-facing triangles, red squares and black circles represent the observed output
	rotation curves derived from each of the associated input rotation curves.}
\end{figure}

\texttt{GALMOD} produces model data cubes given these galaxy parameters. We smoothed these
data cubes to the $13''$ resolution of HERACLES, and computed the $\her3$ velocity fields and
rotation curves from the ``observed'' galaxy cubes. The results are presented in Figure
\ref{fig:beam-smearing}. 

This shows that while beam smearing has a large effect on the observed rotation curve when the input
rotation curve rises to a maximum within a resolution element, it becomes negligible when the input
rotation curve rises to a maximum at radii larger than two resolution elements, as shown in Figure
\ref{fig:beam-smearing}. This is the case for all the galaxies in our sample. Furthermore, the
inclination of the galaxies in our sample are less than $70^\circ$, except for NGC 7331. Beam
smearing is thus expected to play only a negligible here.

\section{Results and Discussion - Kinematics}
\label{sec:rotcurs}
\begin{figure*}
	\centering
	\includegraphics[origin=lb,width=17cm]{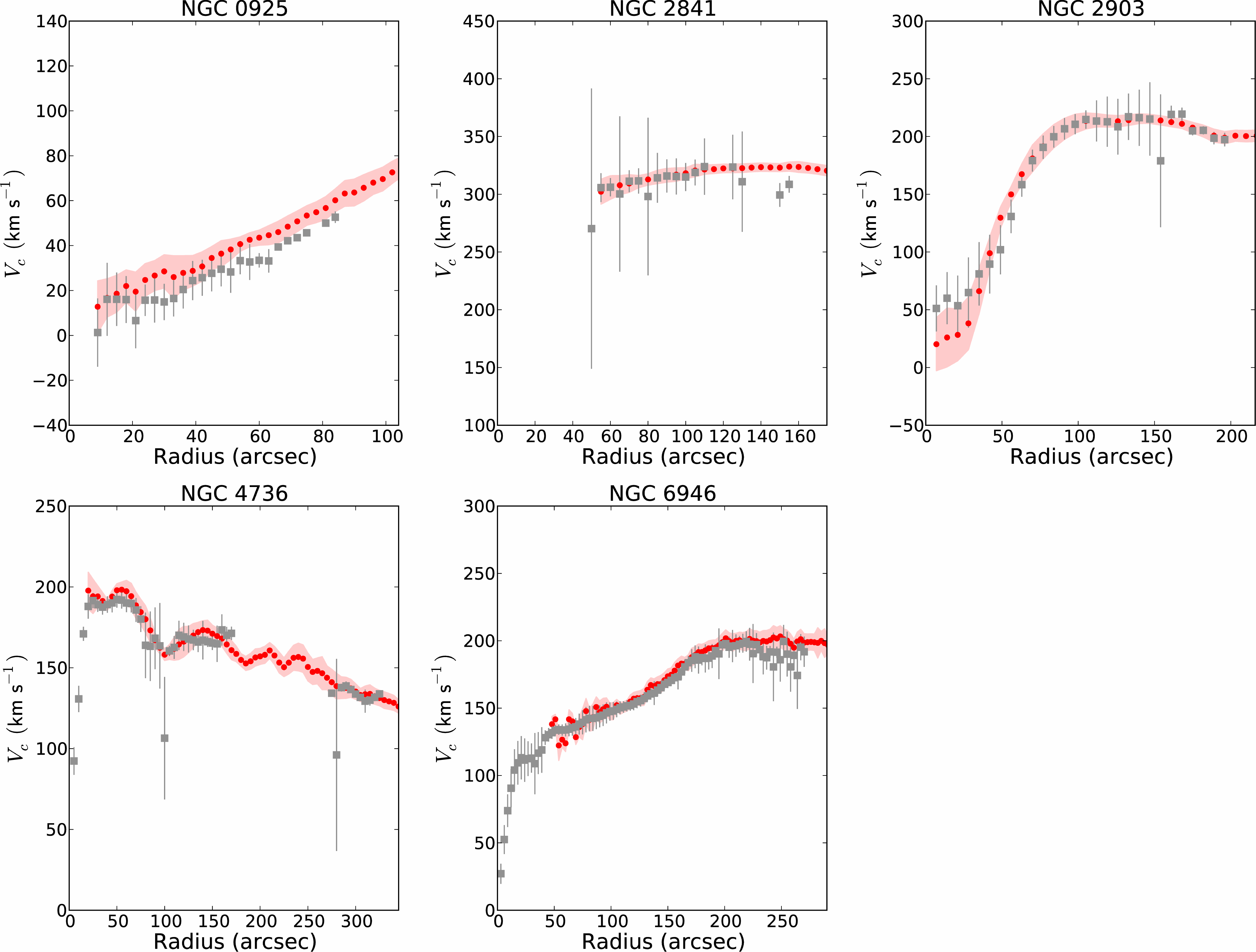}
	\caption{\label{fig:all-tm-rotcurs}CO and H\,{\sc i} rotation curves for
	NGC 925, NGC 2841, NGC 2903, NGC 3627, NGC 4736 and NGC 6946. The H\,{\sc i} rotation curves
	from dB08 are plotted as filled red circles; the associated errors are plotted as a filled
	red region.  The CO rotation curves are computed using the TM, i.e., using the tilted-ring
	model parameters based on the H\,{\sc i} data, and are plotted as filled grey squares
	with errorbars.}
\end{figure*}

\begin{figure*}
	\centering
	\includegraphics[origin=lb,width=17cm]{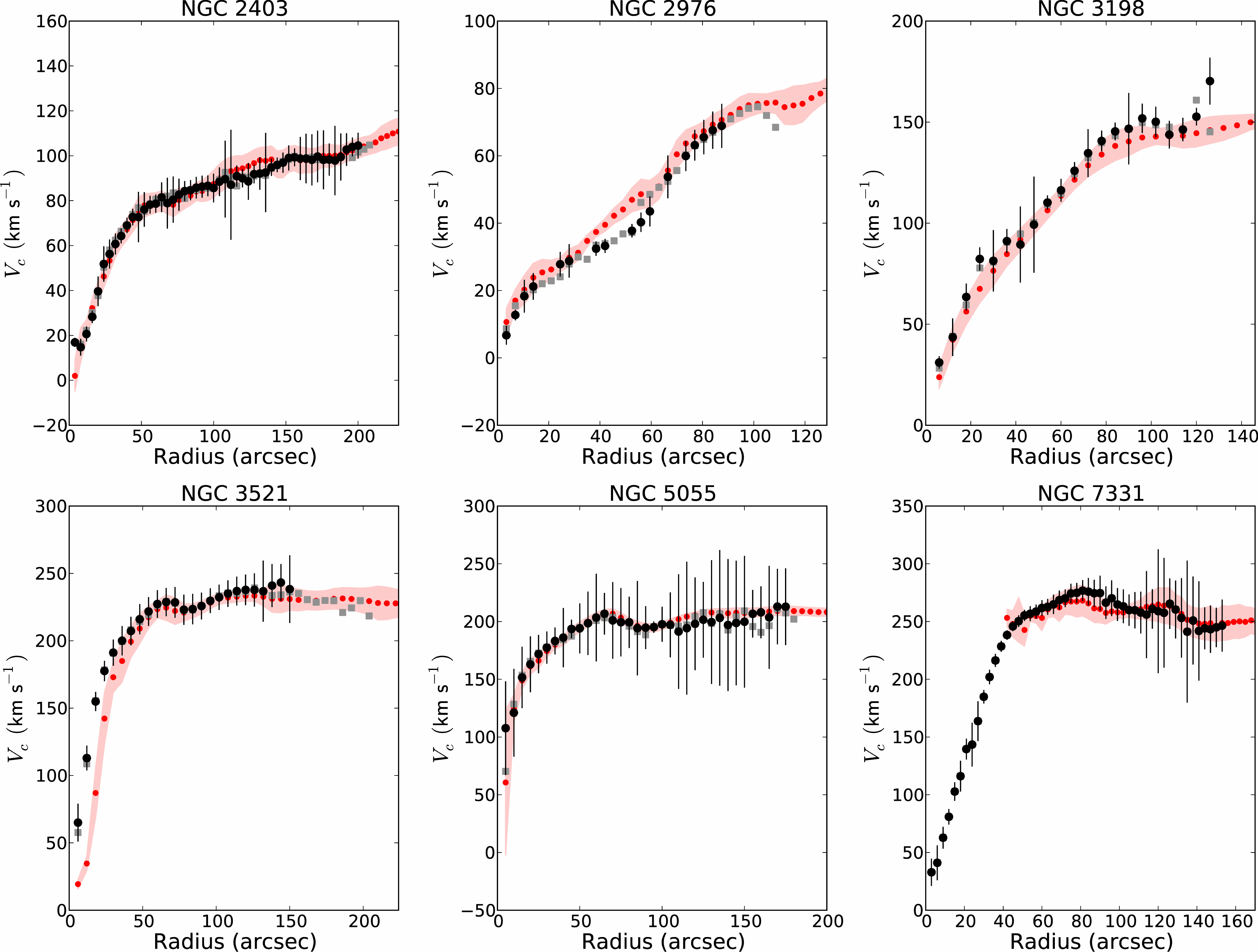}
	\caption{\label{fig:all-hm-rotcurs}CO and H\,{\sc i} rotation curves for NGC 2403, NGC 2976,
	NGC 3198, NGC 3521, NGC 5055 and NGC 7331. The H\,{\sc i} rotation curves from dB08 are
	plotted as filled red circles; the associated errors are plotted as a filled red region. The
	CO rotation curves are computed using the THINGS-Model (TM) and the independently determined
	HERACLES-Model (HM). The TM rotation curves are plotted as filled grey squares, HM
	rotation curves are plotted as filled black circles with errorbars.}
\end{figure*}

The rotation curves are plotted in Figures \ref{fig:all-tm-rotcurs} (TM) and
\ref{fig:all-hm-rotcurs} (HM) respectively. The corresponding HM tilted-ring parameters for NGC
2403, NGC 2976, NGC 3198, NGC 3521, NGC 5055 and NGC 7331 are presented in Table
\ref{tab:the-hera-mod-tilted-ring}. In the appendix we present our full results: H\,{\sc i} and CO
velocity fields (IWM and $\her3$); CO integrated surface brightness contours overlaid on the Spitzer
Infrared Nearby Galaxy Survey \citep[SINGS]{2003PASP..115..928K} 3.6$\mu$m images, \textit{pV}
major- and minor-axis diagrams and H\,{\sc i} and CO TM/HM rotation curves. For each of the
tilted-ring models considered in this work we compute an associated model velocity field and we
calculate the residuals by taking the difference between the model and observed velocity fields.  As
the residuals generally have a Gaussian distribution (except where there are major non-axisymmetric
features, such as bars), we fit a Gaussian function to the normalised histogram of the residuals to
estimate the mean $\mu$ and the standard deviation $\sigma$. These values are presented in Table
\ref{tab:residual-summary}. In the appendix we also provide a brief description of the rotation
curves for each galaxy. We also fit a functional form to the CO rotation curves presented in this
work, which is also presented in the appendix. Using such a functional form allows for the
computation of the derivative of the rotation curve which is important in determining the star
formation threshold, for example \citep[see, e.g., ][]{2008AJ....136.2782L}.

\begin{table*}
	\caption{\label{tab:residual-summary}Statistics from the tilted-ring model
	fits to the $\mathrm{Her_{3}}$ velocity fields derived from the HERACLES
	data.}
	\centering
		\begin{tabular}{c r r r r r}
		\hline \hline
                Name &  $\sigma_{\mathrm{TM}}$ & $\mu_{\mathrm{TM}}$  & $\sigma_{\mathrm{HM}}$ & $\mu_{\mathrm{HM}}$  \\
		\	& $(\kmss)$ & $(\kmss)$ &  $(\kmss)$ & $(\kmss)$  \\
		\hline
		NGC 0925 & 5.26 & 2.46 & $\ldots$ & $\ldots$   \\
		NGC 2403 & 5.51 & 0.03 & 5.72 & -1.47   \\
		NGC 2841 & 7.02 & 1.61 & $\ldots$ & $\ldots$   \\ 
		NGC 2903 & 12.7 & -2.06 & $\ldots$ & $\ldots$   \\
		NGC 2976 & 3.85 & -3.44 & 2.84 & 0.66   \\ 
		NGC 3198 & 4.38 & -3.04 & 4.93 & -1.27   \\ 
		NGC 3521 & 7.92 & 0.57 & 5.45 & -1.58   \\ 
		NGC 3627 & 13.3 & -2.57	& $\ldots$ & $\ldots$ \\
		NGC 4736 & 5.19 & 1.42 & $\ldots$ & $\ldots$   \\ 
		NGC 5055 & 6.31 & -2.40 & 6.09 & -3.34   \\ 
		NGC 6946 & 5.83 & -1.41 & $\ldots$ & $\ldots$   \\ 
		NGC 7331 & 14.7 & -3.37 & 11.96 & -1.17   \\ 
		\hline
	\end{tabular}
	\tablecomments{The symbols $\mu$ and $\sigma$ denote the mean and standard deviation of the
	best fitting Gaussian fit to a histogram of the residuals, respectively. TM denotes the
	model from dB08 as derived from the THINGS data. HM denotes the tilted-ring models derived in this
	work from the HERACLES data.}
\end{table*}

\begin{table*}
	\caption{The HERACLES-Model tilted-ring parameters\label{tab:the-hera-mod-tilted-ring}}
	\centering
	\begin{tabular}{c c c c c c}
		\hline \hline
		Name & $\alpha\,(2000)$ & $\delta(\,2000)$ & $V_{\mathrm{sys}}$ & $\langle i \rangle $ & $\langle pa \rangle$ \\
			& $(\mathrm{^h\,^m\,^s})$& $(\mathrm{^{^\circ}\,^{'}\,^{''}})$ & ($\kmss$) & ($^\circ$) & $(^\circ)$ \\
		\hline 
		NGC 2403 & 07 37 15.8 & +65 30 17.1 & 132.9 & 59.2 & 120.2\\ 
		NGC 2976 & 09 47 26.4 & +67 51 04.2 & 1.1 & 65.7 & 338.8\\ 
		NGC 3198 & 10 19 42.7 & +45 36 40.3 & 659.9 & 68.3 & 211.1\\ 
		NGC 3521 & 11 06 01.3 & +00 02 25.7 & 802.3 & 67.4 & 120.0\\ 
		NGC 5055 & 13 15 15.7 & +42 04 16.8 & 497.0 & 63.6 & 98.5\\ 
		NGC 7331 & 22 37 05.3 & +34 30 07.7 & 822.8 & 72.0 & 167.1\\ 
		\hline 
		\end{tabular}
	\tablecomments{(1) Name of galaxy; (2) Right Ascension (J2000);
	(3) Declination (J2000), Centre positions as determined in this
	work; (4) Systemic velocity as determined in this work; (7) Average
	value of the inclination; (8) Average value of the position-angle
	of the receding side, measured from north to east and in the plane
	of the sky.}
\end{table*}
For NGC 2403, NGC 2841, NGC 3627, NGC 4736, NGC 6946 and NGC 7331 there is excellent agreement
between the H\,{\sc i} and CO rotation curves.  For NGC 925, NGC 2903, NGC 2976, NGC 3198, NGC 3521
and NGC 5055 there are some differences which can be explained by the presence of bars and the
lopsided emission of the CO in comparison to the H\,{\sc i}.  For NGC 4736, NGC 6946 and NGC 7331
the CO rotation curves also covers the inner part of the galaxy not traced by the H\,{\sc i}
emission. We conclude that CO is a good tracer of the rotation curve in the inner part of galaxies.
For NGC 2903 there is a clear indication of the effect of the bar on the velocity field. In the
following sub-section we present a brief analysis of the non-circular motions in NGC 2903. 

Detailed comments about the kinematics and rotation curves for each galaxy are presented the Appendix.

\subsection{Non-circular Motions in NGC 2903}
\label{sec:N2903-df}
NGC 2903 hosts a strong bar, which is closely aligned to the major axis of the disk. The inner part
of the  H\,{\sc i} distribution is strongly affected by the non-circular streaming motions due to
the bar, as noted in \citet{TRACHTERNACH:2008FK}. They present an analysis of non-circular motions
in the THINGS galaxies and  estimate the amplitude of the higher order coefficients in a Fourier
series description of the velocity field.  These coefficients can be interpreted as perturbations
due to physical effects, such as radial motions or streaming motions - as due to a bar, for example.
While the amplitudes of the non-circular motions are estimated by \citet{TRACHTERNACH:2008FK},
they do not discuss how these might affect the circular rotation curve. \citet{SPEKKENS_2007} use a
Fourier series expansion to describe the radial and tangential components of non-circular motions as
higher order perturbations, and explicitly correct for these effects on the rotation curve. This
formalism was coded into the software tool \texttt{VELFIT}.  \citet{SELLWOOD_SZ_2010} used
\texttt{VELFIT} to derive a model velocity field and corrected rotation curve using a bar-like
($m=2$ in the Fourier expansion) perturbation to the THINGS H\,{\sc i} velocity field and the
BH$\alpha$Bar $\mathrm{H\alpha}$ velocity field \citep{2005MNRAS.360.1201H} for NGC 2903. We used
an updated version of the \texttt{VELFIT} tool called
\texttt{DISKFIT}\footnote{http://www.physics.rutgers.edu/$\sim$spekkens/diskfit/}
\citep{2012MNRAS.427.2523K,2015arXiv150907120S} to model the HERACLES velocity field of NGC 2903.  

Using the CO total intensity image and velocity field we estimated the radial extent of the bar to be 
approximately $56''$. We solve for a bar-like perturbation corresponding
to the $m=2$ mode within $56''$ (in addition to circular rotation) and we assume that the gas is
regularly rotating at larger radii. We also solve for the systemic velocity, disk position angle
$\phi'_\mathrm{d}$, inclination angle and central position. The resultant velocity field is further
characterized by the bar position angle in the disk-plane (denoted as $\phi_\mathrm{d}$) and the sky-plane
(denoted as $\phi'_\mathrm{b}$).  

Because of its higher signal-to-noise, we used the IWM velocity field to solve for the non-circular
motions. Tests show the output rotation curves when using either the IWM and $\her3$
velocity fields are identical.

\begin{figure}
	\centering
	\resizebox{\hsize}{!}{\includegraphics{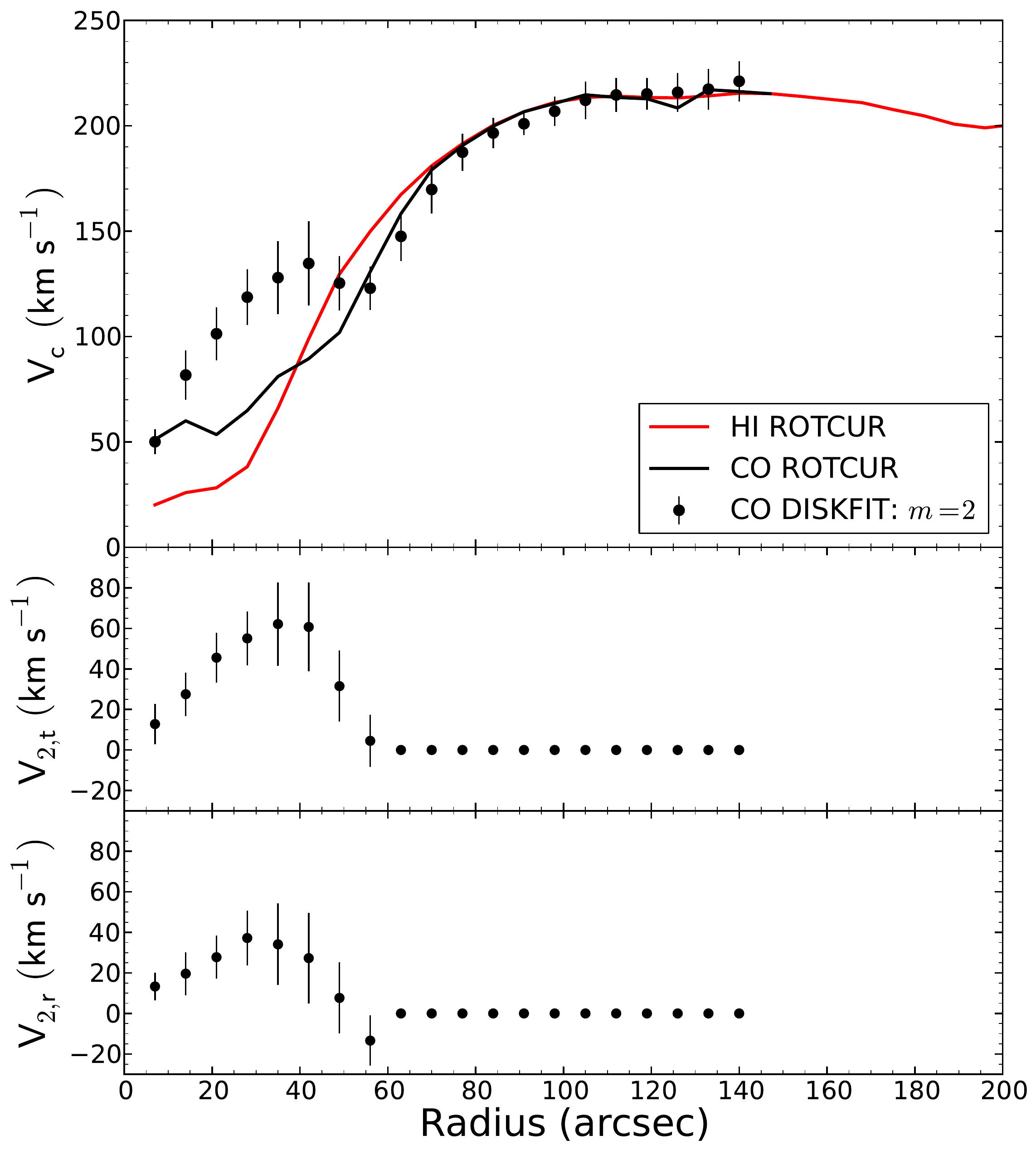}}
	\caption{\label{fig:diskfit}\texttt{DISKFIT} model rotation curves for NGC 2903 for an $m=2$
	perturbation. In the top panel we plot the observed rotation curves from the THINGS data
	(red-line) and the HERACLES data (black-line) using \texttt{ROTCUR}. In this model we solve
	and correct for the $m=2$ modes in the velocity field by assuming that the non-circular
	motions are dominant within $56''$. The black filled circles with errorbars correspond to
	the \texttt{DISKFIT} rotation curve. In the second panel we plot the tangential component for the
	$m=2$ mode, denoted as $V_\mathrm{2,t}$. In the third panel we plot the radial component for the
	$m=2$ mode, denoted as $V_\mathrm{2,r}$.}
\end{figure}

The resulting best fit parameters and the associated uncertainties from our \texttt{DISKFIT} run are
presented in Table \ref{tab:diskfit}, along with the values presented in \cite{SELLWOOD_SZ_2010}. We plot the
rotation curve, $V_\mathrm{t}$, and the tangential and radial components of the $m=2$ mode, denoted as
$V_\mathrm{2,t}$ and $V_\mathrm{2,r}$ respectively, in Figure \ref{fig:diskfit}.

We note a relatively good agreement between our parameters and those presented in
\cite{SELLWOOD_SZ_2010}. The best fit inclination and disk position angle are consistent with the
THINGS values, and the systemic velocity is slightly higher than the THINGS value and the
\cite{SELLWOOD_SZ_2010} values. The best fit central position is identical to the coordinates
presented in Table \ref{tab:The-THINGS-Model-Tilted-Ring}. The bar position angles are different
from the values in \cite{SELLWOOD_SZ_2010}, but agree within the uncertainties.
\cite{SELLWOOD_SZ_2010} noted that the close alignment of the bar axis with the major axis of the
disk made the modelling of the bar perturbation difficult and, therefore, the bar position
angles solved for in this work still have a large uncertainty. We will not consider the bar in our
further models.

\begin{table*}
	\caption{\label{tab:diskfit}\texttt{DISKFIT} results for NGC 2903.}
	\centering
		\begin{tabular}{c c c c}
		\hline
		Parameter & HERACLES\tablenotemark{a} & THINGS\tablenotemark{b} & BH$\alpha$Bar\tablenotemark{b}  \\
		\hline \hline
		$V_{\mathrm{sys}}\,(\kmss)$ & $558.8\pm1.2$ & $549.9\pm0.3$ & $554.1\pm0.5$\\
		$i\ (^{\circ})$ 		& $64.5\pm4.8$ & $64.0\pm1.0$ & $66.0\pm3$ \\
		$\phi'_\mathrm{d}\ (^{\circ})$ 	& $206.1\pm1.9$ & $201.5\pm0.5$ & $204.0\pm1$ \\
		$\phi_\mathrm{b}\ (^{\circ})$ 	& $17\pm8$ & $6\pm14$ & $-12\pm8$ \\
		$\phi'_\mathrm{b}\ (^{\circ})$ 	& $214\pm7$ & $204\pm6$ & $199\pm4$ \\
		\hline
		\end{tabular}
	\tablecomments{Comparison of the best fitting bar and disk parameters solved in this
	work with those presented in \cite{SELLWOOD_SZ_2010}. 
	\tablenotemark{a}{As determined in this work}
	\tablenotemark{b}{From \cite{SELLWOOD_SZ_2010}}}
\end{table*}

\section{The Tully-Fisher Relation}
\label{sec:tfr}
As in Section \ref{sec:the-data}, we again consider all galaxies overlapping in the THINGS/HERACLES
sample detected CO.  We present a comparison of the CO and H\,{\sc i} TFRs for the galaxies.   Since
the galaxies studied here are comparatively massive, we restrict our study to the classical TFR and
not the baryonic TFR \citep{2000ApJ...533L..99M}. \citet{BLOK:2014AA} demonstrated how the
distribution of the gas tracer leads to different global profiles and hence different TFRs even for
identical rotation curves. In general, the CO has a compact distribution, while the more constant
H\,{\sc i} surface density extends to beyond the optical radius of the galaxy.  This can also be
seen in the radial mass surface densities of the molecular gas and the H\,{\sc i}, as shown in
\citet{2008AJ....136.2782L} - the shape of the $\mathrm{H_2}$ distribution closely follows the
exponential stellar distribution, while the H\,{\sc i} distribution remains relatively constant over
a large range of radii. Here we use the CO linewidths as plotted in Figure \ref{fig:hico-profiles}
to show the effect of these differences on the TFR.

Many profiles are double-horned with steep sides, which is indicative of massive spiral
galaxies. We exclude two profiles for two galaxies where the linewidths cannot be used as a proxy
for the maximum rotational velocity --- NGC 925 and NGC 3077. For NGC 925 the emission is one-sided,
and for NGC 3077 the gas is tidally disturbed.

We make an inclination correction to the CO and H\,{\sc i} global profiles. For the H\,{\sc i}
profiles we simply use the THINGS inclinations presented in \citet{FABIAN-WALTER:2008ZL}, except for
NGC 5194, for which we use the improved inclination from \citet{2014APJ...784....4C}. For the CO profiles we
also use these inclinations except for the galaxies for which we have HM models. For these galaxies
we use the average inclinations given in Table \ref{tab:the-hera-mod-tilted-ring}. The inclination
corrected rotation velocities are denoted by $V^i$, assuming that the linewidths can be used as a proxy for
the rotational velocity through $V^i=\frac{W^i}{2}$.  

\citet{2001APJ...563..694V} calculated the TFR for galaxies in the Ursa Major Cluster of galaxies,
and found a reduced scatter in the $K'$-band. We therefore calculate the infrared TFR with data
from the Spitzer Survey of Stellar Structure in Galaxies \citep[S$^4$G, ][]{Sheth:2010jh}, and
compare this with the TFR presented in \citet{Sorce:2014by}.

Using the distances from \citet{FABIAN-WALTER:2008ZL} we convert the $3.6 \mu \mathrm{m}$ apparent
magnitudes from  S$^4$G to absolute magnitudes. NGC 2403, NGC 6946 and NGC 7331 were not part of the
S$^4$G sample. We therefore exclude them from the TFR presented here. Our tests of the $B$-band TFR
show that omitting these galaxies does not significantly affect the scatter. 

\begin{figure*}
	\centering
	\includegraphics[width=17cm]{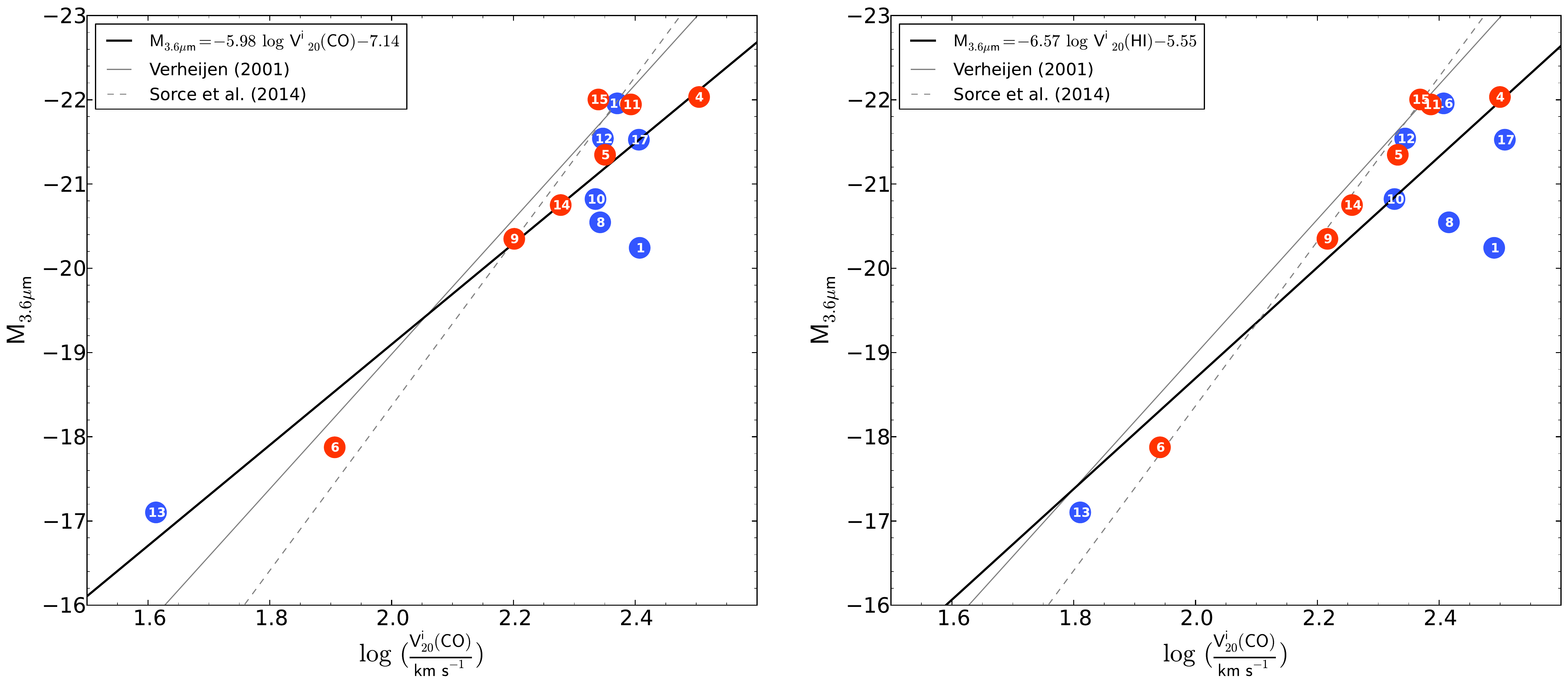}
	\caption{\label{fig:hico-tfr}  A comparison of the CO
	(left-subplot) and H\,{\sc i} (right-subplot) TFR for the HERACLES/THINGS sample. The red
	filled circles denote galaxies whose rotation curves are analysed in this work; the blue
	filled circles denote the other galaxies in the THINGS/HERACLES sample.  The number on each
	point corresponds to galaxies as numbered in Figure \ref{fig:hico-w20w50}. In both subplots
	we plot the H\,{\sc i} TFRs from \citet{2001APJ...563..694V} and \citet{Sorce:2014by} as a
	thin grey solid and a dashed line, respectively, for comparison. The best fitting lines from
	the CO and H\,{\sc i} data are plotted in black (thick solid), and the best fit parameters
	are indicated in the individual subplot legend.}
\end{figure*}

In Figure \ref{fig:hico-tfr} we plot the resultant TFR for the galaxies in our sample.  The CO-TFR
is in the left panel; the H\,{\sc i}-TFR is in the right panel. We plot the H\,{\sc i}-TFR from
\citet{2001APJ...563..694V} and \citet{Sorce:2014by} for comparison in both plots. We derive the
Tully-Fisher relation of the form $M_\mathrm{3.6\mu m}=a\log V^i_{20}+b$ by fitting a line to the data
using a least squares algorithm. The resultant CO-TFR derived is:

\begin{equation}
	M_\mathrm{3.6 \mu m}=-5.98\log V^i_{20}-7.14
\end{equation}

This is shallower than the H\,{\sc i}-TFR that we have derived from our data:

\begin{equation}
	M_\mathrm{3.6 \mu m}=-6.57\log V^i_{20}-5.55
\end{equation}

This shows that even using double-horned CO profiles, the resulting TFR is shallower as compared
with the H\,{\sc i}-TFR. This shows that using CO a tracer of the velocity width leads to a
shallower TFR as compared to the H\,{\sc i}-TFR as described in \citet{BLOK:2014AA}.

\section{Mass Modeling}
\label{sec:massmodels}
Mass models of galaxies are used to quantify the contribution of the different
constituents to the dynamics, thereby allowing us to model how much dark matter may be
present. Mass models are simply the decomposition of the observed rotation curve 
into the predicted contributions of the visible components combined with a particular dark
matter potential.

Consequently, components in mass models traditionally comprise four ingredients - the observed
rotation curve, the predicted rotation curves from the stellar and neutral gas components, and a
dark matter component usually described by a corresponding halo parameterization. The dark matter
halo parameters are adjusted as free parameters to reach the best fit, usually under specific
assumptions of the masses of the stellar component (through the stellar mass-to-light ratio,
hereafter denoted as $\Upsilon_*$) and the gas disk.

Sometimes it is possible to solve for all these parameters simultaneously, i.e.,  the halo and the
disk parameters. In practice this means including $\Upsilon_*$ as a free parameter in the fit.
However, the uncertainties are large and generally values of $\Upsilon_*$ are fixed to reduce the
degeneracies in the fit.  Since our goal in this section is to quantify the relative impact of the
inclusion of molecular gas in mass models, so we do not focus on the pros and cons of particular
choices for $\Upsilon_*$ or halo models.  We refer to dB08 for a full description of the
$\Upsilon_*$ profiles.  We fix $\Upsilon_*$ for the disk parameters and solve only for the halo
parameters corresponding to the dark matter profiles described below.

We use the CO observations as a proxy for the $\mathrm{H_2}$ by converting the CO luminosity to a
molecular gas mass surface density through the use of the conversion factor $\mathrm{\alpha_{CO}}$.
Radial profiles for $\mathrm{\alpha_{CO}}$ based on a dust-to-gas analysis have been calculated in
\citet{2013APJ...777....5S}. We therefore consider two different conversion factors - the commonly
used conversion factor based on the Milky Way ($\amw$), and the values presented in
\citet{2013APJ...777....5S} where available, which we denote as $\ad2g$. The values of $\ad2g$ were
derived using observations of CO, dust mass surface density and H\,{\sc i}. For some the galaxies
the average value $\aad2g$ is significantly different than the Milky Way value, as discussed in
Section \ref{sec:molgas}. 

We extend the analysis of the dynamics performed in dB08 by including the contribution of the
molecular gas into the mass models of the rotation curve sample indicated in Table
\ref{tab:The-THINGS-Model-Tilted-Ring}. As was done in dB08, we exclude NGC 3627 since it hosts
a bar and shows signs of tidal interactions with the neighbouring galaxies.


\subsection{Method}
\label{sec:Method}
We use the analysis done in dB08 as a template for this work. However, we only use
the photometrically determined $\Upsilon_*$ from dB08 and we do not solve for it as a
free parameter. 

The atomic gas surface density is determined from the integrated H\,{\sc i} column density map from
the THINGS data, as described in dB08. We use the predicted rotation curves calculated in dB08 as
inputs to the mass models in this work. This calculation includes a factor of 1.36 to correct for
the presence of Helium. We denote these rotation curves as $V_{\mathrm{g,A}}$.

The predicted stellar rotation curve is calculated using the stellar mass surface density. In some
cases this is the sum in quadrature of the disk and bulge components. This is converted from the
stellar luminosity profile using the corresponding mass-to-light ratio $\Upsilon_{*}$. We provide a
brief discussion of this conversion in Section \ref{sec:starmass}. The corresponding predicted
rotation curve is denoted as $V_{*}$.

The $\mathrm{H_{2}}$ mass surface density is converted from the observed CO luminosity, and we
discuss this procedure in Section \ref{sec:molgas}. The resulting predicted rotation curve is
denoted as $V_{\mathrm{g,M}}$.

The rotation curve due to the dark matter halo is usually parameterised by a halo model. We discuss
the details related to the halo parameters in Section \ref{sec:halos}, and we denote the halo
rotation curves as $V_{\mathrm{Halo}}$.

We use the mass surface density of the stars and the respective gas components to calculate
predicted rotation curves.  These predicted rotation curves show the
rotational velocity (as a function of radius) that a test particle would experience due to that
particular component alone. We subtract the predicted curves from the observed rotation curve, and
fit a halo rotation curve to the residual curve, with the halo parameters as free parameters. We can
therefore construct a mass-model relating the contribution of each component to the predicted and
observed rotation curves by using the following equation:

\begin{equation}
	V_{\mathrm{Obs}}^{2}=V_{\mathrm{g,A}}^{2}+V_{\mathrm{g,M}}^{2}+V_{*}^{2}+V_{\mathrm{Halo}}^{2}\label{eq:mass-model}
\end{equation}

where $V_{\mathrm{obs}}$ denotes the observed rotation curve.

In practice, all the terms on the right hand side of Equation \ref{eq:mass-model} produce a total
rotation curve, and this total rotation curve is fitted to the observed rotation curve using a least
squares algorithm in the \texttt{GIPSY} task \texttt{ROTMAS}. By varying the characteristic
parameters of the dark matter halo parameter a best fitting curve is then derived. 
\subsection{Stellar Mass Distribution}
\label{sec:starmass}
The stellar mass surface densities are derived using the prescription in dB08. We provide a brief
summary here.  The $\sings$-derived luminosity profiles from the Spitzer Infrared Nearby Galaxy
Survey \citep[SINGS]{2003PASP..115..928K} are used to determine the stellar mass surface density
using the $\sings$ mass-to-light ratio $\Upsilon_{*}^{3.6}$. The $K$-band mass-to-light ratio
$\Upsilon_{*}^{K}$ using the method from \citet{Oh:2008ys}.\textbf{ $\Upsilon_{*}^{K}$ }is
determined from the $J-K$ colors from the 2MASS Large Galaxy Atlas \citep{Jarrett:2003uq} assuming
the models from \citet{Bell:2001zr}.  The derived stellar mass surface density profiles depend on
the choice of the initial mass function (IMF). In general, $\Upsilon_{*}^{3.6}$ shows slight radial
gradients. 

Here we only consider the stellar mass surface densities calculated using the Kroupa IMF
\citep{Kroupa:2001kx}. We do not derive fits with  $\Upsilon_{*}$ as a free parameter, nor do we
adjust the fixed values given in dB08. Changes in $\Upsilon_{*}$ tend to have a large impact on the
halo parameters, and these would detract from the more subtle changes due to the inclusion of
$\mathrm{H_2}$ in the mass models. In that sense we are not trying to improve on the dB08 models by
deriving an updated ``best'' model.  Our main goal is to quantify how the (quality of the) fit
changes for each galaxy when $\mathrm{H_2}$ is included. 
\subsection{Molecular Gas Distribution}
\label{sec:molgas}
In this section we describe the calculation of the molecular gas mass surface density from the CO
observations, using two different conversion factors - the constant Milky Way value and the radially
varying values presented in \citet{2013APJ...777....5S}.

The observed CO luminosity $I_{\mathrm{CO}}\,(\mathrm{K\,\kmss})$ is converted to
$\mathrm{H_2}\,(\msunpc)$ mass surface density using the following relation:

\begin{equation}
\label{eq:con}
	\Sigma_{\mathrm{H_2}}=\alpha_{\mathrm{CO}}I_{\mathrm{CO}}
\end{equation}

This is analogous to the more familiar expression
$N_{\mathrm{H_2}}(\mathrm{cm^{-2}})=X_{\mathrm{CO}}I_{\mathrm{CO}}$ used to convert CO luminosity to
molecular gas column density. The typical value for $X_{\mathrm{CO}}$ for the CO $J=1\rightarrow0$
transition in the Milky Way is $2\times10^{20}\,\mathrm{cm^{-2}\,(K\,\kmss)^{-1}}$ \citep{Dame:2001bg}. The corresponding value for $\alpha_{\mathrm{CO}}$, for the CO
$J=2\rightarrow1$ transition is $6.3\,\msunpc\,(\mathrm{K}\,\kmss)^{-1}$, which we denote as $\amw$.
This assumes a constant  CO $J=2\rightarrow1$ to $J=1\rightarrow0$ conversion ratio of $0.7$
\citep{2013AJ....146...19L,2013APJ...777....5S}. The $\alpha_{\mathrm{CO}}$ value quoted here
corresponds to the CO $J=2\rightarrow1$ transition.

\citet{2013APJ...777....5S} used the HERACLES, THINGS and KINGFISH \citep[Key Insights into Nearby
Galaxies: A Far-Infrared Survey with Herschel]{2011PASP..123.1347K} surveys to simultaneously solve
for $\alpha_{\mathrm{CO}}$ and the dust-to-gas (D2G) ratio in nearby galaxies, including many that
are in our sample.  All derived values of $\alpha_{\mathrm{CO}}$ contain a correction of $1.36$ in
order to account for the presence of Helium. This correction is also present in the computation of
the atomic mass surface density. 

We denote the \citet{2013APJ...777....5S} radially varying conversion factors as $\ad2g$; the
corresponding mean value will be denoted as $\aad2g$. The values for $\mathrm{\alpha_{CO}}$ vary
substantially in the sample and can be very different from the Milky Way value $\amw=6.3\,\auni$.
The average values vary from $\aad2g=1.4\,\auni$ to $\aad2g=15.7\,\auni$. In addition, the radial
profiles of $\ad2g$ presented in \citet{2013APJ...777....5S} are not flat but usually show a gradual
radial increase in the $\ad2g$ value.

\begin{figure}
	\centering \resizebox{\hsize}{!}{\includegraphics{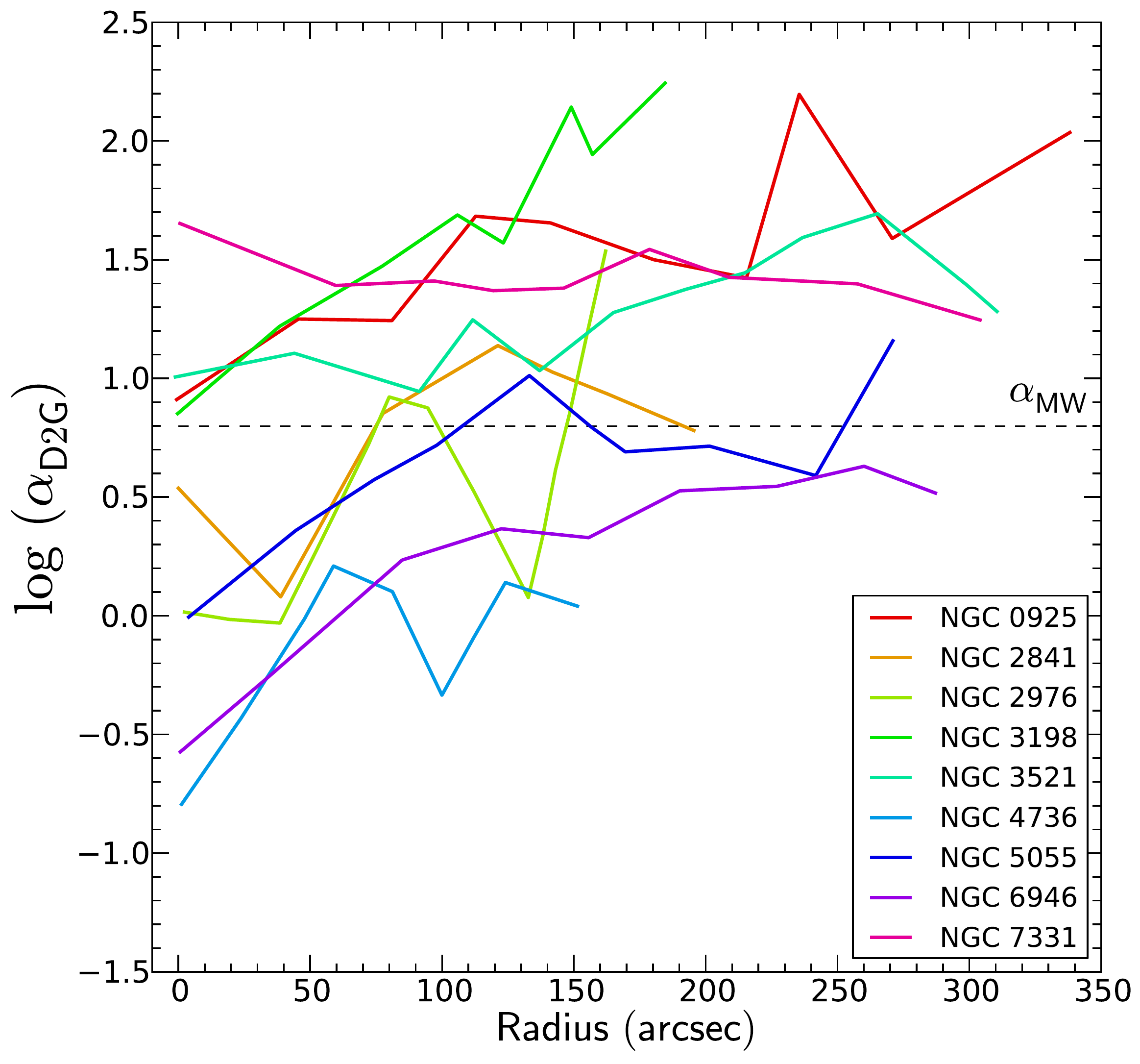}}
	\caption{\label{fig:ad2g}Radial profiles for $\log\,(\ad2g)$ used to convert
	$I_{\mathrm{CO}}$ to $\Sigma_{\mathrm{H_2}}$ for galaxies in this work. The CO
	$J=2\rightarrow1$ values for $\ad2g$ are plotted here. The Milky Way value of
	$\amw=6.3\,\auni$ is indicated as a dashed horizontal line. For NGC 2976, NGC 4736, NGC 5055
	and NGC 6946 we use the profiles corresponding to those presented in Figure 7 of
	\citet{2013APJ...777....5S}. For NGC 0925, NGC 2841, NGC 3198, NGC 3521 and NGC 5055 we
	derive profiles by binning the individual data points in Figure 22 in
	\citet{2013APJ...777....5S} in increments of $R_{25}$, where $R_{25}$ is the $B$-band
	isophotal radius at $25\,\mathrm{mag\,arcsec^{-2}}$ presented in that paper.}
\end{figure}

In Figure \ref{fig:ad2g} we plot the radial profiles for $\ad2g$ used in this work, based on the data
presented in \citet{2013APJ...777....5S}. These radial profiles will be used to calculate the
molecular gas mass surface density. In Table \ref{tab:galaxies} we list the average
weighted mean $\aad2g$ values for the galaxies in our sample.

Binned radial profiles for NGC 2976, NGC 4736, NGC 5055 and NGC 6946 were presented in
\citet{2013APJ...777....5S}. For NGC 0925, NGC 2841, NGC 3198, NGC 6946, NGC 7331 we derive radial
profiles for $\ad2g$ by binning the individual measurements presented in \citet{2013APJ...777....5S} in increments of
$0.1\,R_{25}$ where $R_{25}$ is the $B$-band isophotal radius at $25\,\mathrm{mag\,arcsec^{-2}}$
presented in that work.

We calculate molecular gas mass surface densities using both the radially dependent $\ad2g$ and the
constant  Milky Way value for the galaxies in our sample. For NGC 2403 and NGC 2903 we only consider
the constant Milky Way value, since these galaxies were not studied in \citet{2013APJ...777....5S}.
There is a comparatively higher uncertainty in the $\ad2g$ value for galaxies at higher
inclinations, which could be due to opacity effects and the ambiguity in associating specific dust
and gas features along the line of sight.

To calculate the molecular gas mass surface density, we use the HERACLES integrated intensity map to
calculate the radial surface brightness distribution. For this we use the THINGS tilted-ring
geometry from dB08, and we apply an inclination correction. This is used as the input in  Equation
\ref{eq:con} when calculating the molecular gas mass surface density. 

\begin{table}
	\caption{\label{tab:galaxies}The average value of the 
	CO-to-$\mathrm{H_2}$ conversion factors $\aad2g$.}
	\centering
	\begin{tabular}{l c}
		\hline
		Galaxy & $\mathrm{\aad2g}$ \\
		 & $\auni$ \\
		\hline \hline
		NGC 0925& 14.3 \\
		NGC 2841& 7.1 \\
		NGC 2976& 4.7 \\
		NGC 3198& 15.7 \\
		NGC 3521& 10.9 \\
		NGC 4736& 1.4 \\
		NGC 5055& 5.3 \\
		NGC 6946& 2.9 \\
		NGC 7331& 14.0 \\
		\hline
	\end{tabular}
	\tablecomments{Values are from \citet{2013APJ...777....5S}. The galaxies NGC 2403 and NGC 2903 were 
	not part of the \citet{2013APJ...777....5S} sample, so we use the Milky Way conversion
        factor of $6.3\,\msunpc\,(\mathrm{K}\,\kmss)^{-1}$ for these galaxies.}
\end{table}
\subsection{Putting it all together}
The mass surface densities computed using the methods described above are then used to calculate the
predicted rotation curves (i.e., $V_\mathrm{g,A}$, $V_\mathrm{g,M}$ and $V_{*}$) by using the
\texttt{GIPSY} task \texttt{ROTMOD}.  \citet{BLOK:2008UQ} assume an infinitely thin disk for the
H\,{\sc i} and a $\mathrm{sech^{2}}$ distribution for the stellar component, and we do the same
here. The stellar predicted rotation curves were derived using mass surface densities calculated
from photometrically determined $\Upsilon_{*}$. We assume that the CO is also distributed in an
infinitely thin disk. The predicted rotation curves are inserted into the mass-model in Equation
\ref{eq:mass-model}. The parameterised halo rotation curve is then fitted to the observed rotation
curve. Therefore, the only free parameters in our fits are the dark-matter halo parameters,
discussed below. 
\subsection{Dark Matter mass models}
\label{sec:halos}
We compute mass-models using both the Navarro-Frenk-White (NFW) halo \citep{NFW_1996,NFW_1997} and
the observationally motivated pseudo-isothermal (ISO) halo. 

Following \citet{NFW_1996,NFW_1997} the NFW mass-density distribution
has the form

\begin{equation}
	\rho_\mathrm{NFW}(R)=\frac{\rho_\mathrm{i}}{(R/R_\mathrm{s})(1+R/R_\mathrm{s})^{2}}\label{eq:massmodel-nfw}
\end{equation}

where $R_\mathrm{s}$ is the scale radius of the halo (and $\rho_\mathrm{i}$
is proportional to the density of the universe at the time of collapse
of the dark matter halo. This leads to a halo rotation curve \citep{NFW_1996}
given by: 

\begin{equation}
	V(R)=V_{200}\left[\frac{\ln(1+cx)-cx/(1+cx)}{x[\ln(1+c)-c/(1+c)]}\right]^{1/2}
\end{equation}

where $x=R/R_{200}$, $c=R_{200}/R_\mathrm{s}$ is the concentration parameter and $V_{200}$ is the
characteristic velocity at radius $R_{200}$, the radius where the density contrast relative to the
critical density of the universe exceeds $200$. Cosmologically motivated values for the halo
parameters can be deduced using the simulations from \citet{2001MNRAS.321..559B} and the models from
\citet{2007ApJS..170..377S}. We solve for $c$ and $V_{200}$ by fitting $V(R)$ to Equation
\ref{eq:mass-model}.  

The ISO mass-density distribution has the form 

\begin{equation}
	\rho_{ISO}(R)=\rho_{0}\left[1+\left(\frac{R}{R_{c}}\right)^{2}\right]^{-1}
\end{equation}

where $\rho_{0}$ denotes the central density of the halo and $R_\mathrm{c}$
is the so-called core radius. This leads to a halo rotation curve
given by:

\begin{equation}
	V(R)=\left[4\pi G\rho_{0}R_{c}^{2}\left(1-\frac{R_{c}}{R}\arctan\left(\frac{R}{R_{c}}\right)\right)\right]^{1/2}
\end{equation}

where the asymptotic velocity of the halo, $V_{\infty}$ is given
by: 

\begin{equation}
	V_{\infty}=\sqrt{4\pi G\rho_{0}R_{c}^{2}}
\end{equation}

For the ISO case we can directly solve for $\rho_{0}$ and $R_\mathrm{c}$
by fitting $V(R)$ to the parent Equation \ref{eq:mass-model}. 

We use the \texttt{GIPSY} task \texttt{ROTMAS} to fit the respective halo rotation curves to the
mass model in \ref{eq:mass-model}.  We use the observed rotation curve error-bars to weight the
fits. This is done to keep our results consistent with dB08. Adopting a different weighting scheme,
such as uniform errorbars for all points has little impact on the outcomes of this study.

\section{Results - Mass Models}
\label{sec:massmodelresults}
In this section we present the mass models for each galaxy in our sample, assuming either an NFW or
ISO halo form. The corresponding halo parameters are presented in Tables \ref{tab:NFW} and
\ref{tab:ISO}. 

For a few galaxies the HERACLES rotation curve is either steeper in the inner parts than the THINGS
curve (e.g., NGC 5055) or fills in the inner part of the rotation curve where no H\,{\sc i} has been
detected (e.g., NGC 4736 --- see Appendix for full description). In these cases we also explore fits
to a hybrid rotation curve, where we use the HERACLES rotation curve in the inner few kpc and the
THINGS rotation curve at larger radii. In the text we refer to the hybrid rotation curves as the
HERACLES/THINGS or COH\,{\sc i} rotation curves. We refer to the observed H\,{\sc i} rotation curves
as the THINGS rotation curves, and to the observed CO rotation curves as the HERACLES rotation
curves. 

In each case we compare our results with those from dB08. It is important to note that dB08
considered mass models comprising stellar rotation curves predicted using both ``diet'' Salpeter
\citep{1955ApJ...121..161S,Bell:2001zr} and Kroupa IMFs. For a given mass-to-light ratio
$\mathrm{M/L_*}$ the difference between the Kroupa and diet Salpeter is $\sim0.15$ dex. While dB08 list
derived halo parameters for both these IMFs, their figures only show the diet Salpeter mass models.
Here we only consider Kroupa IMF based models, so this difference should be kept in mind when
comparing the mass models presented in this work with those in dB08.

When dealing with the rotation curves $V_\mathrm{g,A}$ and $V_\mathrm{g,M}$, we adopt the convention
of plotting negative values of $V^{2}$ as negative values of $V$. Such values can occur since, in
the presence of central under-densities, test particles in or near such an under-density will
experience an outward force.
\subsection{NGC 925\label{sec:NGC-925}}

\begin{figure*}
	\centering
	\includegraphics[origin=lb,width=17cm]{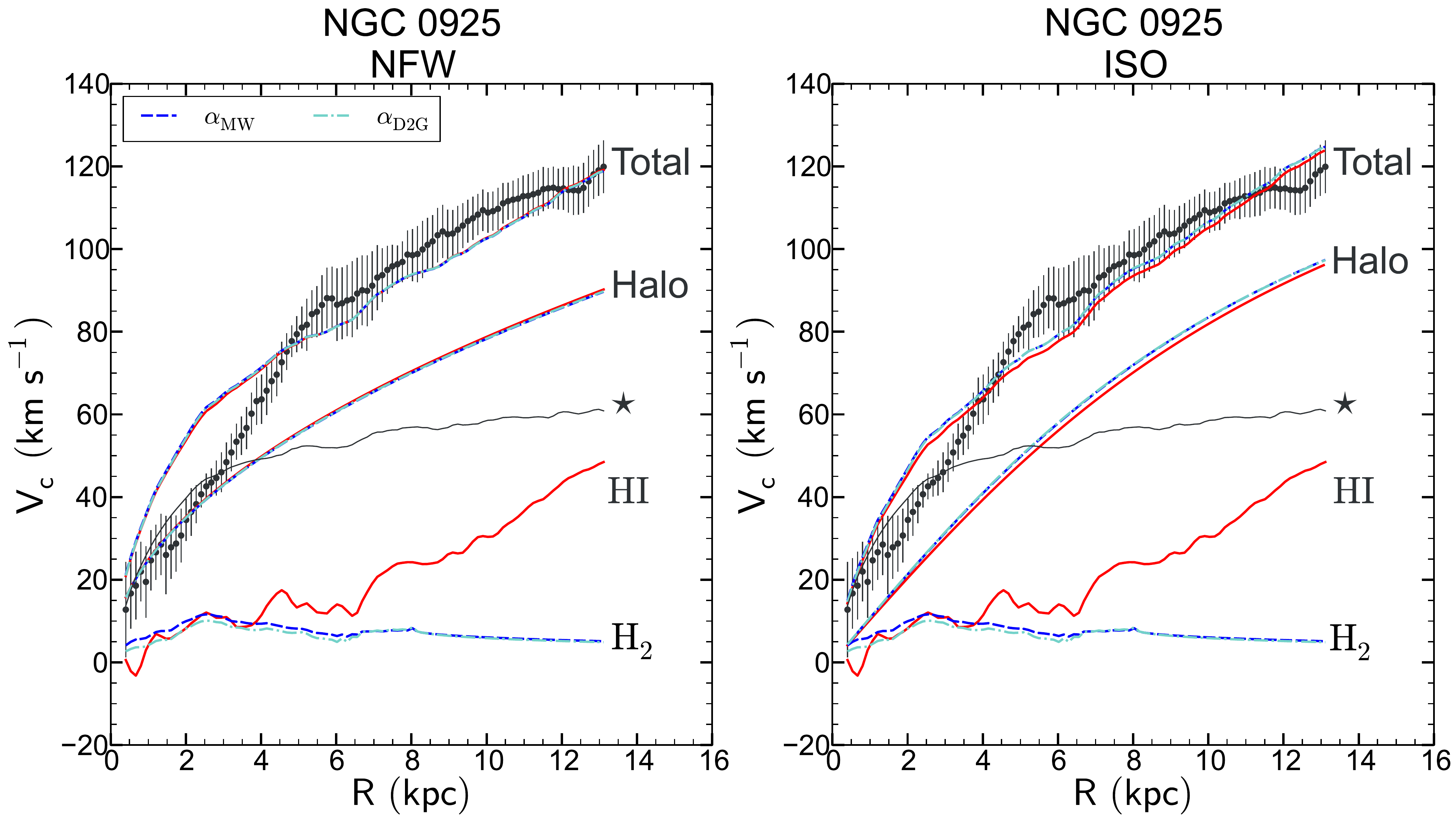}
	\caption{\label{fig:N0925-MMs}NGC 0925 mass models using the THINGS rotation curve. Results
	using the NFW model are in the left panel, results using the ISO model are in the
	right panel. The observed rotation curve is plotted as grey filled circles with
	errorbars.  The predicted stellar rotation curve $V_*$ is plotted as a thin grey
	line, and is labelled by the $\star$ in the figure.  The predicted atomic gas
	rotation curve $V_\mathrm{g,A}$ is plotted as a thicker red line, and is labelled as
	$\mathrm{HI}$ in the figure. The predicted molecular gas rotation curves
	$V_\mathrm{g,M}$ are plotted as dashed-dotted cyan ($\alpha_{D2G}$) and dashed blue
	lines ($\alpha_{MW}$), and are labelled as $\mathrm{H_2}$ in the figure. The dark
	matter halo rotation curves $V_\mathrm{Halo}$ and the resultant rotation curves fits
	to $V_\mathrm{Obs}$ for either choice of $\mathrm{\alpha_{CO}}$ are plotted using
	this convention, labelled as Halo and Total in the figure, respectively. The
	$V_\mathrm{Halo}$ rotation curve and the consequent fit to $V_\mathrm{Obs}$ from dB08 (i.e.,
	no molecular gas) are both plotted as red solid lines.}
\end{figure*}

In Figure \ref{fig:N0925-MMs} we plot the best fitting models using the THINGS rotation curve.  The
stellar mass distribution comprises only a single component, and is the dominant component within
$5\kpc$. At greater radii the dark matter distribution becomes more important.  

The value $\aad2g=14.3\,\auni$ is much larger than the Milky Way value. The predicted molecular gas
rotation curves using $\amw$ and $\ad2g$ do not exceed a maximum of $\sim10\kmss$ at $\sim 3 \kpc$.
The $\ad2g$ curve is slightly higher than the $\amw$ curve, but not sufficiently to change the
predicted molecular gas rotation curve due to the small amount of molecular gas detected.  Although
fitting the NFW rotation curve produces a halo model which appears reasonable, the fit yields
unrealistic halo parameters: $c<1$ and correspondingly high values for $V_{200}>200\kmss$, which
is substantially different from the expected range \citep{2001MNRAS.321..559B}. The NFW fit is
therefore shown for illustrative purposes and will not be considered in further analysis.

\subsection{NGC 2403}
\label{sec:NGC-2403}

\begin{figure*}
	\centering
	\includegraphics[width=17cm]{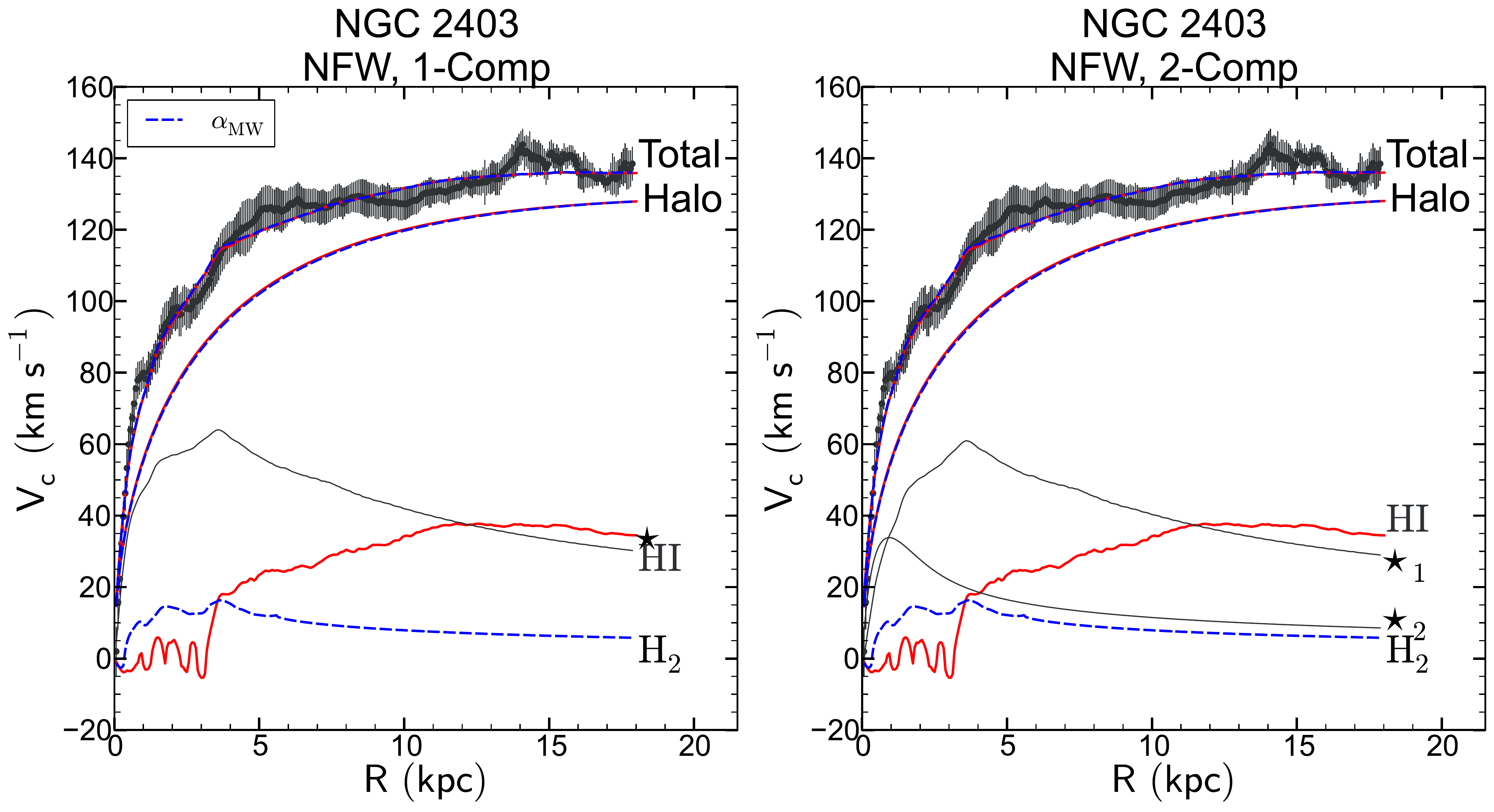}		
	\caption{\label{fig:N2403-NFW}NGC 2403 mass models using the
	HERACLES/THINGS data with an NFW halo. Results using 1- and 2-component stellar
	distributions are plotted in the left and right panels respectively.  Line styles and colours
	are as in Figure \ref{fig:N0925-MMs}.}
\end{figure*}

\begin{figure*}
	\centering
	\includegraphics[width=17cm]{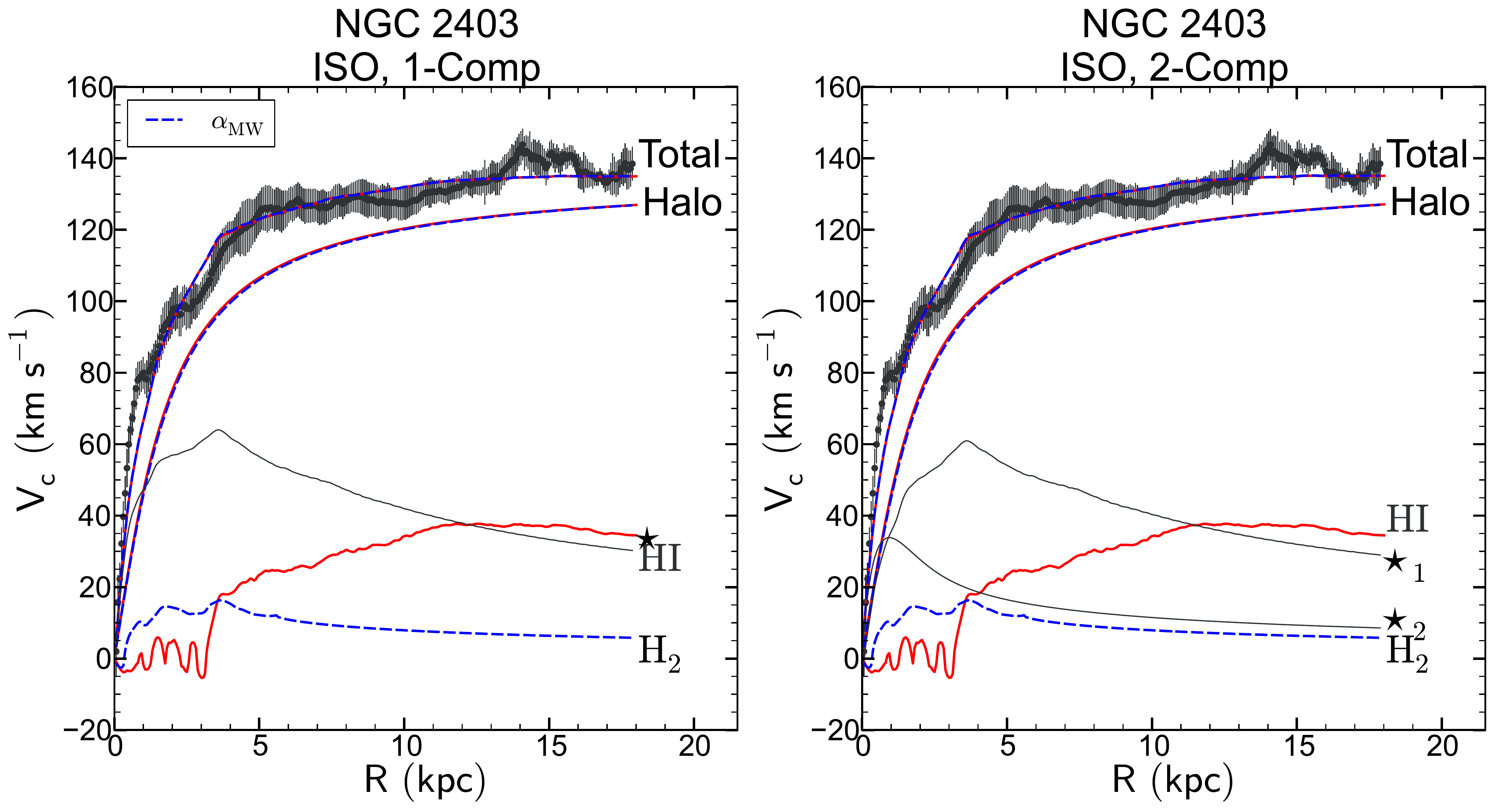}
	\caption{\label{fig:N2403-ISO}NGC 2403 mass models using the
	HERACLES/THINGS data with an ISO halo. Results using 1- and 2-component stellar
	distributions are plotted in the left and right panels respectively. Line styles and colours
	are as in Figure \ref{fig:N0925-MMs}.}
\end{figure*}

In Figures \ref{fig:N2403-NFW} and  \ref{fig:N2403-ISO} we plot the results of the mass modelling
for NGC 2403. NFW and ISO halo models were fitted to the THINGS H\,{\sc i} observed rotation curve.   

The stellar mass distribution can be described by either a 1- or 2-component decomposition. We fit
halo rotation curves using both stellar distributions, treating the 1- and 2-component models separately. 

\citet{2013APJ...777....5S} did not solve for a $\ad2g$ value for NGC 2403.  We therefore only
consider a predicted molecular gas rotation curve using $\amw$. This rotation curve reaches a
maximum of $\sim10\kmss$ at $\sim4\kpc$ and declines thereafter. The molecular gas predicted
rotation curve is only slightly higher than the H\,{\sc i} rotation curve within $\sim3\kpc$, but is
overall the smallest contributor to the dynamics of NGC 2403.

For fits using the NFW model the addition of the molecular gas makes no difference to the model and
total rotation curves.  The NFW total rotation curve is slightly lower than the observed rotation
curve at $1\kpc$, but shows an overall good fit to the observed rotation curve.  For fits using the
ISO model the addition of the molecular gas makes no difference to the best fitting model and
consequent total rotation curves. The ISO models do not fit the inner part (within $2\kpc$) of the
observed rotation curve as well as the NFW models.  The quality of the fits for both the NFW  and ISO
fits including molecular gas are no different from the H\,{\sc i}-only case. 

\subsection{NGC 2841}
\label{sec:NGC-2841}
\begin{figure*}
	\centering
	\includegraphics[width=17cm]{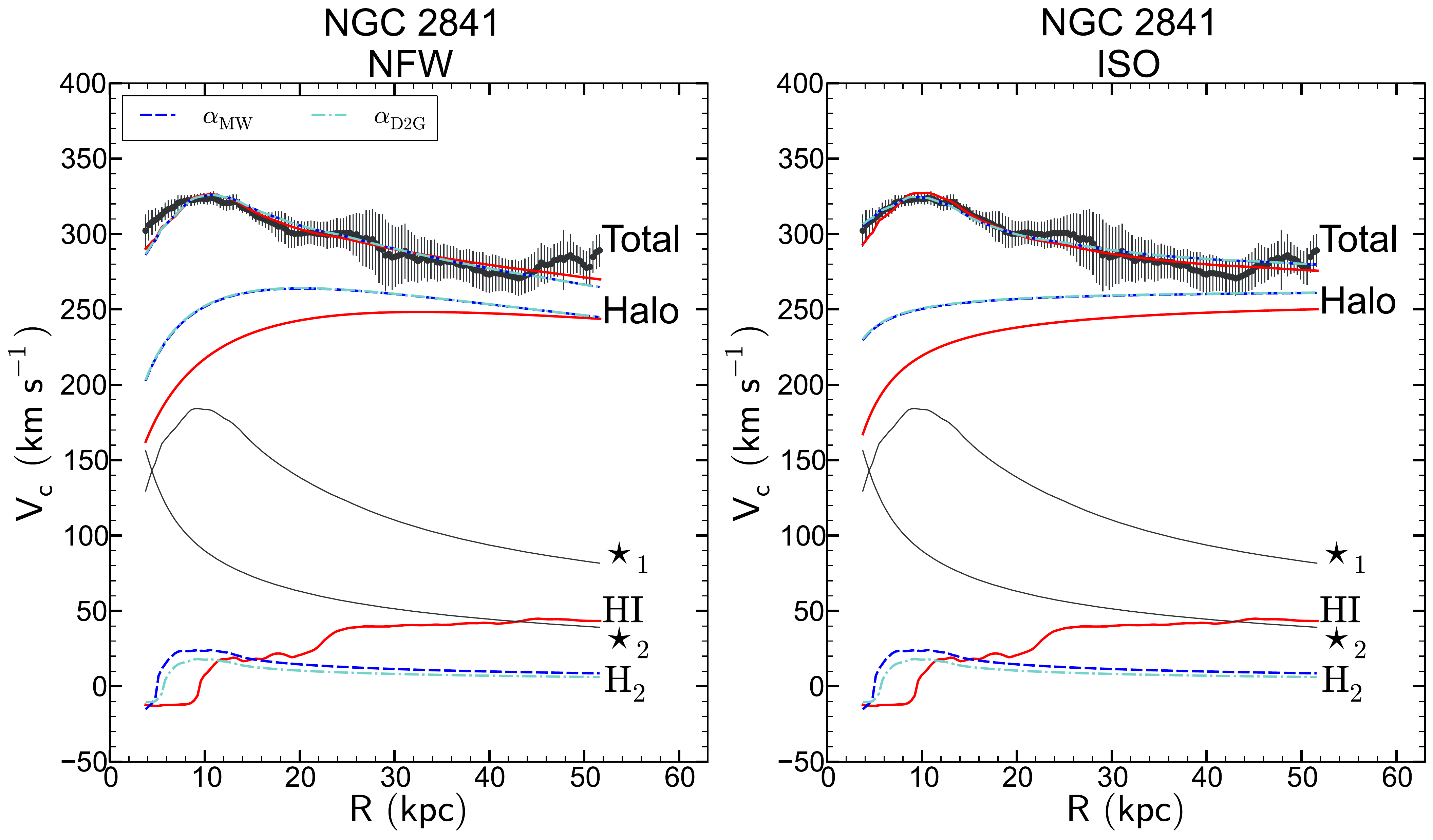}			
	\caption{\label{fig:N2841-MMs}NGC 2841 mass models using the
	HERACLES/THINGS data. Results with the NFW halo are in the left panel, results with the
	ISO halo are in the right panel. Line styles and colours are as in Figure
	\ref{fig:N0925-MMs}.}
\end{figure*}

In Figure \ref{fig:N2841-MMs}  we plot the results of the mass modelling for NGC 2841. ISO and NFW
halo models were fitted using the THINGS H\,{\sc i} observed rotation curve.

We fit models to the THINGS rotation curve, which
starts at $\sim4\kpc$ and extends to $\sim50\kpc$.  The stellar mass distribution is decomposed
into 2-components, which we use to fit the observed rotation curve.

The value $\aad2g=7.1\,\auni$ is very close the Milky Way value. The predicted rotation curves for
both $\ad2g$ and $\amw$ are almost identical, as are the resultant best fitting halo model rotation
curves for both the NFW and ISO cases. The molecular gas mass surface density produces a maximum
velocity of $\sim20\kmss$.  Although the predicted molecular gas rotation velocities are small in
comparison to the other components, the resultant halo model rotation curves which include molecular
gas are different from the H\,{\sc i}-only rotation curves.  The effect of adding molecular gas is
to increase the amplitude of the halo rotation curves.  The total rotation curve with added
molecular gas is almost identical to the H\,{\sc i}-only total rotation curve, showing small
differences at the innermost radii. 

As anticipated, the $c$, $V_{200}$ and $\rchisq$ values for $\ad2g$ and $\amw$ predicted rotation
curves are nearly identical. The values of $c$ are higher than the H\,{\sc i}-only case, while the
$V_{200}$ values are slightly smaller. 

The addition of $\mathrm{H_2}$ slightly increases the $\rchisq$ values when fitting the ISO models. 

\subsection{NGC 2903}
\label{sec:NGC-2903}
\begin{figure}
	\centering
	\resizebox{\hsize}{!}{\includegraphics{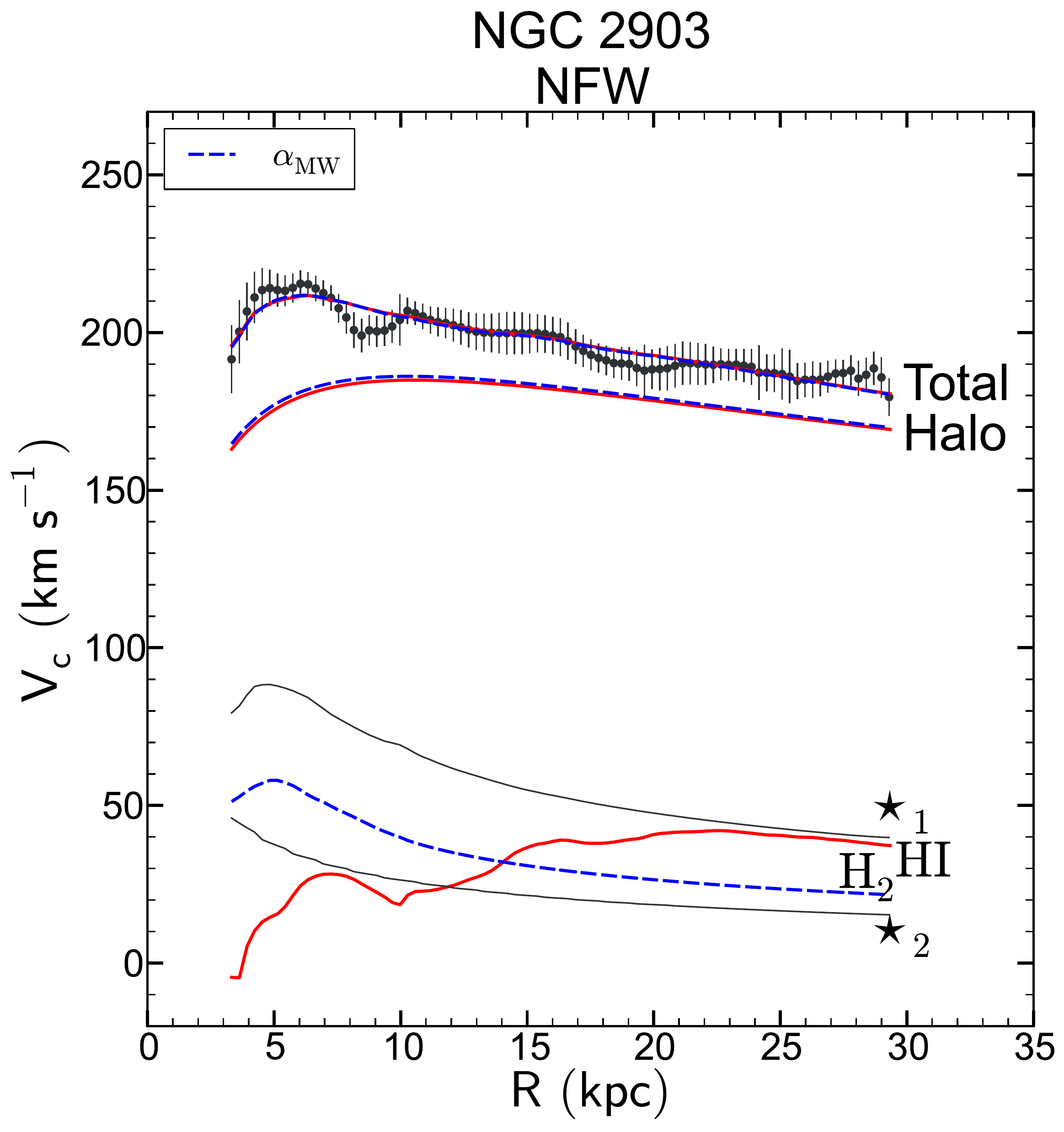}}	
	\caption{\label{fig:N2903-NFW}NGC 2903 mass models
	using the HERACLES/THINGS data with the NFW halo. Line styles and colours are
	as in Figure \ref{fig:N0925-MMs}.}
\end{figure}

The results of the mass modelling for NGC 2903 are plotted in Figure \ref{fig:N2903-NFW}. An NFW
halo model was fitted to the outer regions using the THINGS H\,{\sc i} observed rotation curve, as
described in the Appendix.

NGC 2903 hosts a strong bar and the rotation curve in the inner $3\kpc$ for the H\,{\sc i} and CO
are strongly affected by streaming motions. The stellar mass surface density is described using a
2-component decomposition. 

We do not use the rotation curve corrected for the bar-streaming motions derived in Section
\ref{sec:N2903-df}. In order to do this we would need to do an associated correction of the surface
brightness distribution, which is beyond the scope of this work. We therefore adopt the same
strategy as in dB08, fitting mass models to the observed rotation curves for radii larger than
$3\kpc$.

We plot the results of fitting the NFW halo. Attempting to fit the ISO halo model rotation curve
converges to unrealistic values for the halo parameters - $R_\mathrm{c}\ll1$ and a correspondingly large value
for $\rho_0\gg1000$. This is because there are no constraints from the inner parts of the rotation
curve. 

\citet{2013APJ...777....5S} did not solve for a $\ad2g$ value for NGC 2903.  We therefore only
consider a predicted molecular gas rotation curve using $\amw$. The predicted molecular gas rotation
curve reaches a maximum of $\sim40\kmss$ at a radius $\sim5\kpc$, and declines to approximately
$10\kmss$ at larger radii. 

The velocities of the halo rotation curve are slightly higher than for those of H\,{\sc i}-only
case. The total rotation curve is identical to the H\,{\sc i}-only total rotation curve. The values
for $c$ and $V_{200}$ are almost identical in both cases, as are the $\rchisq$ values. The total
rotation curves produce good fits to the observed THINGS rotation curve. 

\subsection{NGC 2976}
\label{sec:NGC-2976}

\begin{figure}
	\centering
	\resizebox{\hsize}{!}{\includegraphics{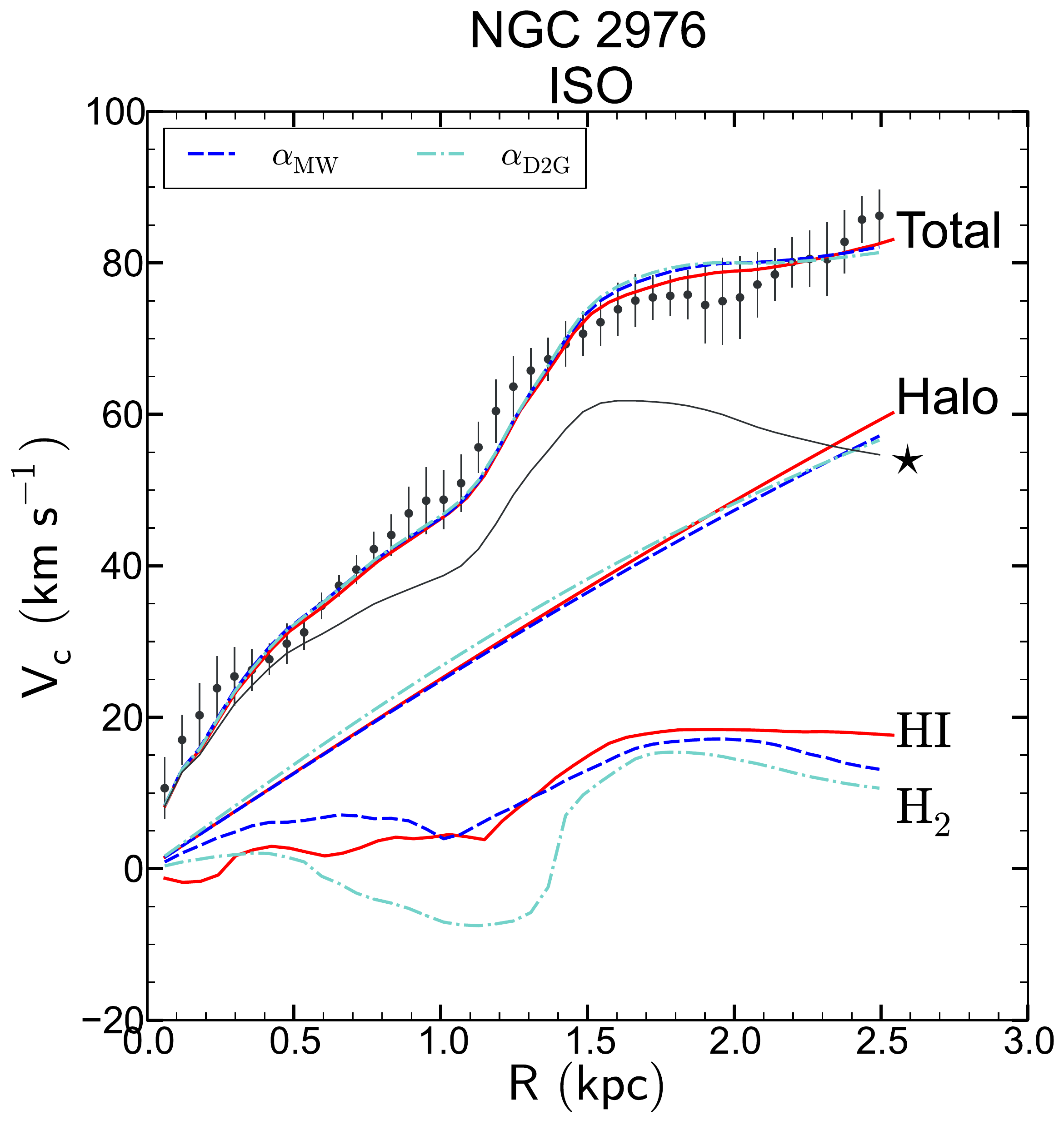}}
	\caption{\label{fig:N2976-ISO}from NGC 2976 mass models using the
	THINGS data with the ISO halo. Line styles and colours for each plot are as in Figure
	\ref{fig:N0925-MMs}.}
\end{figure}

In Figure \ref{fig:N2976-ISO} we plot the results of the mass modelling for NGC 2976. The ISO halo
model was fitted using the THINGS H\,{\sc i} rotation curve.  Fitting the NFW halo model converge
to unrealistic values of halo parameters. This was also the case for the H\,{\sc i}-only fits in
dB08. 

The stellar mass surface density comprises a single component, and we use this to fit our rotation
curves.

While the average value of $\aad2g=4.7\,\auni$ value is very similar to the $\amw$, the radial
profile of $\ad2g$ is steep in the inner part and leads to a much lower molecular gas mass surface
density. As such, the predicted molecular gas rotation curve with $\ad2g$ is very different in
comparison to that calculated with $\amw$. The molecular gas contribution to the dynamics is very
small and the stellar component makes the largest contribution to the model. Therefore, the addition
of the molecular gas does not make an appreciable difference to the halo rotation curves and the
fitted total rotation curves. 

The fitted ISO parameters are similar to the H\,{\sc i}-only case for either choice of
$\mathrm{\alpha_{CO}}$, suggesting that the molecular gas makes a negligible contribution to the
dynamics of NGC 2976.

\subsection{NGC 3198}
\label{sec:NGC-3198}
\begin{figure*}
	\centering
	\includegraphics[width=17cm]{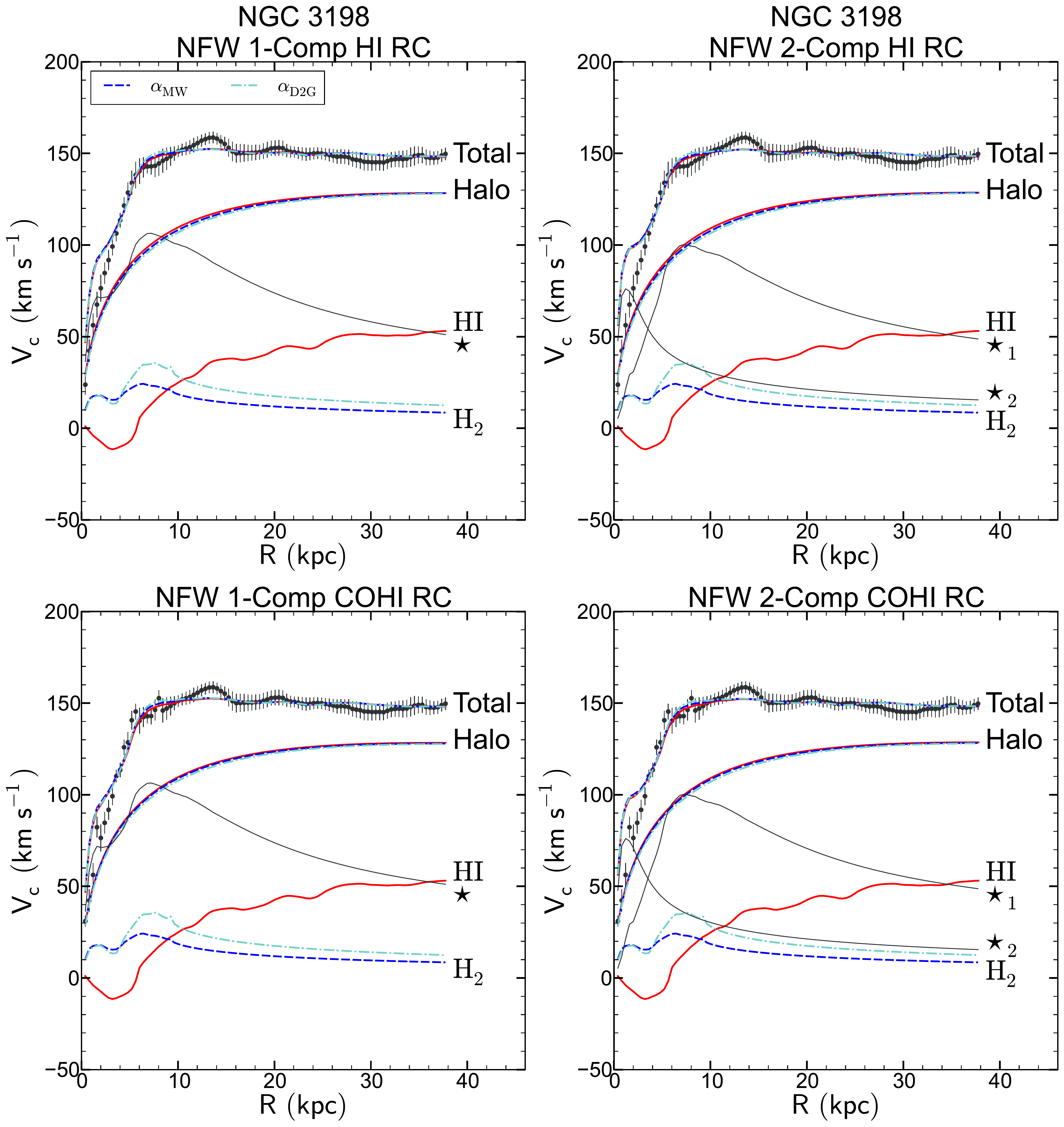}		
	\caption{\label{fig:N3198-NFW}NGC 3198 mass models with the NFW halo.
	Fits using 1- and 2-component stellar distributions are plotted in the left and right panels
	respectively, for each row. Fits to the THINGS rotation curve are plotted in the top row,
	fits to the HERACLES/THINGS rotation curve are plotted in the bottom row. Line styles and
	colours for each plot are as in Figure \ref{fig:N0925-MMs}.}
\end{figure*}

\begin{figure*}
	\centering
	\includegraphics[width=17cm]{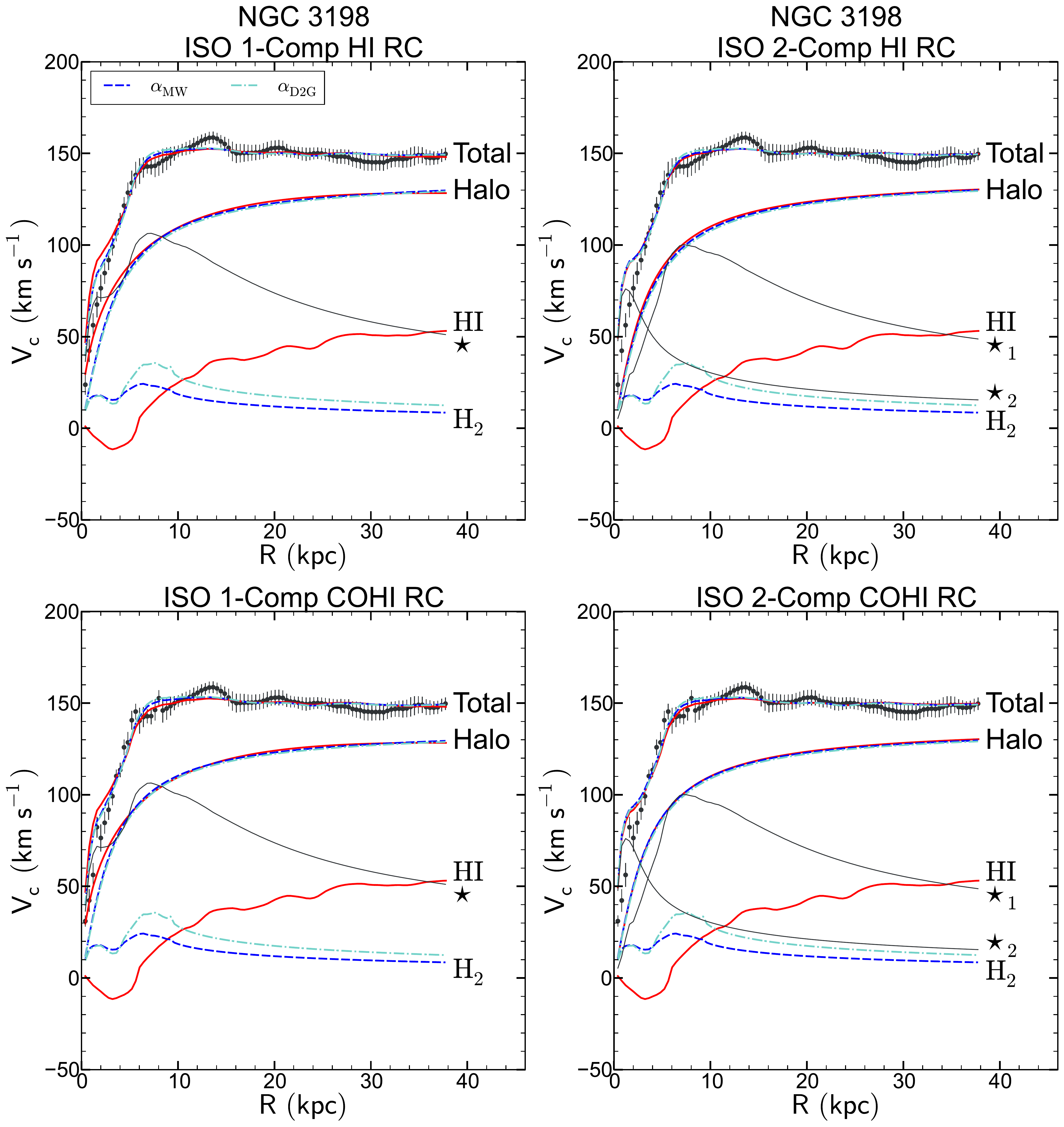}		
	\caption{\label{fig:N3198-ISO}NGC 3198 mass models using the ISO halo.
	Fits using 1- and 2-component stellar distributions are plotted in the left and right panels
	respectively, for each row. Fits to the THINGS rotation curve are plotted in the top row,
	fits to the HERACLES/THINGS rotation curve are plotted in the bottom row. Line styles and
	colours for each plot are as in Figure \ref{fig:N0925-MMs}.}
\end{figure*}

In  Figures \ref{fig:N3198-NFW} and \ref{fig:N3198-ISO} we plot the results of the mass modelling
for NGC 3198.  For NGC 3198 we use 1- and 2-component decompositions of the stellar surface
brightness distribution to derive the predicted stellar rotation curves, and we treat the 1- and
2-component cases separately.  

In the analysis of the observed rotation curves for NGC 3198 presented in the Appendix 
we note a considerable difference in the shapes between the H\,{\sc i} and CO observed rotation
curves within $4\kpc$. We therefore consider two observed rotation curves for NGC 3198 - the THINGS
rotation curve and a hybrid HERACLES/THINGS rotation curve, where we use the HERACLES rotation curve
inside $4\kpc$ and the THINGS rotation curve outside this radius.  Therefore, for both the NFW and
ISO models we have four scenarios, corresponding to the two possible stellar decompositions and the
two observed rotation curves. 

For NGC 3198 the conversion factor $\aad2g=15.7\,\auni$ is considerably higher than the Milky
Way value.  The predicted molecular gas rotation curves are identical within $\sim5\kpc$, but the
$\ad2g$ molecular gas rotation curve is much higher than the $\amw$ rotation curve between
$5\sim10\kpc$. For the NFW case the halo rotation curves with added molecular gas are
identical to the H\,{\sc i}-only case. The NFW halo parameters presented in Table \ref{tab:NFW} are
similar for models with and without molecular gas (within the uncertainties). 

Fitting the ISO halo to the 1-component stellar distribution results in halo rotation curves which
have a slightly different shape than the H\,{\sc i}-only case, and which are steeper than the
H\,{\sc i}-only halo rotation curve within $5\kpc$, and flatter at larger radii. The $\amw$ and
$\ad2g$ predicted molecular gas rotation are identical.  For the fits to the ISO halo with the
2-component stellar distribution the halo rotation curves with added molecular gas are identical to
the H\,{\sc i}-only halo rotation curve. 

In general, better fits were achieved for the 1-component stellar distribution as compared to the
2-component distribution. Fitting the ISO halo produces better fits compared to the NFW halo. 

For all ISO fits the $R_C$ and $\rho_0$ values are fairly similar, while the 1- and 2-component fits
with H\,{\sc i}-only show completely different parameters. For the NFW fits the values for $c$ and
$V_{200}$ are tightly constrained and do not vary much for either of 1- or 2-component stellar
models. The quality of the fits are similar to the H\,{\sc i}-only case.

\subsection{NGC 3521}
\label{sec:NGC-3521}
\begin{figure*}
	\centering
	\includegraphics[width=17cm]{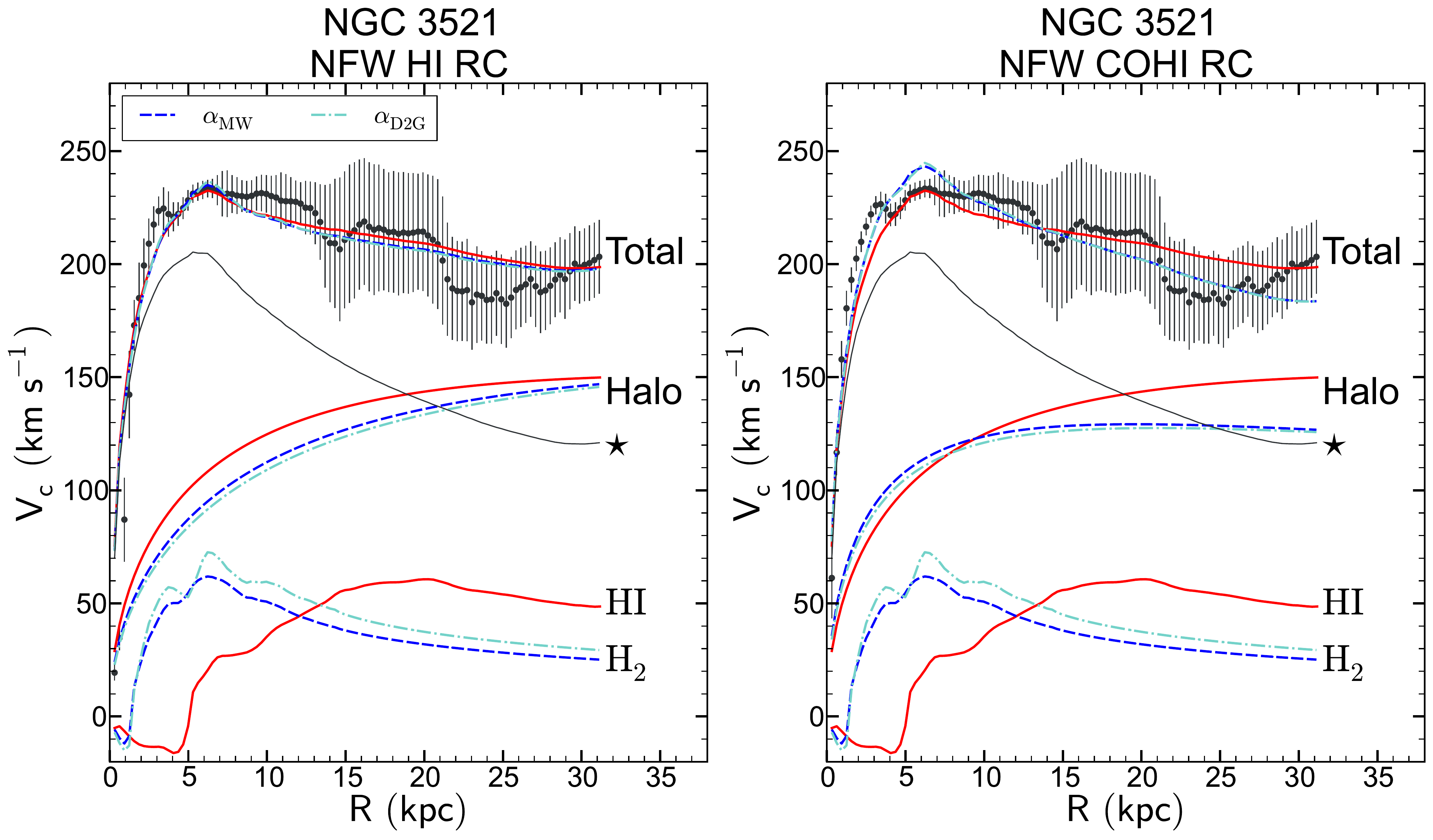}		
	\caption{\label{fig:N3521-NFW}NGC 3521 mass models with an NFW halo.
	Fits to the THINGS rotation curve are plotted in the left panel, fits to the HERACLES/THINGS
	rotation curve are plotted in the right panel. Line styles and colours for each plot are as
	in Figure \ref{fig:N0925-MMs}.}
\end{figure*}

\begin{figure*}
	\centering
	\includegraphics[width=17cm]{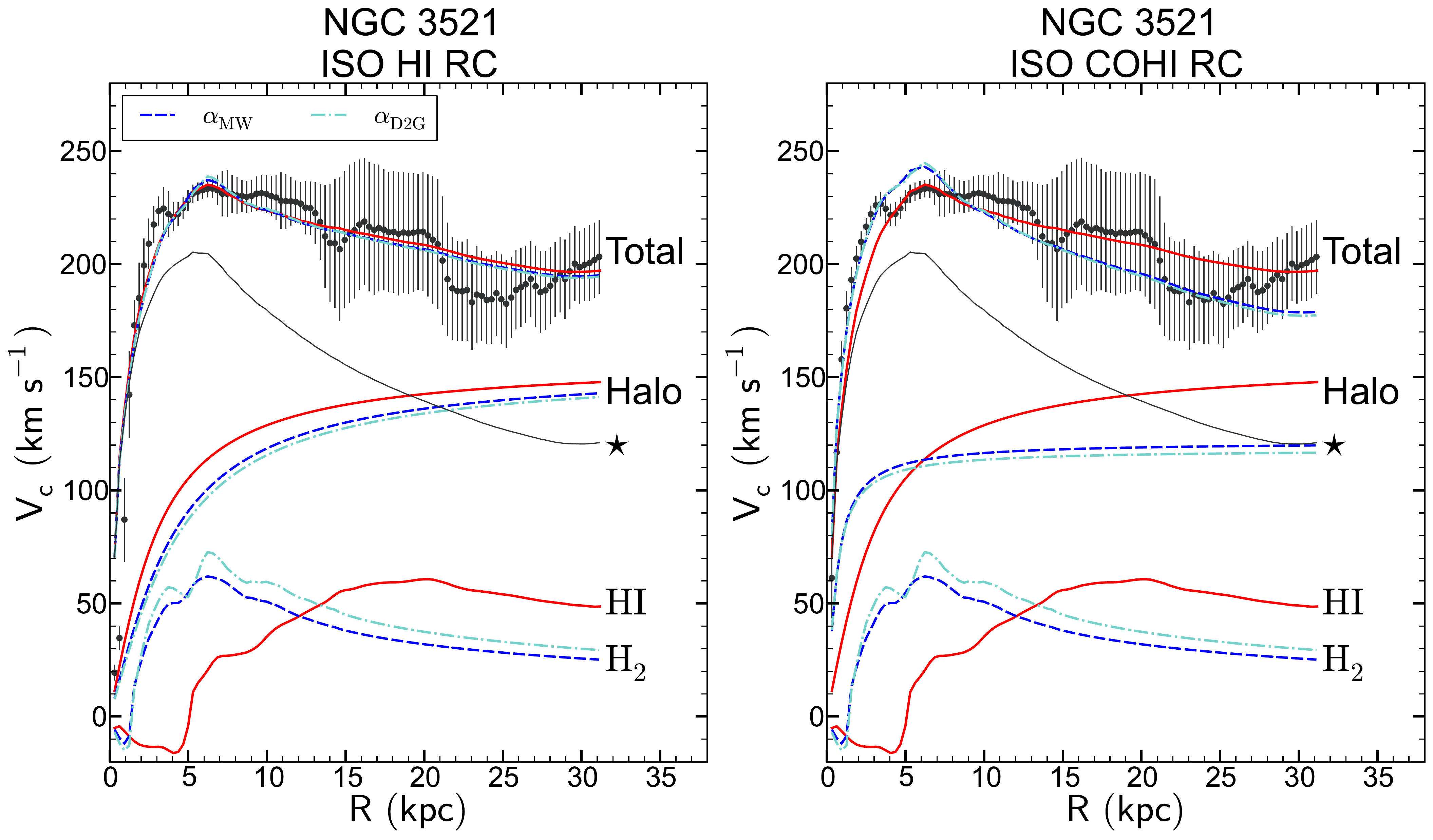}		
	\caption{\label{fig:N3521-ISO}NGC 3521 mass models with an ISO halo.
	Fits to the THINGS rotation curve are plotted in the left panel, fits to the HERACLES/THINGS
	rotation curve are plotted in the right panel. Line styles and colours for each plot are as
	in Figure \ref{fig:N0925-MMs}.}
\end{figure*}

In Figures \ref{fig:N3521-NFW} and \ref{fig:N3521-ISO} we plot the predicted, model and best fitting
rotation curves for NGC 3521.  For NGC 3521 we use a single component stellar decomposition. 

The HERACLES rotation curve is steeper within $4\kpc$. We therefore consider two observed rotation
curves when fitting ISO and NFW models - a THINGS rotation curve and a HERACLES/THINGS rotation
where we use the HERACLES rotation curve within $4\kpc$ and the THINGS rotation curves for larger
radii.

NGC 3521 contains a considerable amount of molecular gas in comparison to the other galaxies in
this sample. In \citet{2008AJ....136.2782L} the mass surface densities are plotted, showing that the
molecular gas surface density reaches $\sim50\,\msunpc$. The resultant predicted molecular gas
rotation curve using $\ad2g$ rises steeply within $5\kpc$ and reaches a maximum of $\sim70\kmss$. 

For both the NFW and ISO fits, the addition of molecular gas leads to better fits as
compared to the H\,{\sc i}-only case (cf. Tables \ref{tab:NFW} and \ref{tab:ISO}). In Figures
\ref{fig:N3521-NFW} and \ref{fig:N3521-ISO}, we see that fitting halo models to the
HERACLES/THINGS rotation curves yield better results than when using the THINGS rotation curve. 

We obtain the best fits  when fitting to the ISO halo. The fits using either $\ad2g$ and the Milky
Way value are similar in this case, and produce halo rotation curves which are considerably
different from the H\,{\sc i}-only case. 

\subsection{NGC 4736}
\label{sec:NGC-4736}
\begin{figure*}
	\centering
	\includegraphics[width=17cm]{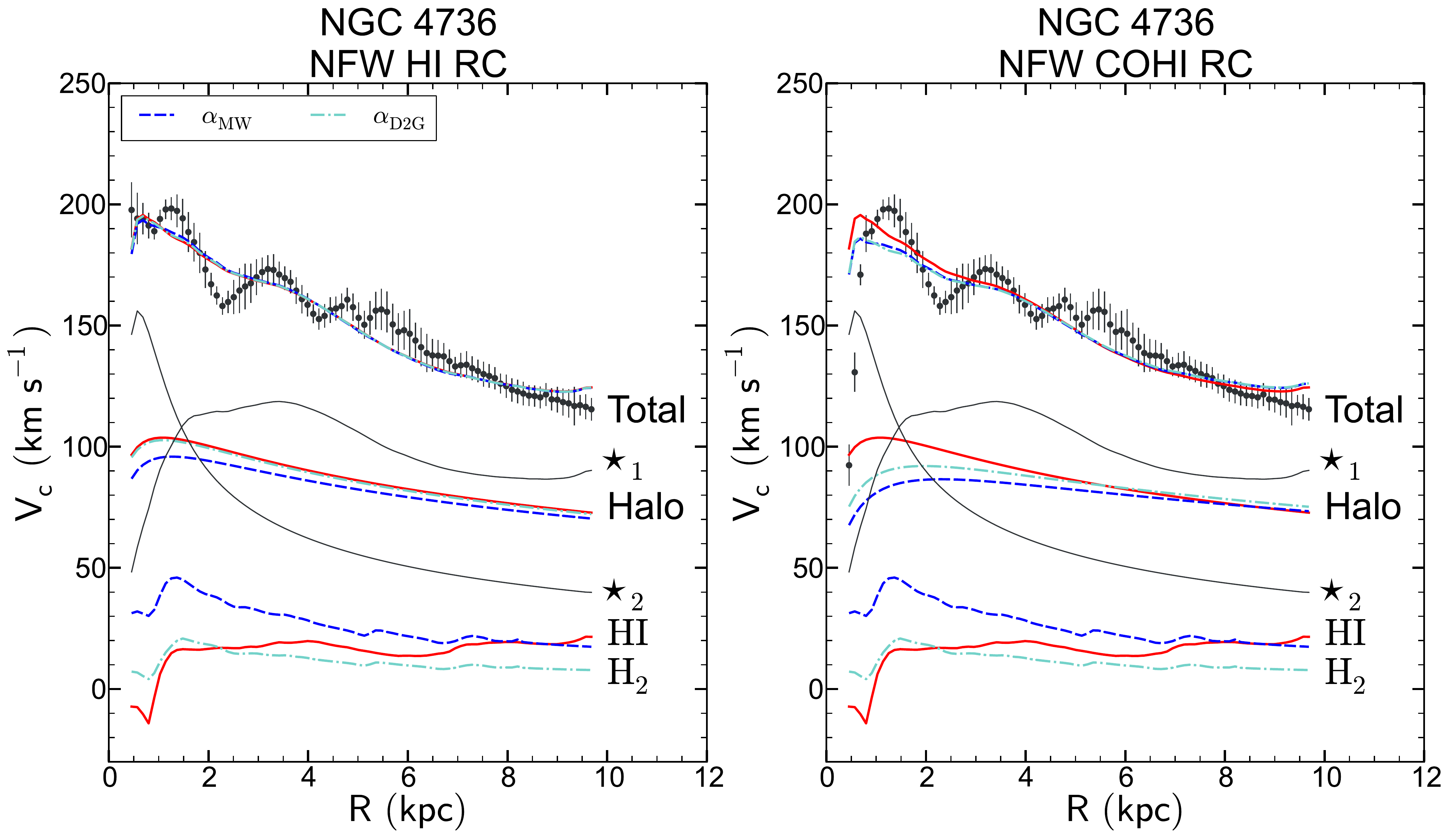}		
	\caption{\label{fig:N4736-NFW}NGC 4736 mass models with an NFW halo.
	Fits to the THINGS rotation curve are plotted in the left panel, fits to the HERACLES/THINGS
	rotation curve are plotted in the right pane. Line styles and colours for each plot are as
	in Figure \ref{fig:N0925-MMs}.}
\end{figure*}

In Figure \ref{fig:N4736-NFW} we plot the results from the mass modelling for NGC 4736. 

Fits using the ISO halos did not converge and resulted in unrealistic fits with $R_c\ll1$ and
$\rho_0\gg1000$, which is most likely due to a sharp decline in the outer part of the rotation
curve. We therefore only include results for the NFW halo.  For NGC 4736 we use a 2-component
stellar decomposition to calculate the stellar predicted rotation curves.

There is a deficiency of H\,{\sc i} in the centre of NGC 4736. Correspondingly, the molecular
gas distribution peaks in the centre of NGC 4736, with the maximum molecular gas mass surface
density reaching $\sim100\,\msunpc$ \citep{2008AJ....136.2782L} using the Milky Way conversion factor.
We therefore consider two observed rotation curves in this work - the THINGS rotation curve which
extends from $\sim0.5\kpc$ outwards and does not track the inner slope of the rotation curve, and a
hybrid HERACLES/THINGS rotation curve where we use the HERACLES rotation curve within $\sim1\kpc$
and the THINGS rotation curve outside this radius.  The average value $\aad2g=1.4\,\auni$ is significantly
smaller than the Milky Way value. This is reflected in Figure \ref{fig:N4736-NFW} - the predicted
molecular gas rotation curve using the Milky Way conversion factor reaches a maximum of
$\sim40\kmss$ while the predicted rotation curve using $\ad2g$ does not exceed $\sim20\kmss$. 

The predicted molecular gas rotation curves are significantly different for the assumed values of
$\mathrm{\alpha_{CO}}$. In addition, mass models using the THINGS rotation curve are significantly better than
those using the HERACLES/THINGS rotation curve.

There are several factors which make interpretation of the results for this galaxy difficult, as
already identified in dB08. Firstly, there is a large uncertainty in the $\Upsilon_{*}$ values for
the central bulge-like component. Secondly, \citet{TRACHTERNACH:2008FK} find evidence for large
non-circular motions in this galaxy, which are also evident as the large spread in velocities along
the minor-axis \textit{pV} diagram in Figure \ref{fig:4736-atlas}. 

Another important factor is the declining rotation curve, which presents difficulties when fitting
either the ISO or the NFW haloes, neither of which were ``designed'' to do so. Fitting the ISO halo
leads to an extremely compact core, and fitting the NFW halo implies a highly concentrated profile. 

\subsection{NGC 5055}
\label{sec:NGC-5055}
\begin{figure*}
	\centering
	\includegraphics[width=17cm]{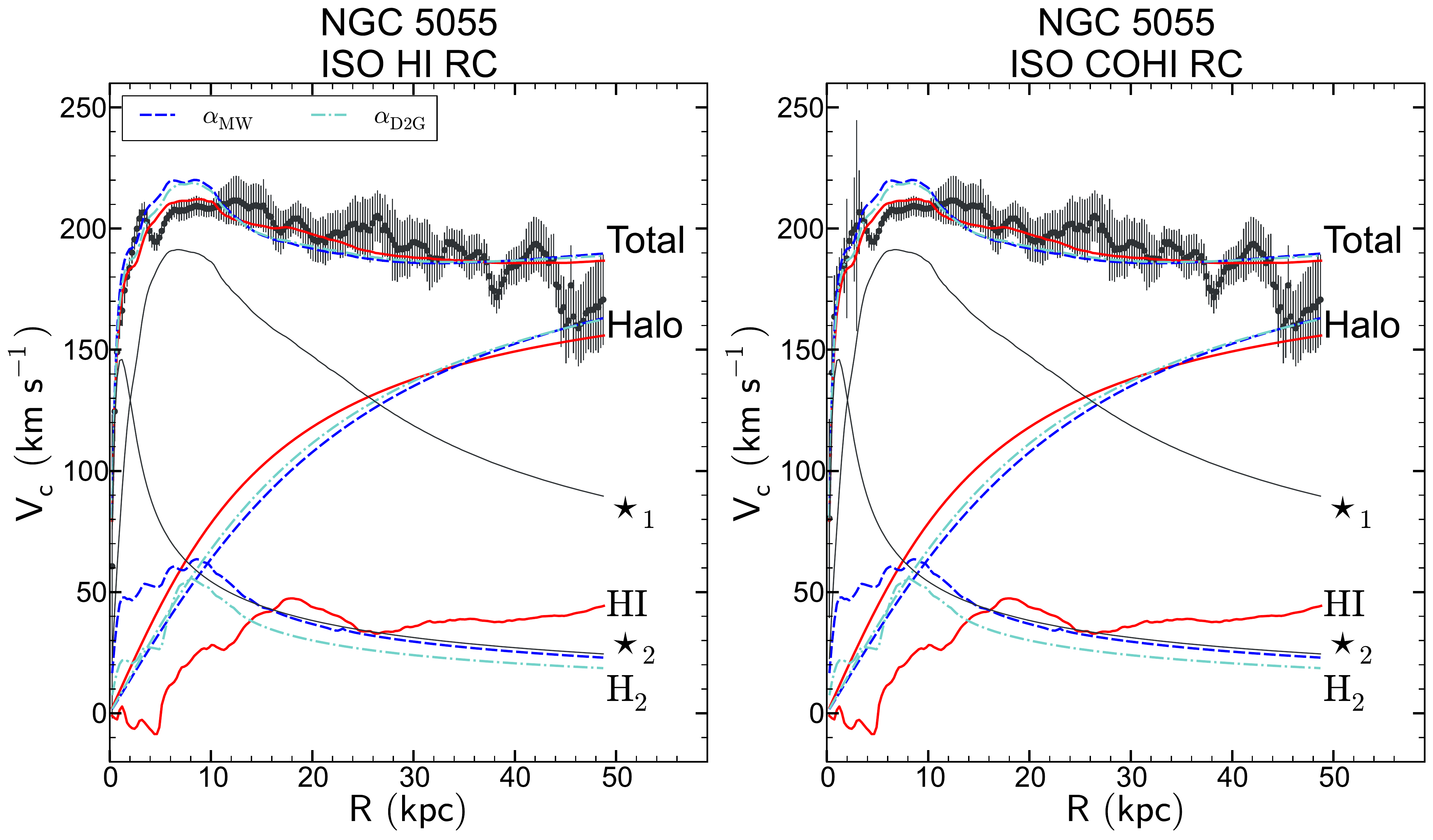}		
	\caption{\label{fig:N5055-ISO}NGC 5055 mass models with an ISO halo.
	Fits to the THINGS rotation curve are plotted in the left panel, fits to the HERACLES/THINGS
	rotation curve are plotted in the right pane. Line styles and colours for each plot are as
	in Figure \ref{fig:N0925-MMs}.}
\end{figure*}

In Figure \ref{fig:N5055-ISO} we plot the results of the mass modelling for NGC 5055.  For NGC 5055
we use a 2-component stellar decomposition to determine the predicted stellar rotation curves. 

For NGC 5055 $\aad2g=5.3\,\auni$, which is very close to the Milky Way value.  However, the radial
profiles for $\ad2g$ shows a depression in the central region, which leads to a substantially
different $\ad2g$ rotation curve in the inner $5\kpc$, as compared to using a single $\amw$ value
for the conversion. 

The HERACLES rotation curve is steeper in the inner $5\kpc$. We therefore consider two observed
rotation curves in our fits --- the THINGS rotation curve and a HERACLES/THINGS rotation curve assembled
by using the HERACLES rotation curve within $5\kpc$ and switching over to the THINGS rotation curve
for larger radii. 

Fits to the NFW halo did not converge to reasonable values. We therefore only plot the fits to the ISO halo. 

Using the ISO profile produces reasonable values of $\rho_0$ and $R_C$.  Fitting to either the
THINGS or HERACLES/THINGS curves yields comparable values of $\rchisq$. The values for $\rho_0$ and $R_C$  are 
different from the H\,{\sc i}-only case, which is evident in the different shapes of the halo
rotation curves. 

\subsection{NGC 6946}
\label{sec:NGC-6946}
\begin{figure*}
	\centering
	\includegraphics[width=17cm]{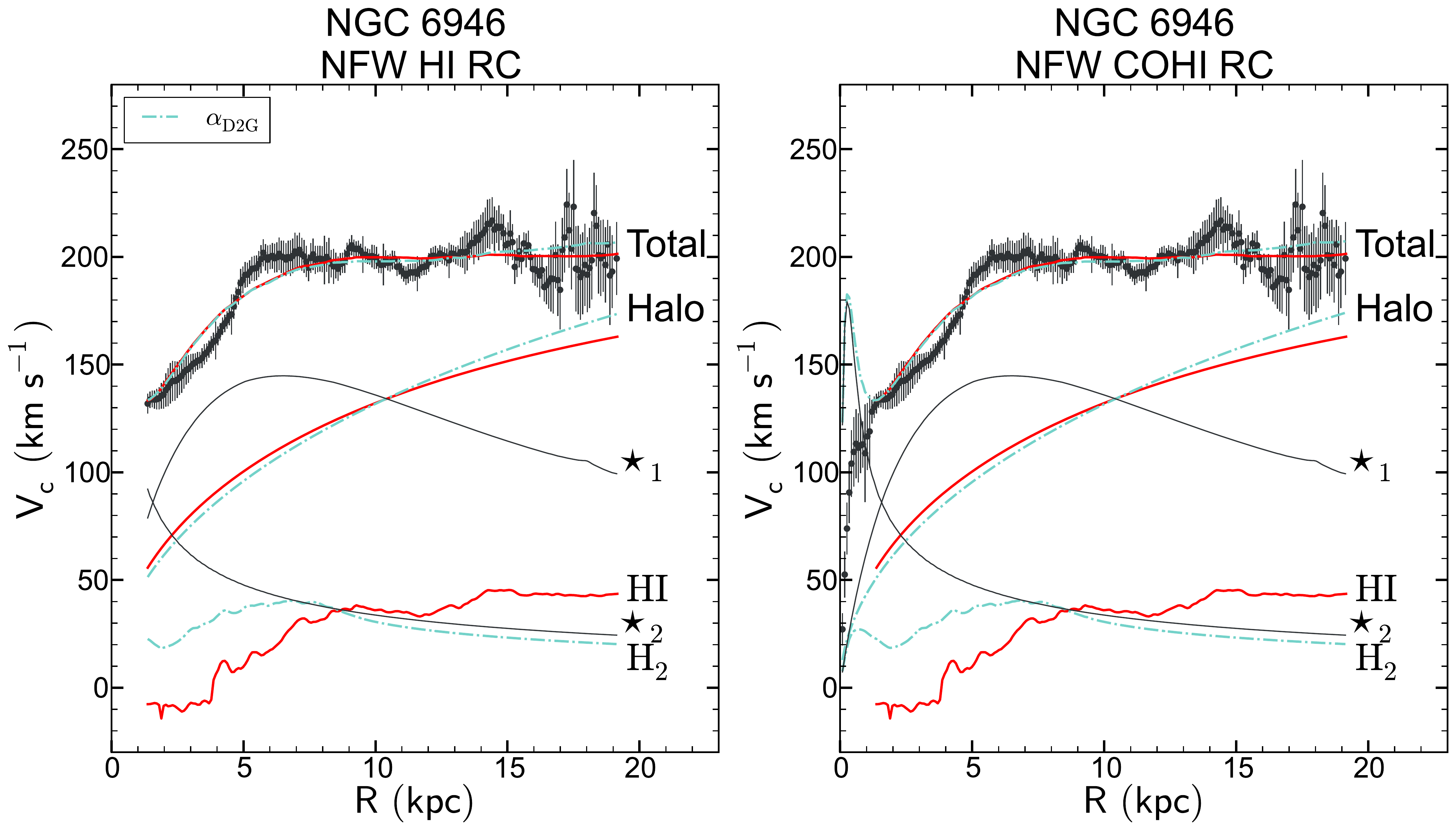}	
	\caption{\label{fig:N6946-NFW}from NGC 6946 mass models with an NFW halo.
	Fits to the THINGS rotation curve are plotted in the left panel, fits to the HERACLES/THINGS
	rotation curve are plotted in the right pane. Line styles and colours for each plot are as
	in Figure \ref{fig:N0925-MMs}.}
\end{figure*}

\begin{figure*}
	\centering
	\includegraphics[width=17cm]{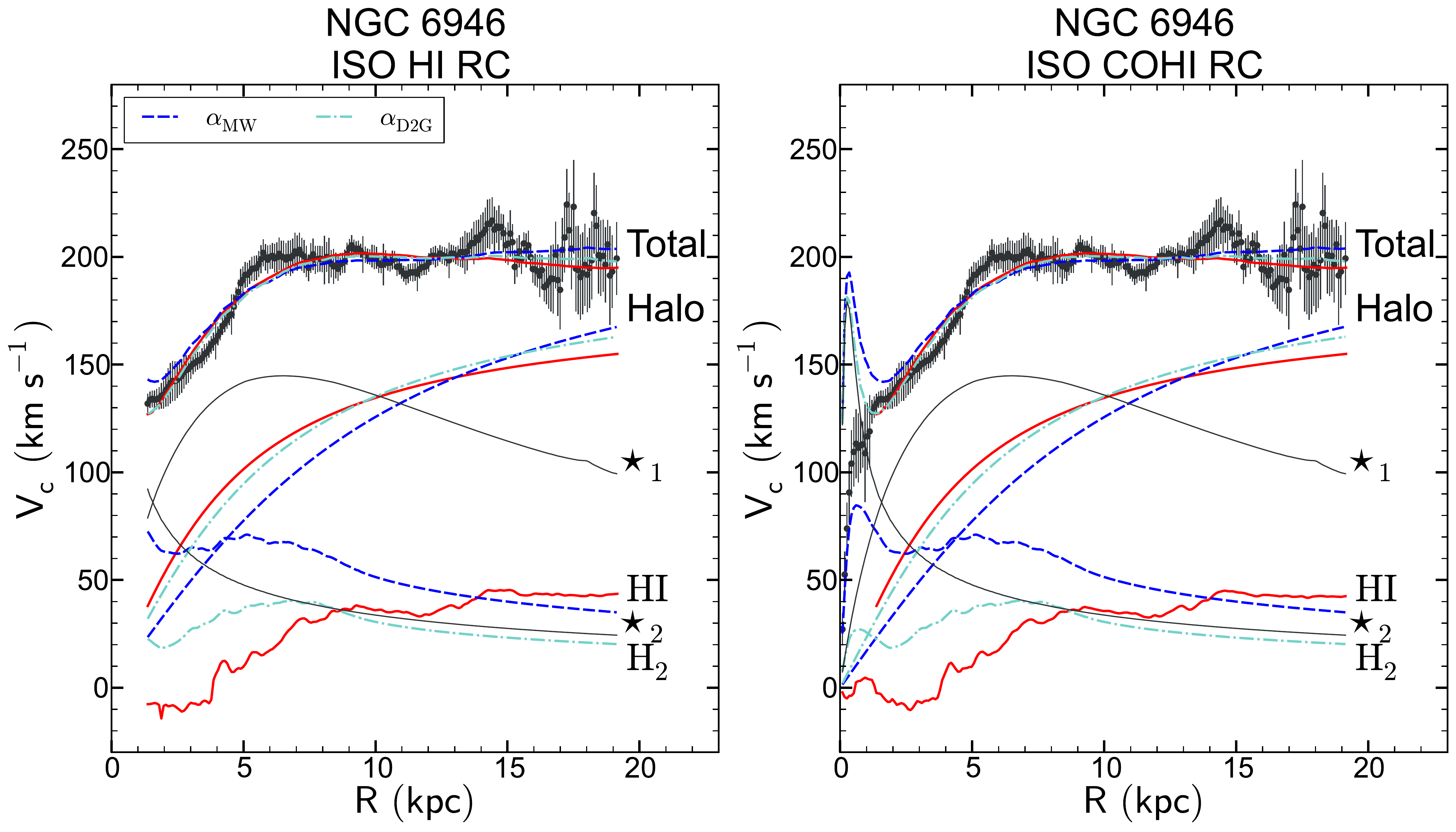}	
	\caption{\label{fig:N6946-ISO}NGC 6946 mass models with an ISO halo.
	Fits to the THINGS rotation curve are plotted in the left panel, fits to the HERACLES/THINGS
	rotation curve are plotted in the right pane. Line styles and colours for each plot are as
	in Figure \ref{fig:N0925-MMs}.}
\end{figure*}

In Figures \ref{fig:N6946-NFW} and \ref{fig:N6946-ISO} we plot the results of the mass modelling for
NGC 6946 for the NFW and ISO halos respectively.  For NGC 6946 we use a 2-component stellar
decomposition to determine the stellar predicted rotation curves. 

NGC 6946 has a large amount of CO detected. Using the Milky Way conversion factor,
\citet{2008AJ....136.2782L} showed that the molecular gas mass surface density is as high as
$\sim400\,\msunpc$. This corresponds to a molecular gas predicted rotation curve which reaches a
maximum of $\sim60\kmss$ at approximately $1\kpc$. The CO emission shows a compact, bulge like
distribution \citep{2008AJ....136.2782L}, similar to the stellar component. 

The H\,{\sc i} observations show a deficiency in the centre of NGC 6946, where we
find an abundance of CO. This allows us to fill in the inner part of the
rotation curve using the HERACLES rotation curve. We therefore
consider two rotation curves: the THINGS rotation curve, which starts from
approximately $1\kpc$ and the HERACLES/THINGS rotation curve where we use
the HERACLES curve in the inner $1\kpc$. 

In both the NFW and ISO cases the fitted total rotation curve significantly
overshoots the inner part of the HERACLES/THINGS rotation curve. As with NGC 4736, the
$\Upsilon_{*}$ value has a large uncertainty. 

For fits using the NFW halo the addition of the molecular gas leads to a large difference in the
derived parameters. Using the predicted molecular gas rotation curve with $\amw$ leads to a poor
fit, so we do not plot the results here, neither do we include them in Table \ref{tab:NFW}. 

For fits using the ISO halo to the THINGS rotation curve the $\rchisq$ is slightly better upon the
addition of $\mathrm{H_2}$. 

\subsection{NGC 7331}
\label{sec:NGC-7331}
\begin{figure}
	\centering
	\resizebox{\hsize}{!}{\includegraphics[width=17cm]{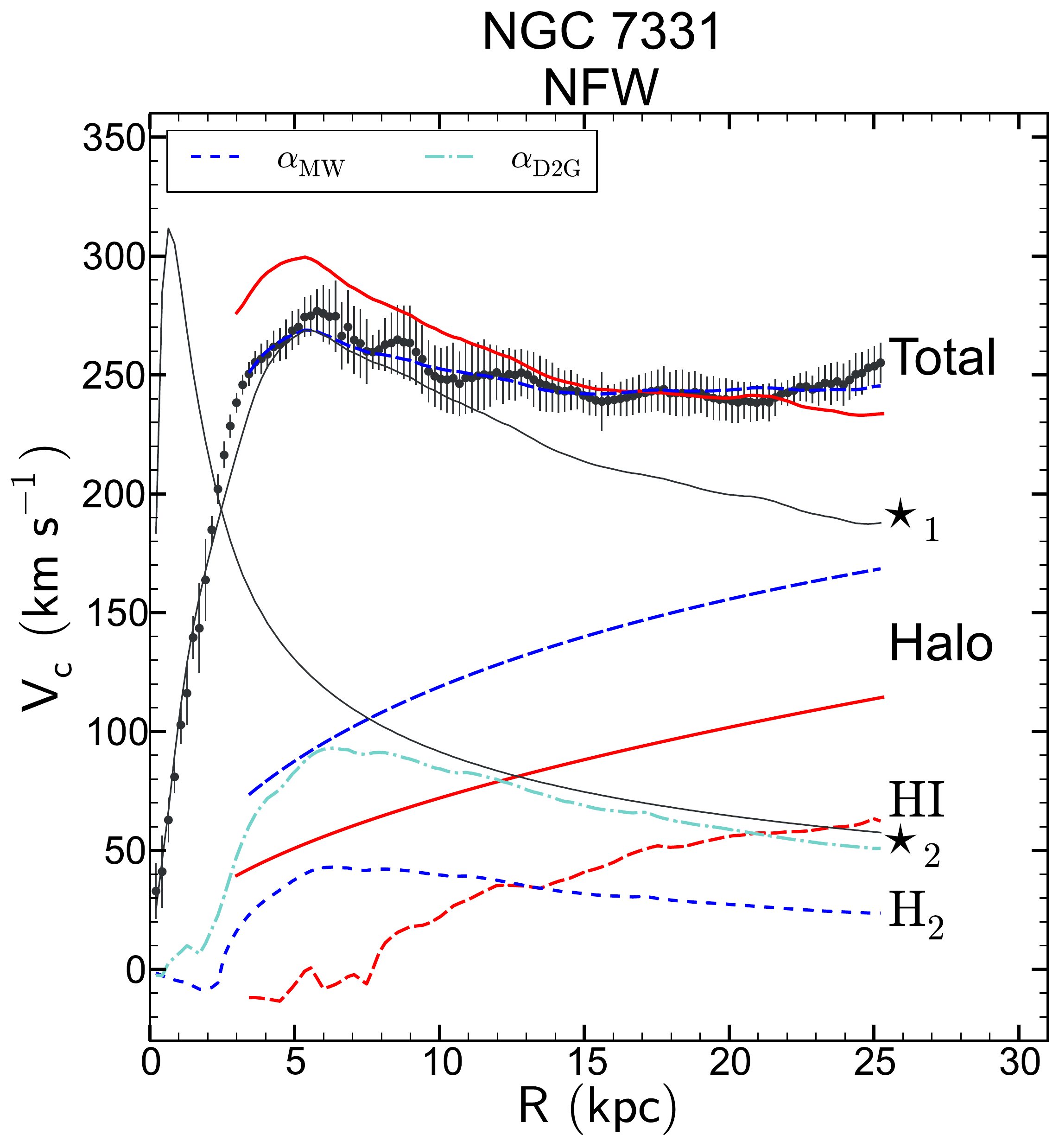}}
	\caption{\label{fig:N7331-NFW}NGC 7331 mass models with an NFW halo
	to the hybrid HERACLES/THINGS rotation curve. Line styles and colours are as in Figure
	\ref{fig:N0925-MMs}.}
\end{figure}

In Figure \ref{fig:N7331-NFW} we plot the results of the mass modelling for NGC 7331 using the NFW
halo.  For NGC 7331 we use a 2-component stellar decomposition to determine the stellar predicted
rotation curves. 

Our general method has been to use a photometrically determined $\Upsilon_{*}$ with a Kroupa IMF to
calculate the stellar mass surface density. However, for NGC 7331 this combination of $\Upsilon_{*}$
and IMF predicts disks which are too massive which leads to predicted stellar rotation curves which
are much higher than the observed rotation curve (see dB08). 

dB08 addressed this by using a radially constant value of $\Upsilon_{*}$ for each stellar component
when determining the stellar mass surface density. This leads to reasonable fits to the observed
rotation curve. We use these stellar mass surface densities to calculate the predicted stellar
rotation curves. 

For NGC 7331 $\aad2g=14.0\,\auni$ is much larger than the Milky Way value.  \citet{2008AJ....136.2782L}
showed that the $\mathrm{H_2}$ mass surface density peaks at slightly more than $20\,\msunpc$ assuming
$\amw$, and is concentrated on a ring at approximately $3\kpc$ away from the centre of the galaxy.
The predicted molecular gas rotation curve  shows a maximum of $\sim70\kmss$ using $\ad2g$, and a
maximum of $\sim40\kmss$ using $\amw$.

The H\,{\sc i} shows a deficiency in the centre, and although the CO shows a similar depression,
there is sufficient emission detected for a CO rotation curve to be derived. We therefore consider
both an H\,{\sc i}-only THINGS rotation curve and a hybrid HERACLES/THINGS rotation curve.

In Figure \ref{fig:N7331-NFW} we show the fit using the NFW model rotation curves using the
predicted molecular gas rotation curve derived using $\amw$.  This shows that the H\,{\sc i}-only
fits overshoot the observed rotation curve within $\sim20\kpc$.

Attempting to fit a mass model using an ISO halo model does not lead to convergence and results in
severely unrealistic values of the halo parameters, and are not plotted here.  

\section{Summary}
\label{sec:discussion-summary}
We summarize the mass model parameters for the NFW and ISO fits in Tables \ref{tab:NFW} and
\ref{tab:ISO} respectively. 

\begin{figure}[H]
	\centering
	\resizebox{\hsize}{!}{\includegraphics{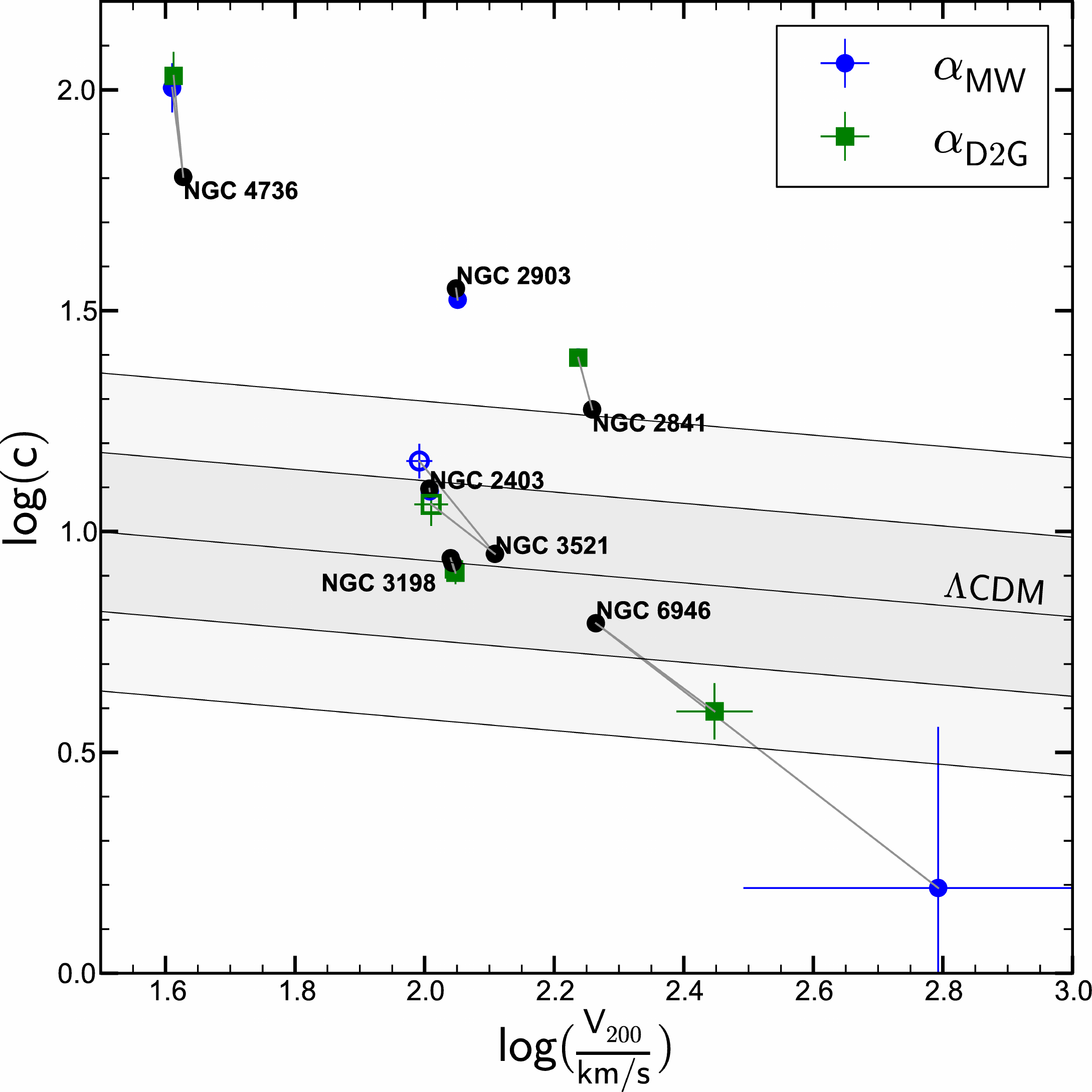}}
	\caption{\label{fig:NFW} Plot of fitted halo parameters for the NFW mass models, $\log\,(c)$
	vs. $\log\,(V_{200}/\kmss)$. Each point corresponds to the NFW model parameters that
	produces the lowest $\rchisq$ value for each galaxy, for each value of
	$\alpha_\mathrm{CO}$.  Green squares correspond to models where we use $\ad2g$; blue
	circles correspond to models where we use the Milky Way value. Filled symbols correspond to
	cases where the best fit was obtained with the THINGS rotation curve; unfilled symbols
	correspond to those with the HERACLES rotation curve. The H\,{\sc i}-only values are
	plotted in black circles, and a line joins the H\,{\sc i}-only value to the
	corresponding values calculated with added molecular gas, for each galaxy. The 1-
	and 2- component models are indicated for NGC 3198.  We plot a line of expected values from
$\mathrm{\Lambda CDM}$, bounded by $1\,\sigma$ and $2\,\sigma$ regions.}
\end{figure}

We plot the NFW parameters from Table \ref{tab:NFW} in Figure \ref{fig:NFW} for both the models with
and without molecular gas. We also plot the expected values using $\mathrm{\Lambda CDM}$ parameters
\citep{2003MNRAS.340..657D} and the $1\,\sigma$ and $2\,\sigma$ scatter from \citet{2001MNRAS.321..559B}.
This plot shows that the NFW halo parameters for NGC 2403, NGC 3198, NGC 3521 and NGC 6946 lie
within the $1\,\sigma$ region for the H\,{\sc i}-only case. The addition of the molecular gas pushes
these values away from the region of expected values. 

\begin{figure}[H]
	\centering
	\resizebox{\hsize}{!}{\includegraphics{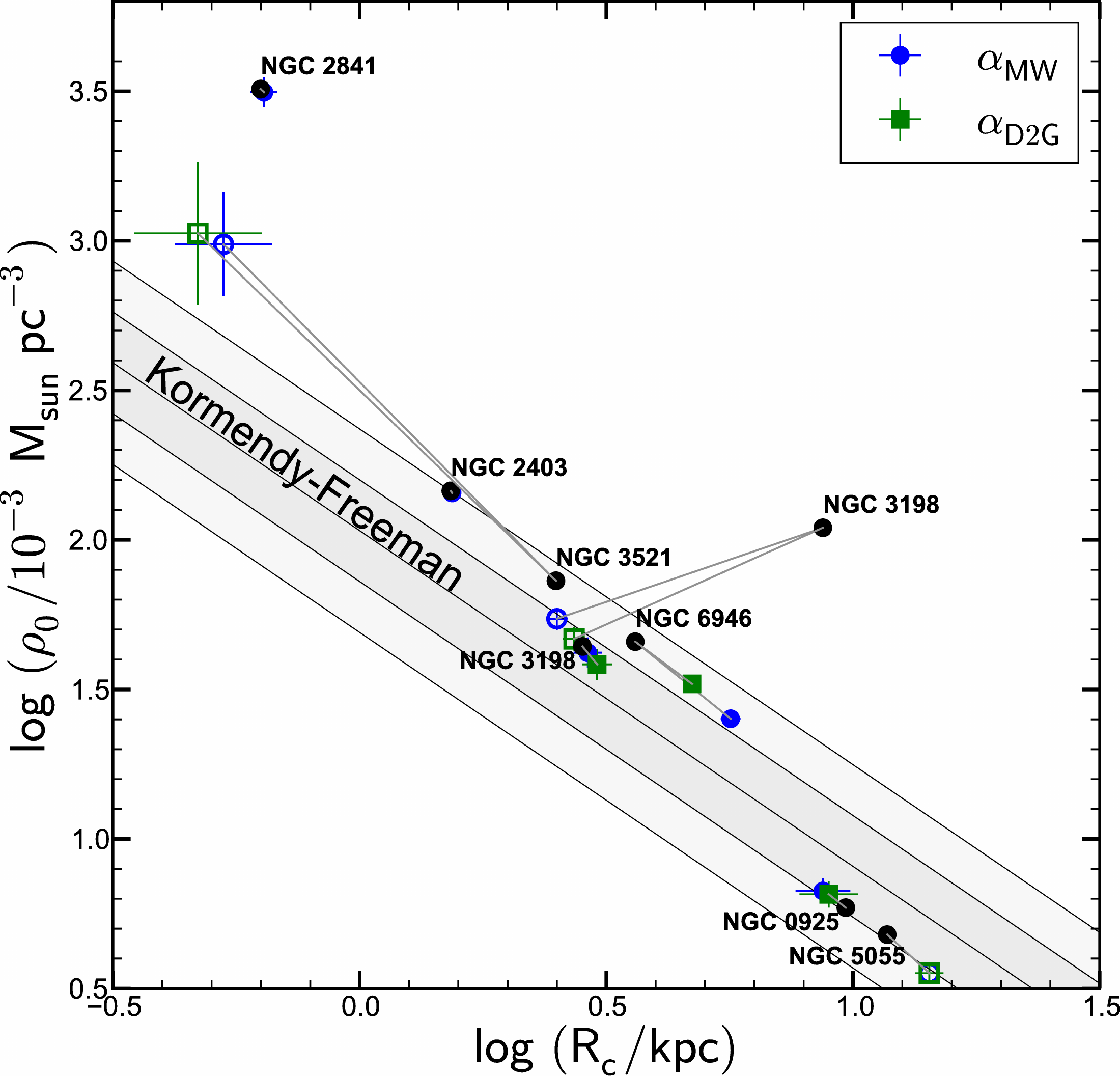}}
	\caption{\label{fig:ISO} Plot of fitted halo parameters for the ISO mass models,
	$\mathrm{\log(\rho_0/10^{-3}M_{\odot}\,pc^{-3})}$ vs. $\mathrm{\log{(R_c/\kpc)}}$. Each
	point corresponds to the ISO model parameters that produces the lowest $\rchisq$ value for
	each galaxy, for each value of $\alpha_\mathrm{CO}$.  Symbols are as in Figure
	\ref{fig:NFW}.  The 1- and 2- component models are indicated for NGC 3198.  We plot a line
	of expected values from \citet{2004IAUS..220..377K}, bounded by $1\,\sigma$ and $2\,\sigma$
	regions from \citet{2001MNRAS.321..559B}}
\end{figure}

We plot the ISO parameters from Table \ref{tab:ISO} in Figure \ref{fig:ISO} for both the models with
and without molecular gas. We  plot the expected values suggested by \citet{2004IAUS..220..377K} and
the region corresponding to a $1\,\sigma$ and $2\,\sigma$ scatter. Here there is no systematic trend
in the parameters upon the addition of molecular gas. 

In Figures \ref{fig:NFW} and \ref{fig:ISO} we plot parameters corresponding to the model which
produces the lowest $\rchisq$ for each galaxy, for each value of $\alpha_\mathrm{CO}$. 

It is important to note the difference in the predicted velocities for the molecular gas rotation
curves that arises from a different choice of $\mathrm{\alpha_{CO}}$. For some galaxies the $\amw$
and $\ad2g$ predicted molecular gas rotation curves are very similar, e.g., NGC 925 and NGC 2841.
For others, the predicted molecular gas rotation curves can be quite different for different choices
of $\mathrm{\alpha_{CO}}$ - especially in the inner $10\kpc$. In this region, the addition of the
molecular gas makes the largest difference.

The results for the galaxies in our sample fall into two groups. Firstly, for NGC 925, NGC 2403,
NGC 2903, NGC 2976 and NGC 3198 the addition of the molecular gas does not make a substantial
difference to the halo model parameters and the shape of the halo rotation curves. This is largely
because the contribution of the molecular gas, in comparison with the halo and the stellar
components, is insignificant. This result is independent of the value of conversion factor
$\mathrm{\alpha_{CO}}$ used to calculate the molecular gas mass surface density.

Secondly, for NGC 2841, NGC 3521, NGC 4736, NGC 5055, NGC 6496 and NGC 7331 the halo parameters and
halo rotation curves change noticeably upon the addition of molecular gas. The value of
$\mathrm{\alpha_{CO}}$ used in converting CO luminosity to molecular gas mass surface density
slightly affects the halo parameters and rotation curve derived in the mass model. 

We have also shown that the CO-TFR is shallower than the H\,{\sc i}-TFR for our sample of galaxies
--- which is due to the molecular gas distribution being more compact than the more extended, flat
atomic gas distribution. For NGC 2903 we have done a brief analysis of the non-circular motions due
to the bar. Our results for this galaxy are in reasonable agreement with previous work, but we
discuss reasons why the corrected rotation curve cannot be used for the mass model analysis
presented here.

This study has only investigated a limited number of galaxies and galaxy types. In addition we have
also considered a limited range in $\Upsilon_{*}$. Future studies with larger samples can
investigate the effect of using different $\mathrm{\alpha_{CO}}$ conversion factors, as well as the
interplay with the stellar mass-to-light ratio $\Upsilon_{*}$.

To conclude, in this work we have, for the first time, included the molecular gas component into the
mass models for a comprehensive sample of nearby galaxies, using high resolution rotation curves and
different values of the $\mathrm{\alpha_{CO}}$ conversion factor. The impact of this addition
changes from galaxy to galaxy depending on the molecular gas content. For the galaxies in our sample
where the molecular gas content is the highest, the impact on the mass models can be significant.

\section{Acknowledgements}
B.S.F. and C.C. acknowledges support provided by the South African Research Chairs Initiative of the
Department of Science and Technology and National Research Foundation. B.S.F. further acknowledges
support from the UCT Science Faculty Research Committee's Postgraduate Publication Incentive (PPI)
funding, and the funding from the European Research Council under the European Union's Seventh
Framework Programme (FP/2007-2013) / ERC Advanced Grant RADIOLIFE-320745.  W.J.G.dB. was supported
by the European Commission (grant FP7-PEOPLE- 2012-CIG 333939). 

\begin{table*}
	\caption{ \label{tab:NFW}Fitted halo parameters for the NFW Model.}
	\centering
	\begin{tabular}{l c c c c c c c c c}
		\hline
		\hline
		Galaxy & Rotcur & $\mathrm{\overline{\alpha}_{CO}}$ \footnote{The $\aad2g$ values are italicized.} & $c$ & $V_{200}$ & $\chi^2_r$ & $c$ & $V_{200}$ & $\chi^2_r$ \\
		 &  & ($\aunif$) &  & ($\kmss$) &  &  & ($\kmss$) &  \\
		 &   &  &  &  &  & H\,{\sc i} only & H\,{\sc i} only & H\,{\sc i} only \\
		\hline
		NGC 2403 (1 comp) 	& H\,{\sc i} 	& 6.3 		& $12.2\pm0.2$ 		& $102.3\pm0.7$ 	& 0.6 		& $12.4\pm0.2$ 	& $101.7\pm0.7$ 	& 0.6 \\
		NGC 2403 (2 comp) 	& H\,{\sc i} 	& 6.3 		& $12.1\pm0.2$	 	& $102.3\pm0.8$ 	& 0.6 		& $12.3\pm0.2$ 	& $102.2\pm0.7$ 	& 0.6 \\
		NGC 2841 		& H\,{\sc i} 	& 6.3 		& $24.7\pm0.4$	 	& $172.7\pm1.0$ 	& 0.6 		& $18.9\pm0.4$ 	& $181.4\pm1.0$ 	& 0.2 \\ 
		NGC 2841 		& H\,{\sc i} 	& \textit{7.1}	& $24.8\pm0.4$	 	& $172.6\pm1.0$ 	& 0.6 		& 	& 	& \\
		NGC 3198 (1 comp) 	& COH\,{\sc i} 	& 6.3 		& $8.6\pm0.4$ 		& $109.7\pm1.8$ 	& 1.5 		& $8.7\pm0.4$ 	& $109.7\pm1.7$ 	& 1.3 \\
		NGC 3198 (1 comp) 	& H\,{\sc i} 	& 6.3 		& $8.3\pm0.4$ 		& $110.7\pm1.8$ 	& 1.4 		&  	&  	&  \\
		NGC 3198 (1 comp) 	& COH\,{\sc i} 	& \textit{15.7}	& $8.3\pm0.4$ 		& $110.3\pm1.8$ 	& 1.4 		&  	& 	& \\
		NGC 3198 (1 comp) 	& H\,{\sc i} 	& \textit{15.7}	& $8.0\pm0.4$ 		& $111.3\pm1.9$ 	& 1.4 		& 	& 	& \\
		NGC 3198 (2 comp) 	& COH\,{\sc i} 	& 6.3 		& $8.4\pm0.5$ 		& $110.5\pm2.4$ 	& 2.5 		& $8.5\pm0.5$ 	& $110.4\pm2.2$ 	& 2.1 \\
		NGC 3198 (2 comp) 	& H\,{\sc i} 	& 6.3 		& $8.2\pm0.5$ 		& $111.3\pm2.4$ 	& 2.2 		&  	&  	& \\
		NGC 3198 (2 comp) 	& COH\,{\sc i} 	& \textit{15.7}	& $8.1\pm0.5$ 		&  $111.2\pm2.5$ 	& 2.4 		&  	&  	& \\
		NGC 3198 (2 comp) 	& H\,{\sc i} 	& \textit{15.7}	& $7.9\pm0.5$ 		& $112.0\pm2.4$ 	& 2.1 		&  	&  	&  \\
		NGC 3521 		& COH\,{\sc i} 	& 6.3 		& $14.5\pm1.3$	 	& $98.1\pm4.5$ 		& 1.2 		& $8.9\pm2.0$ 	& $128.4\pm16.4$ 	& 5.6 \\
		NGC 3521 		& H\,{\sc i} 	& 6.3 		& $6.4\pm1.8$ 		& $139.6\pm24.2$ 	& 5.2 		&  	&  	&  \\
		NGC 3521 		& COH\,{\sc i} 	& \textit{10.9}	& $13.6\pm1.3$	 	& $98.5\pm4.9$ 		& 1.3 		&  	&  	&  \\
		NGC 3521 		& H\,{\sc i} 	& \textit{10.9}	& $5.8\pm1.8$ 		& $143\pm27$ 		& 5.1 		&  	&  	&  \\
		NGC 4736 		& COH\,{\sc i} 	& 6.3 		& $54.8\pm10.5$ 	& $43.9\pm2.0$ 		& 3.4 		& $63.5\pm24.2$ 	& $42.4\pm1.7$ 	& 1.4 \\
		NGC 4736 		& H\,{\sc i} 	& 6.3		& $91.9\pm12.3$ 	& $40.5\pm1.0$ 		& 1.5 		& 	&  	&  \\
		NGC 4736 		& COH\,{\sc i} 	& \textit{1.4} 	& $64.8\pm11.6$ 	& $44.0\pm1.8$ 		& 3.5 		&  	&  	& \\
		NGC 4736 		& H\,{\sc i} 	& \textit{1.4}	& $108.3\pm13.7$ 	& $40.9\pm1.0$ 		& 1.5 		& 	& 	&  \\
		NGC 6946 		& COH\,{\sc i} 	& \textit{2.9}	& $3.4\pm1.0$ 		& $313.0\pm84.3$ 	& 3.2           & $6.2\pm0.5$ & $183.8\pm11.1$ & 1.03 \\
		NGC 6946 (outer) 	& H\,{\sc i} 	& \textit{2.9}	& $3.6\pm0.6$ 		& $296.2\pm239.6$ 	& 1.2   \\
		NGC 7331 (outer) 	& H\,{\sc i} 	& 6.3 		& $4.0\pm0.4$ 		& $223.0\pm14.1$ 	& 0.2           & $4.9\pm0.4$ & $200.0\pm10.7$ & 0.24 \\ 
		\hline
	\end{tabular}
	\tablecomments{NFW Parameters: Fitted parameters  $c$ and $V_{200}\,(\kmss)$  and associated uncertainties for
	the NFW halo. For each galaxy we indicate the number of stellar components used, the
	rotation curve and the value of $\mathrm{\overline{\alpha}_{CO}}$ used to compute the
	molecular gas mass surface density. We also show the corresponding H\,{\sc i}-only values
	and errors for each set of parameters where appropriate. We denote HERACLES/THINGS rotation curves
	as COH\,{\sc i}, and THINGS rotation curves as H\,{\sc i}.}
\end{table*}


\begin{table*}
	\caption{\label{tab:ISO}Fitted halo parameters for the ISO Model.}
	\centering
	\begin{tabular}{l c c c c c c c c c}
		\hline
		\hline
		Galaxy & Rotcur & $\mathrm{\overline{\alpha}_{CO}}$ \footnote{The $\aad2g$ values are italicized.}  & $R_\mathrm{c}$ & $\rho_0$ & $\chi^2_r$ & $R_\mathrm{c}$ & $\rho_0$ & $\chi^2_r$ \\
		 &  & ($\aunif$) & ($\kpc$) & $\mathrm{(10^{-3}\,\msun\,pc^{-3})}$ &  & ($\kpc$) & $\mathrm{(10^{-3}\,\msun\,pc^{-3})}$ & \\
		 &   &  &   &  &  & H\,{\sc i} only & H\,{\sc i} only & H\,{\sc i} only \\
		\hline
		NGC 0925 		& H\,{\sc i} 	& 6.3 	& $9.0\pm1.3$ 	& $6.4\pm0.7$ 	& 1.5 & $9.7\pm1.3$ & $5.9\pm0.5$ & 1.1 \\
		NGC 0925 		& H\,{\sc i} 	& \textit{14.3} & $8.9\pm1.2$ 	& $6.5\pm0.7$ 	& 1.5 \\
		NGC 2403 (1 comp) 	& H\,{\sc i} 	& 6.3 	& $1.5\pm0.1$ 	& $144.4\pm7.1$ & 1.0 & $1.49\pm0.05$ & $152.8\pm7.5$ & 1.0 \\
		NGC 2403 (2 comp) 	& H\,{\sc i} 	& 6.3 	& $1.6\pm0.1$ 	& $138.0\pm6.6$ & 1.0 & $1.52\pm0.04$ & $145.8\pm6.9$ & 1.0 \\
		NGC 2841	 	& H\,{\sc i} 	& 6.3 	& $0.7\pm0.1$ 	& $3035\pm335$ 	& 0.2 & $0.63\pm0.04$ & $3215.3\pm371.8$ & 0.2 \\
		NGC 2841	 	& H\,{\sc i} 	& \textit{7.1} 	& $0.6\pm0.1$ 	& $3163\pm358$ 	& 0.2 &   \\
		NGC 2976	 	& H\,{\sc i} 	& 6.3 	& $4.1\pm1.9$ 	& $35.3\pm4.0$ 	& 0.7 &  $5.1\pm2.5$ & $35.5\pm3.1$ & $0.5$ \\
		NGC 2976	 	& H\,{\sc i}	& \textit{4.7} 	& $2.6\pm0.7$ 	& $42.8\pm4.9$ 	& 0.8 &   \\
		NGC 3198 (1 comp) 	& COH\,{\sc i}  & 6.3 	& $2.7\pm0.1$ 	& $48.5\pm4.3$ 	& 0.9 & $2.7\pm0.1$ & $46.9\pm4.0$ & 0.8 \\
		NGC 3198 (1 comp) 	& H\,{\sc i}	& 6.3 	& $2.9\pm0.2$ 	& $42.2\pm3.8$ 	& 0.9  \\
		NGC 3198 (1 comp) 	& COH\,{\sc i}  & \textit{15.7} & $2.8\pm0.1$ 	& $45.2\pm4.0$ 	& 0.9  \\
		NGC 3198 (1 comp) 	& H\,{\sc i} 	& \textit{15.7} & $3.0\pm0.2$ 	& $39.1\pm3.5$ 	& 0.9  \\
		NGC 3198 (2 comp) 	& COH\,{\sc i}  & 6.3 	& $2.8\pm0.2$ 	& $45.1\pm5.5$ 	& 1.7 & $2.8\pm0.1$ & $44.0\pm5.1$ & 1.4 \\
		NGC 3198 (2 comp) 	& H\,{\sc i} 	& 6.3 	& $3.0\pm0.2$ 	& $39.7\pm4.7$ 	& 1.5 \\
		NGC 3198 (2 comp) 	& COH\,{\sc i} 	& \textit{15.7} & $2.9\pm0.3$ 	& $41.9\pm5.1$ 	& 1.7  \\
		NGC 3198 (2 comp) 	& H\,{\sc i} 	& \textit{15.7} & $3.1\pm0.2$ 	& $36.8\pm4.3$ 	& 1.5  \\
		NGC 3521 		& COH\,{\sc i} 	& 6.3 	& $0.5\pm0.1$ 	& $973\pm388$ 	& 1.0 & $2.5\pm0.7$ & $73.0\pm30.6$ & 4.8 \\
		NGC 3521 		& H\,{\sc i} 	& 6.3 	& $3.4\pm0.9$ 	& $39.0\pm15.3$ & 4.7  \\
		NGC 3521 		& COH\,{\sc i} 	& \textit{10.9} & $0.5\pm0.1$ 	& $1024\pm451$ 	& 1.1  \\
		NGC 3521 		& H\,{\sc i} 	& \textit{10.9} & $3.6\pm1.0$ 	& $33.4\pm13.1$ & 4.6 \\
		NGC 5055 		& COH\,{\sc i} 	& 6.3 	& $18.6\pm2.0$ 	& $2.6\pm0.3$ 	& 2.5 & $11.7\pm0.7$ & $4.8\pm0.4$ & 1.0 \\
		NGC 5055 		& H\,{\sc i} 	& 6.3 	& $18.6\pm2.0$ 	& $2.6\pm0.3$ 	& 2.7 \\
		NGC 5055 		& COH\,{\sc i} 	& \textit{5.3} 	& $16.7\pm1.4$ 	& $3.0\pm0.3$ 	& 1.8  \\
		NGC 5055 		& H\,{\sc i} 	& \textit{5.3} 	& $16.6\pm1.5$ 	& $3.1\pm0.3$ 	& 1.9  \\
		NGC 6946 		& COH\,{\sc i} 	& 6.3 	& $7.8\pm0.8$ 	& $16.3\pm1.7$ 	& 3.8 & $3.6\pm0.2$ & $45.7\pm3.0$ & 1.0 \\
		NGC 6946 (outer) 	& H\,{\sc i} 	& 6.3 	& $7.8\pm0.5$ 	& $16.4\pm1.0$ 	& 1.3  \\
		NGC 6946		& COH\,{\sc i} 	& \textit{2.9} 	& $4.8\pm0.4$ 	& $31.4\pm3.0$ 	& 2.9  \\
		NGC 6946 (outer) 	& H\,{\sc i} 	& \textit{2.9} 	& $4.8\pm0.2$ 	& $31.4\pm1.7$ 	& 0.9  \\
		\hline
	\end{tabular}
	\tablecomments{ISO Parameters : fitted parameters
	$\rho_0\,\mathrm{(10^{-3}\,\msun\,pc^{-3})}$ and $R_\mathrm{c}\,(\kpc)$  and associated uncertainties
	for the ISO halo. For each galaxy we indicate the number of stellar components used, the
	rotation curve and the value of $\mathrm{\overline{\alpha}_{CO}}$ used to compute the molecular gas
	mass surface density. We also show the corresponding H\,{\sc i}-only values and errors from dB08 for
	each set of parameters where appropriate. We denote HERACLES/THINGS rotation curves as COH\,{\sc i},
	and THINGS rotation curves as H\,{\sc i}.}
\end{table*}


\appendix
\section{Velocity Fields and Tilted-ring Models}
\label{app:velfields}
For each galaxy we present the IWM and $\her3$ velocity fields derived in
dB08 and in this work. We plot the CO distribution overlaid on the SINGS
$3.6\mu$m image. We plot major- and minor-axis \textit{pV} diagrams, where we have
used the position angle from the THINGS data to make the \textit{pV} slices. We also
plot the H\,{\sc i} and CO TM and HM rotation curves. These are all plotted
in Figures \ref{fig:925-atlas}-\ref{fig:7331-atlas}. Each plot is organized
as follows:

\textit{Top Row.} Left: H\,{\sc i} IWM velocity field. The THINGS
systemic velocity $V_{\mathrm{sys}}$ is plotted with a thick black contour, the
approaching contours are plotted in white, receding contours are plotted in
black. The velocity increments $\Delta V$ between contours are indicated in the figure
caption. Centre: The H\,{\sc i} $\her3$ velocity field contours are the
same as for velocity field in the left panel. The colour bar on the right
of this panel denotes the velocity spread for all the velocity fields in
these plots in units of $\kmss$. Right: CO integrated surface brightness contours are plotted in
red on top of the SINGS $3.6\mu$m image (greyscale); the contour levels are
generally given by $I_\mathrm{CO}=I_0,I_0+\Delta I,...\,\mathrm{K}\,\kmss$ where
$I_0$ and $\Delta I$ are denoted in the caption. For some galaxies (e.g., NGC
2903) levels are given by $I_\mathrm{CO}=2^N\,\mathrm{K}\,\kmss$ where the values
of $N$ are denoted in the figure caption. The CO and $3.6\mu$m levels are denoted
in each caption. 

\textit{Middle Row.} Left: CO IWM velocity field. The
systemic velocity $V_{\mathrm{sys}}$ is plotted with a thick black contour, the
approaching contours are plotted in white, receding contours are plotted in
black. The velocity increments $\Delta V$ are indicated in the figure
caption. Middle: CO $\her3$ velocity field which was used to calculate the
tilted-ring models, contours are the same as for velocity field in the left
panel. Right: Difference velocity field, showing the difference between the
H\,{\sc i} and CO $\her3$ velocity field. The velocity colour scale
corresponds to the colour bar to the left of this panel in units of
$\kmss$. Black contours correspond to $\pm5\kmss$, thicker dark grey
contours correspond to $\pm10\kmss$.

\textit{Bottom Row.} Left: Position
velocity (\textit{pV}) major-axis diagram; the H\,{\sc i} is plotted in filled
contours from $\sigma_{\mathrm{HI}}$ upwards in multiples of $\sigma_{\mathrm{HI}}$ and the
CO is plotted in red contours from $\sigma_{\mathrm{CO'}}$ upwards in multiples of
$\sigma_{\mathrm{CO'}}$. These values are tabulated in Table \ref{tab:noise}. The
major-axis position angle is indicated in the panel title.  Middle: \textit{pV}
minor-axis diagram. Right: Rotation curves presented in this work. The
H\,{\sc i} rotation curves derived in dB08 are plotted in red unfilled
circles. For galaxies where we only calculate TM rotation curves, we plot
the CO rotation curve in filled dark-grey squares with error-bars. For
galaxies where we calculate TM and HM rotation curves, we plot the TM
rotation curves in unfilled light-grey squares, and we plot the HM rotation
curves in filled black circles with errorbars. 

\subsection{NGC 925}

\begin{figure*}[H]
	\centering
	\resizebox{\hsize}{!}{\includegraphics[origin=lb,width=22cm]{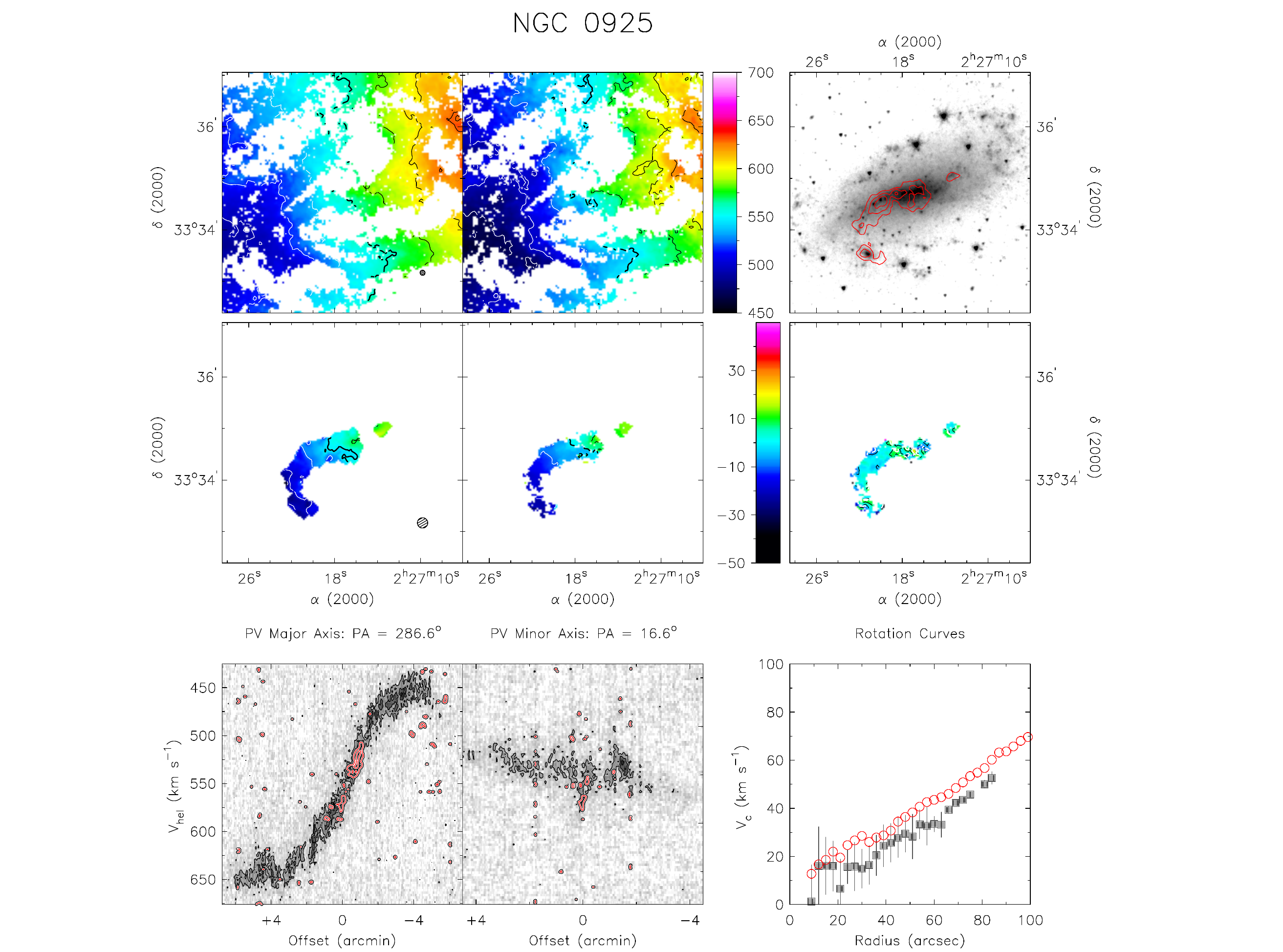}}
	\caption{\label{fig:925-atlas}Multi-panel plots for NGC 925. See
	the Appendix for a description of each panel. $V_{\mathrm{sys}}=546.3\kmss$, $\Delta V=25\kmss$. 
	SINGS $3.6\mu$m grayscale image: $0.1\times2^{N}$ MJy/sr,
	$N=0,0.2,...,4$; CO Moment-0 contours: 0 to 4 K$\,\kmss$ in stpdf
	of 1 K$\,\kmss$.}
\end{figure*}

NGC 925 is an SABd galaxy, and shows the least CO emission of the sample galaxies.
In Figure \ref{fig:925-atlas} we plot the H\,{\sc i} and CO velocity fields and
the \textit{pV} diagrams. We see that the CO is concentrated along the approaching side of the
galaxy. The CO emission is insufficient to constrain an independent model.
We therefore use the TM to compute the rotation curve, which is plotted in 
Figure \ref{fig:all-tm-rotcurs}. The CO rotation curve is lower than the H\,{\sc i} 
rotation curve. This is due to the lopsided emission of the CO, which
introduces a bias towards the approaching side when computing the rotation
curve. This can be seen for the THINGS rotation curve in dB08, where it is
shown that the rotation curve on the approaching side is lower than the total
rotation curve within a radius of $\sim80''$.

\subsection{NGC 2403}
NGC 2403 is a late-type SABcd galaxy, and its H\,{\sc i} emission has been
studied extensively, e.g., early observations by \citet{1973A&A....24..405S},
more recent observations include \citet{1538-3881-123-6-3124}.
\citet{SOFUE_1997} used the observations from \citet{THORNLEY_WILSON_1995}
to derive the CO rotation curve using the envelope tracing method.

\begin{figure*}[H]
	\centering
	\resizebox{\hsize}{!}{\includegraphics[origin=lb,width=22cm]{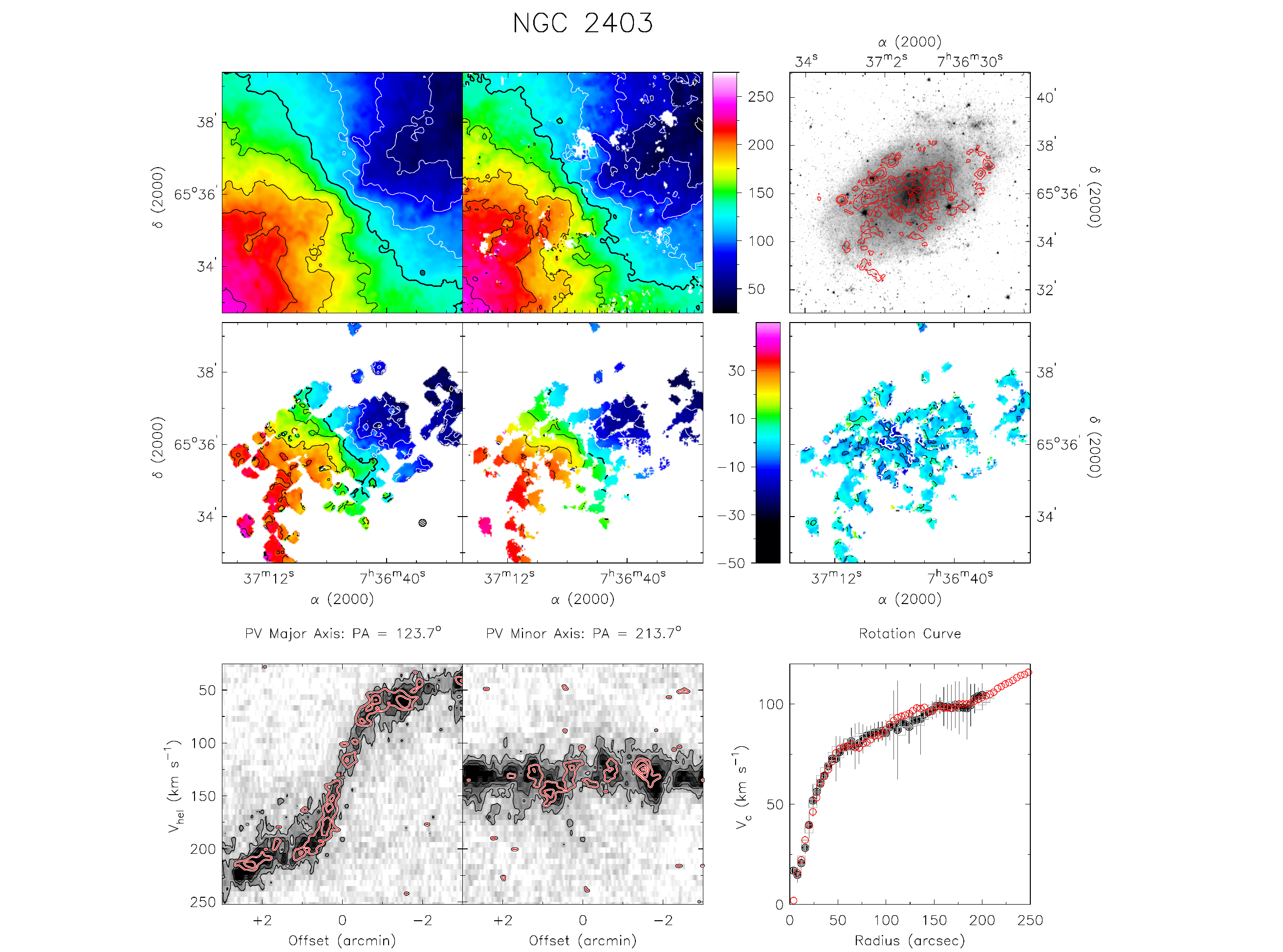}}
	\caption{\label{fig:2403-atlas}Multi-panel plots for NGC 2403. See
	the Appendix for a description of each panel.  $V_{\mathrm{sys}}=132.8\kmss$, $\Delta V=25\kmss$.
	SINGS $3.6\mu$m grayscale image: $0.1\times2^{N}$ MJy/sr,
	$N=0,0.5,...,5.0$; CO Moment-0 contours: 0, 1.5 and 3
	K$\,\kmss$.}
\end{figure*}

In Figure \ref{fig:2403-atlas} we present the velocity fields and
\textit{pV} diagrams of the
H\,{\sc i} and CO distributions of NGC 2403. The CO emission is patchy and the \textit{pV}
diagrams show a good correspondence between the H\,{\sc i} and CO emission. The CO
emission is sufficient to calculate an independent HM rotation curve, which
we plot in Figure \ref{fig:all-hm-rotcurs}. The H\,{\sc i} and CO rotation curves
are in good agreement, and only show slight differences for the very inner
radius. 

\subsection{NGC 2841}
\begin{figure*}[H]
	\centering
	\resizebox{\hsize}{!}{\includegraphics[origin=lb,width=22cm]{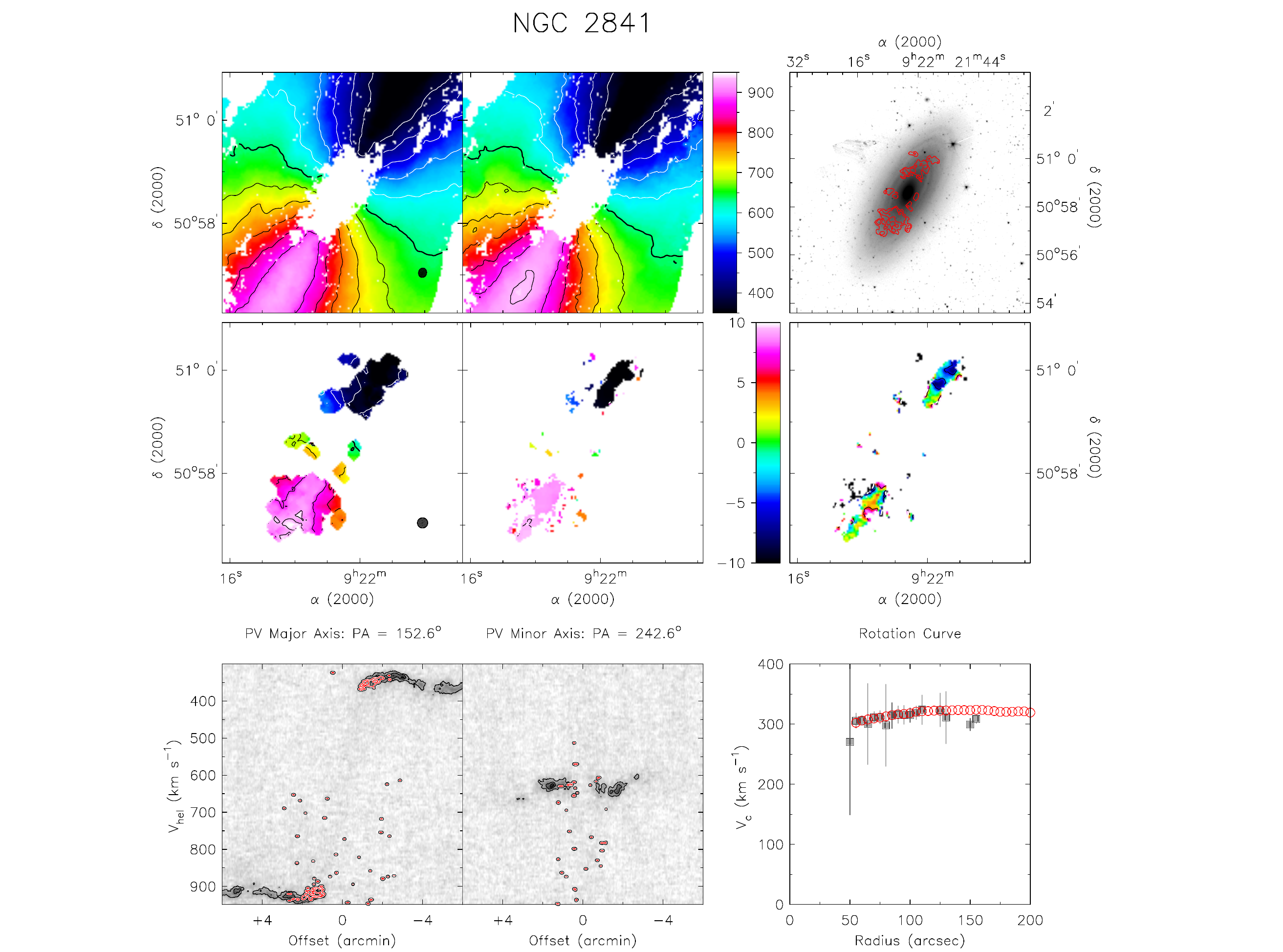}}
	\caption{\label{fig:2841-atlas}Multi-panel plots for NGC 2841. See
	the Appendix for a description of each panel. $V_{\mathrm{sys}}=632.7\kmss$, $\Delta V=50\kmss$.
	SINGS $3.6\mu$m grayscale image: $0.1\times2^{N}$ MJy/sr,
	$N=0,0.1,...,6.0$; CO Moment-0 contours: 0, 2 and 4
	K$\,\kmss$.}
\end{figure*}

NGC 2841 is an SAb galaxy. In Figure \ref{fig:2841-atlas} we plot the
velocity fields and \textit{pV} diagrams for the H\,{\sc i} and CO data. The H\,{\sc i} shows a hole
in the centre, which we also observe in the CO. We see a good
correspondence between the H\,{\sc i} and the CO in the \textit{pV} diagrams. The CO
emission is not sufficient to constrain an independent tilted-ring model,
so we compute the rotation curve using the TM. The rotation curve is
plotted in Figure \ref{fig:all-tm-rotcurs}.

\subsection{NGC 2903}
\label{sec:2903}
NGC 2903 is an SABbc type galaxy. The BIMA-SONG \citep{BIMA-SONG-II} observations show that the CO
$J=1\rightarrow0$ emission is concentrated along a molecular bar. NGC 2903 has also been observed as
part of the survey described in \citet{2007PASJ...59..117K}. There is an excellent correspondence
between the major-axis \textit{pV} diagram and the velocity field presented therein, with those
presented in this work. 

\begin{figure*}[H]
	\centering
	\resizebox{\hsize}{!}{\includegraphics[origin=lb,width=22cm]{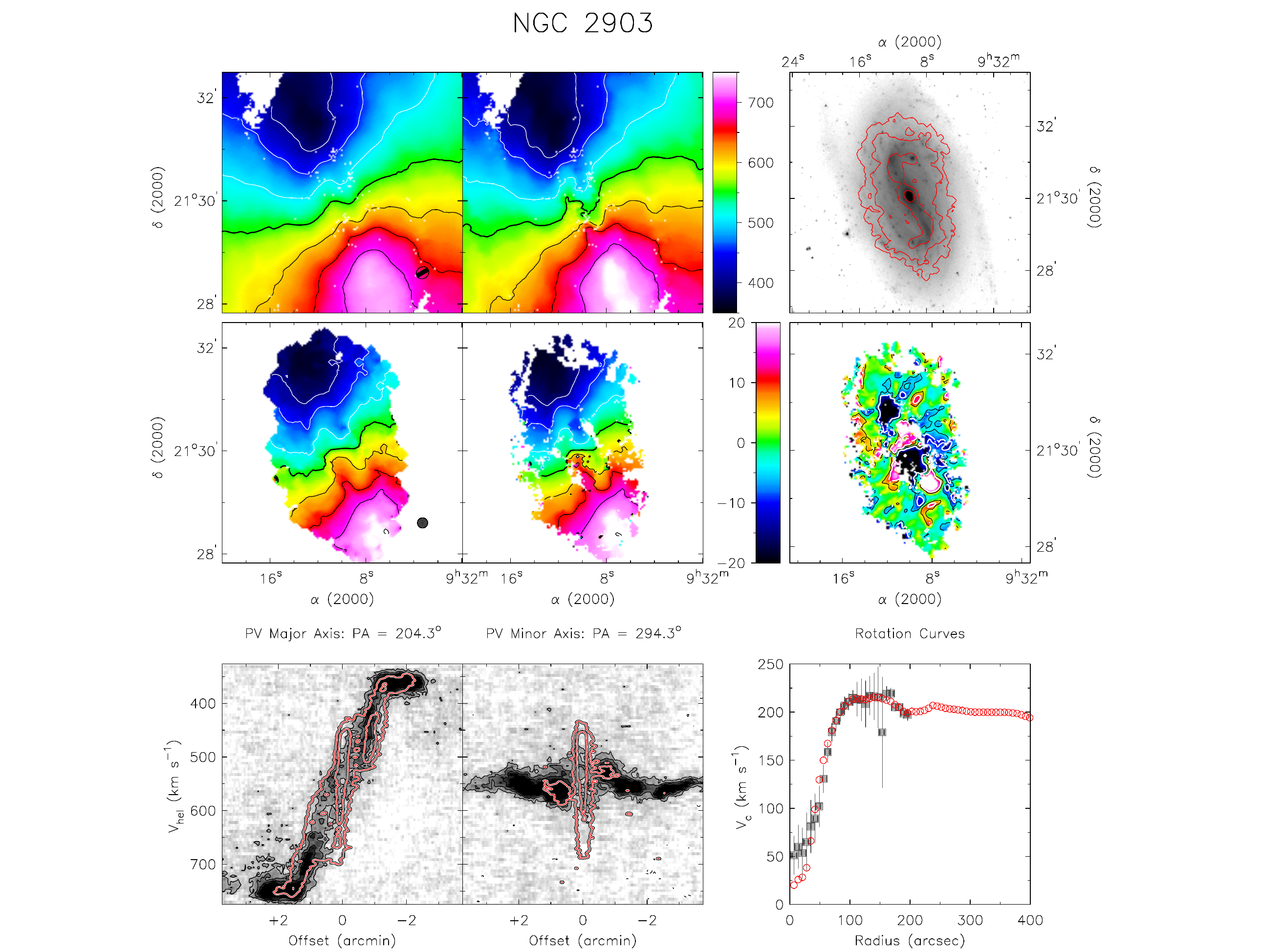}}
	\caption{\label{fig:2903-atlas}Multi-panel plots for NGC 2903. See
	the Appendix for a description of each panel.
	$V_{\mathrm{sys}}=555.6\kmss$,, $\Delta V=40\kmss$.
	SINGS $3.6\mu$m grayscale image: $0.1\times2^{N}$ MJy/sr,
	$N=0,0.1,...,8.0$; CO Moment-0 contours: $2^{N}$ K$\,\kmss$,
	$N=0,2,...20$. \textit{pV} major- and minor-axis diagrams: H\,{\sc i} contours are
	as designated in the appendix, the CO contours are
	$\sigma_{\mathrm{CO'}}\times2^{N}$ K$\,\kmss$, $N=0,2,...,10$.}
\end{figure*}

In Figure \ref{fig:2903-atlas} we plot the velocity fields and \textit{pV} diagrams
of the H\,{\sc i} and CO data. The CO/$3.6\mu$m overlay shows that the
CO $J=2\rightarrow1$ emission is concentrated along a bar. There
is also low surface density CO across the disk of the galaxy. The
kinks in the iso-velocity contours in the CO velocity fields are indicative of the non-circular
streaming motions which dominate in the central regions. The difference
velocity field shows a large difference, indicating that the H\,{\sc i} is not
sensitive to the small scale bar perturbations. This is also noted in 
\citet{SELLWOOD_SZ_2010}, who found the H$\alpha$ to be a more reliable
tracer of the bar than the H\,{\sc i} data.

The presence of the bar makes it difficult to compute an independent HM
model. We therefore use the TM with the understanding that the inner parts
(within $\sim 100''$)
of the CO rotation curve is dominated by non-circular motions, and
therefore cannot be used to trace the rotational velocity of the galaxy.
The final rotation curve is plotted in Figure \ref{fig:all-tm-rotcurs}.

\subsection{NGC 2976}
\begin{figure*}[H]
	\centering
	\resizebox{\hsize}{!}{\includegraphics[origin=lb,width=22cm]{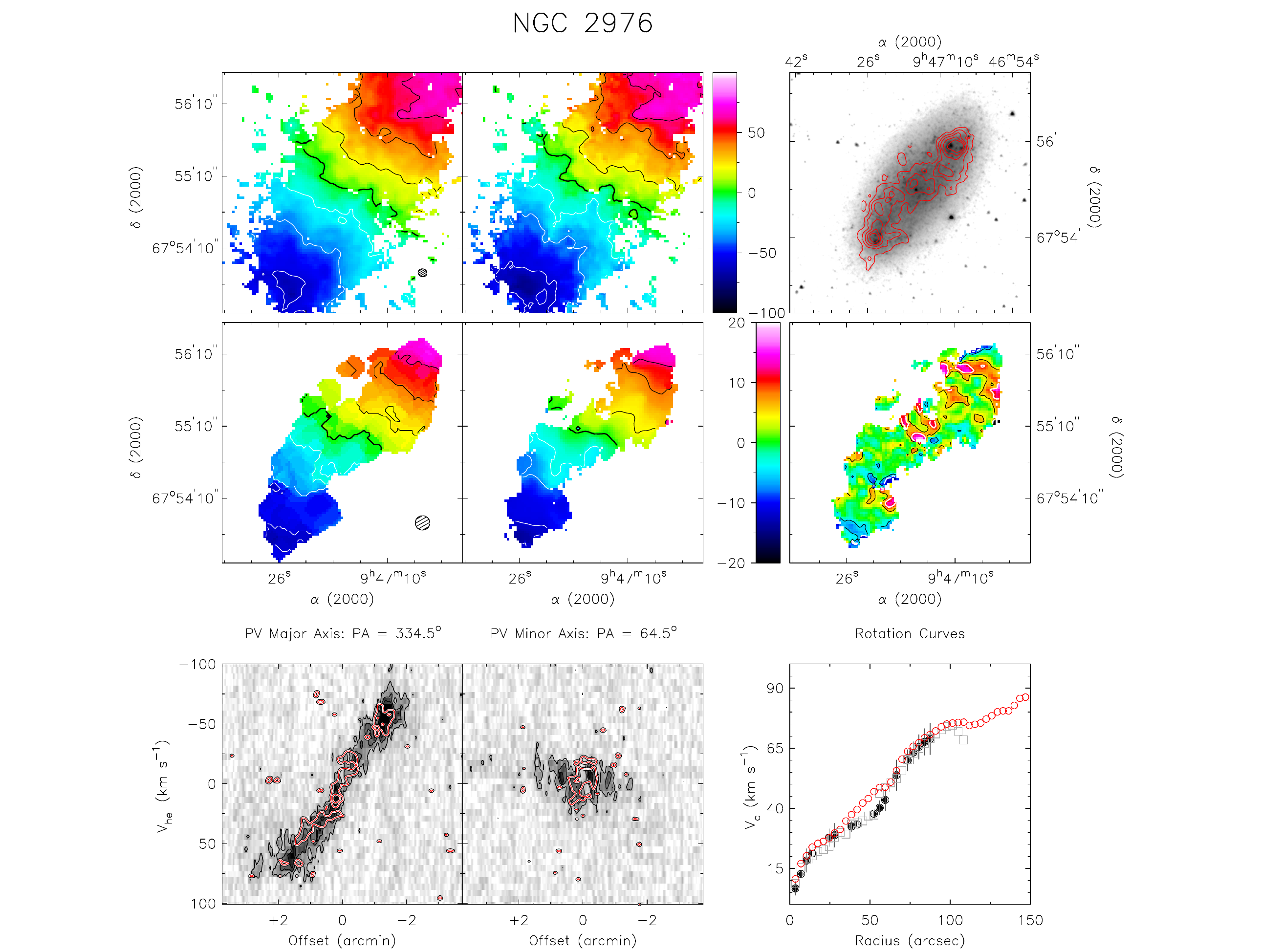}}
	\caption{\label{fig:2976-atlas}Multi-panel plots for NGC 2976. See
	the Appendix for a description of each panel.
	$V_{\mathrm{sys}}=1.1\kmss$, $\Delta V=20\kmss$.
	SINGS $3.6\mu$m grayscale image: $0.1\times2^{N}$ MJy/sr,
	$N=0,0.1,...,6.0$; CO Moment-0 contours: 1.6,3.2,...,8.0
	K$\,\kmss$.}
\end{figure*}

NGC 2976 is an SAc galaxy. \citet{SIMON_2012} obtained CO $J=1\rightarrow0$
observations of NGC 2976 using the BIMA array. In Figure
\ref{fig:2976-atlas} we plot the CO and H\,{\sc i} velocity fields, \textit{pV} diagrams,
rotation curves and the CO distribution overlaid on the associated SINGS 
map. The CO distribution is almost as extended as the H\,{\sc i}. The \textit{pV} diagrams
show a good agreement between the H\,{\sc i} and the CO. The distribution of the CO
is sufficient to calculate an independent HM model. The HM and TM
rotation curve is plotted in Figure \ref{fig:all-hm-rotcurs}.

\citet{SIMON_2012} show that the CO $J=1\rightarrow0$
surface density falls off rapidly at about $40''$ away from the 
centre of the galaxy. They therefore use their $\halpha$ data to
determine the rotation curve out to the edge of the disk. Our rotation
curve is in good agreement with their hybrid rotation curve; their $\halpha$ 
rotation curve also tracks the deviation from a linear rise at approximately $60''$.

\citet{SPEKKENS_2007} perform an analysis of the non-circular motions
in NGC 2976 using their package VELFIT, which includes the $m=1$
and $m=2$ modes in an explicit fit to the entire velocity field (as
opposed to the radially independent tilted-ring model). They fit an
$m=2$ (as due to a bar) Fourier mode to the CO $J=1\rightarrow0$
velocity field from \citet{SIMON_2012}. They show that the deviation
in the CO rotation curve from a linear rise at approximately $60''$
is principally due to the bar-streaming motion. It additionally seems
likely that the $m=2$ radial and tangential variations are responsible for
the fact that the CO rotation curve is consistently lower than the
THINGS rotation curve. 

\subsection{NGC 3198}
\label{app:3198}
\begin{figure*}[H]
	\centering
	\resizebox{\hsize}{!}{\includegraphics[origin=lb,width=22cm]{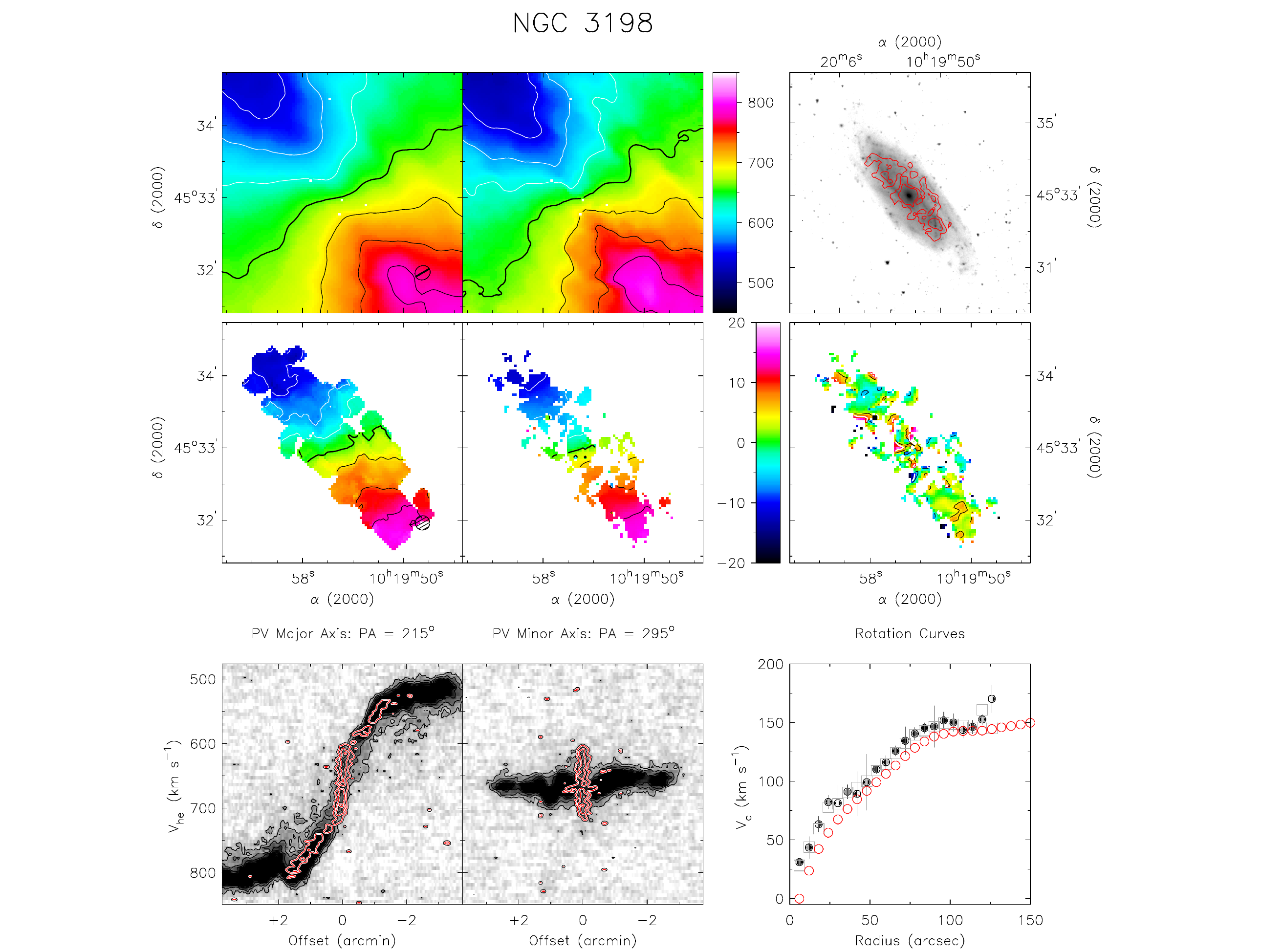}}
	\caption{\label{fig:3198-atlas}Multi-panel plots for NGC 3198. See
	the Appendix for a description of each panel.
	$V_{\mathrm{sys}}=660.7\kmss$, $\Delta V=40\kmss$.
	SINGS $3.6\mu$m grayscale image: $0.1\times2^{N}$ MJy/sr,
	$N=0,0.1,...,6.0$; CO Moment-0 contours: 1.5 and 3.0 K$\,\kmss$.}
\end{figure*}

NGC 3198 is an SBc type galaxy. In Figure \ref{fig:3198-atlas}
we plot the velocity fields derived in this work. The velocity fields
show the paucity of the CO emission. The difference velocity field
in Figure shows that the H\,{\sc i} and CO velocity fields
are in good agreement. The major-axis \textit{pV} diagram shows a good correspondence
between the CO and the H\,{\sc i}, while the minor-axis \textit{pV} diagram shows possible
kinematic evidence for streaming motions, which can be associated with the
weak bar \citep{2002A&A...393..453B}. The CO emission is sufficient for us to
compute an HM model, which we plot in Figure \ref{fig:all-hm-rotcurs}.

While the CO and H\,{\sc i} tilted-ring models are in generally good
agreement, we note two large differences between the CO and H\,{\sc i}
rotation curves. The first difference is at approximately $30''$.  The
second is at approximately $90''$.  The difference at $30''$ can be related
to the presence of a weak bar. The effect of a bar was noted by
\citet{2002A&A...393..453B}, who speculate that the bar would be oriented
parallel to the line of sight (aligned with the observed minor-axis) based
on the $\halpha$ morphology.  They use the results of
\citet{1985MNRAS.212..257T} to provide a qualitative interpretation of the
effect of the bar - which is to increase the apparent rotational velocity
within $30''$. This is consistent with our results.  The difference at
$90''$ can be related to the low-filling factor of the CO at these radii. The 
filling factor drops to less than $5\%$ beyond $\sim100''$.

\subsection{NGC 3521}
\begin{figure*}[H]
	\centering
	\resizebox{\hsize}{!}{\includegraphics[origin=lb,width=22cm]{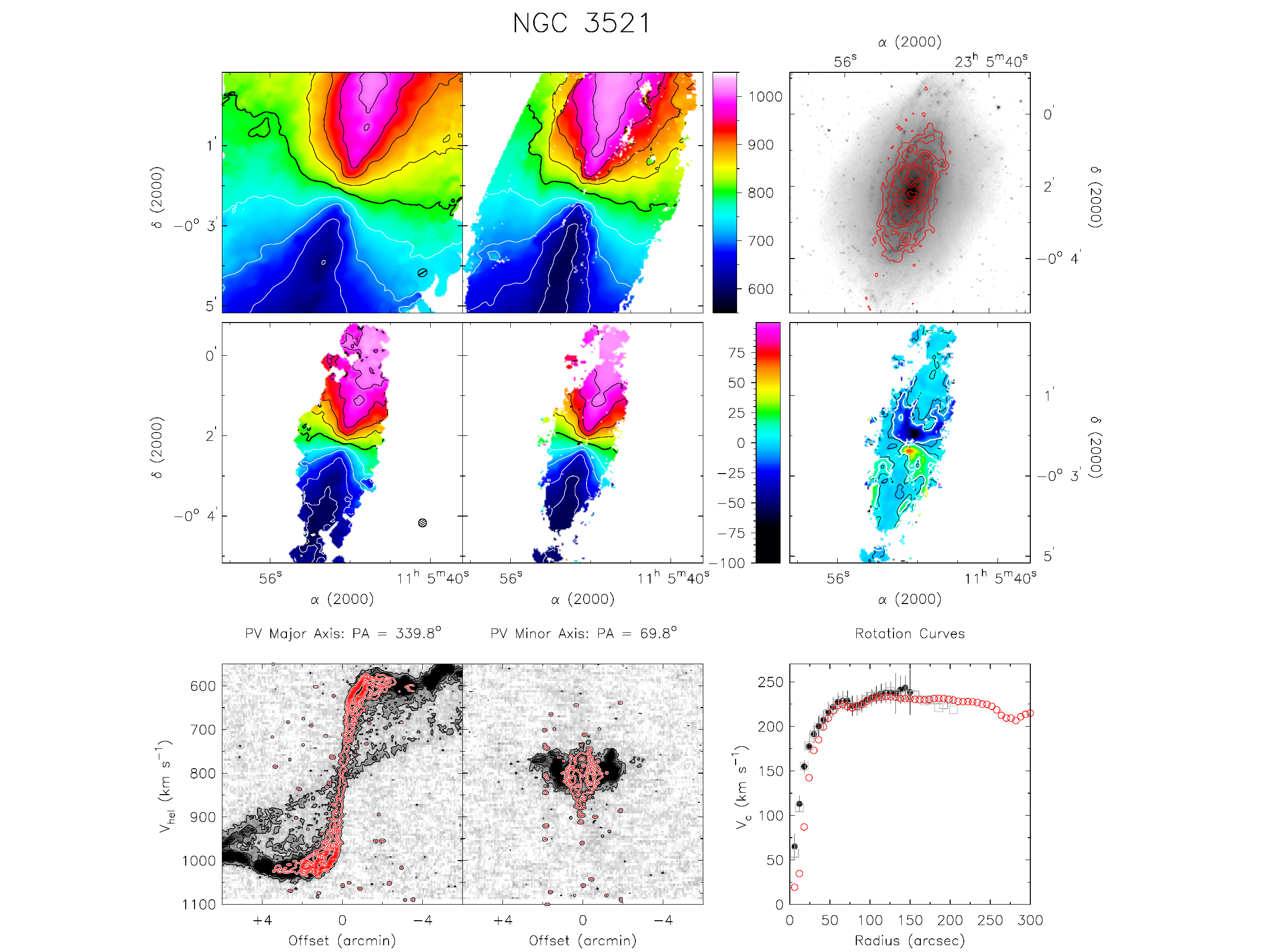}}
	\caption{\label{fig:3521-atlas}Multi-panel plots for NGC 3521. See
	the Appendix for a description of each panel.  $V_{\mathrm{sys}}=803.5\kmss$, $\Delta V=50\kmss$.
	SINGS $3.6\mu$m grayscale image: $0.1\times2^{N}$ MJy/sr,
	$N=0,0.1,...,8.0$; CO Moment-0 contours: 5,10,...,35
	K$\,\kmss$.}
\end{figure*}

NGC 3521 is an SABbc galaxy, and has also been been observed as part
of the \citet{2007PASJ...59..117K} survey.  Their results also show a
depletion of CO in the centre of the galaxy. The comparison of the H\,{\sc i} and
CO velocity fields are shown in Figure \ref{fig:3521-atlas}.  The H\,{\sc
i}-CO difference velocity field  is large in the inner part of the galaxy,
which corresponds to the difference in the \textit{pV} diagrams for both
components. This corresponds to the warp in the H\,{\sc i}
distribution, along the line of sight, which affects the derived velocity
field. 

We derive an HM rotation curve from the CO data. This is plotted in Figure
\ref{fig:all-hm-rotcurs}. The CO rotation curve is higher than
the H\,{\sc i} within $\sim 60''$ - the CO rotation curve is almost
50$\kmss$ higher than the H\,{\sc i} rotation curve for the innermost four
points. It is likely that the H\,{\sc i} velocities at these radii are
affected by the presence of emission due to the warp in the line of sight.
\subsection{NGC 3627}
NGC 3627 is an SABb galaxy and is part of the Leo Triplet. Notable studies of the molecular gas
content of NGC 3627 are \citet{1983ApJ...269..136Y} and \citet{2007PASJ...59..117K}. Studies of the
kinematics of the ionized gas \citep{2003A&A...405...89C} and the H\,{\sc i}
\citep{TRACHTERNACH:2008FK} show strong signs of non-circular motions, which are due to the bar
streaming motions as well as tidal interactions within the group.

\begin{figure*}[H]
	\centering
	\resizebox{\hsize}{!}{\includegraphics[origin=lb,width=22cm]{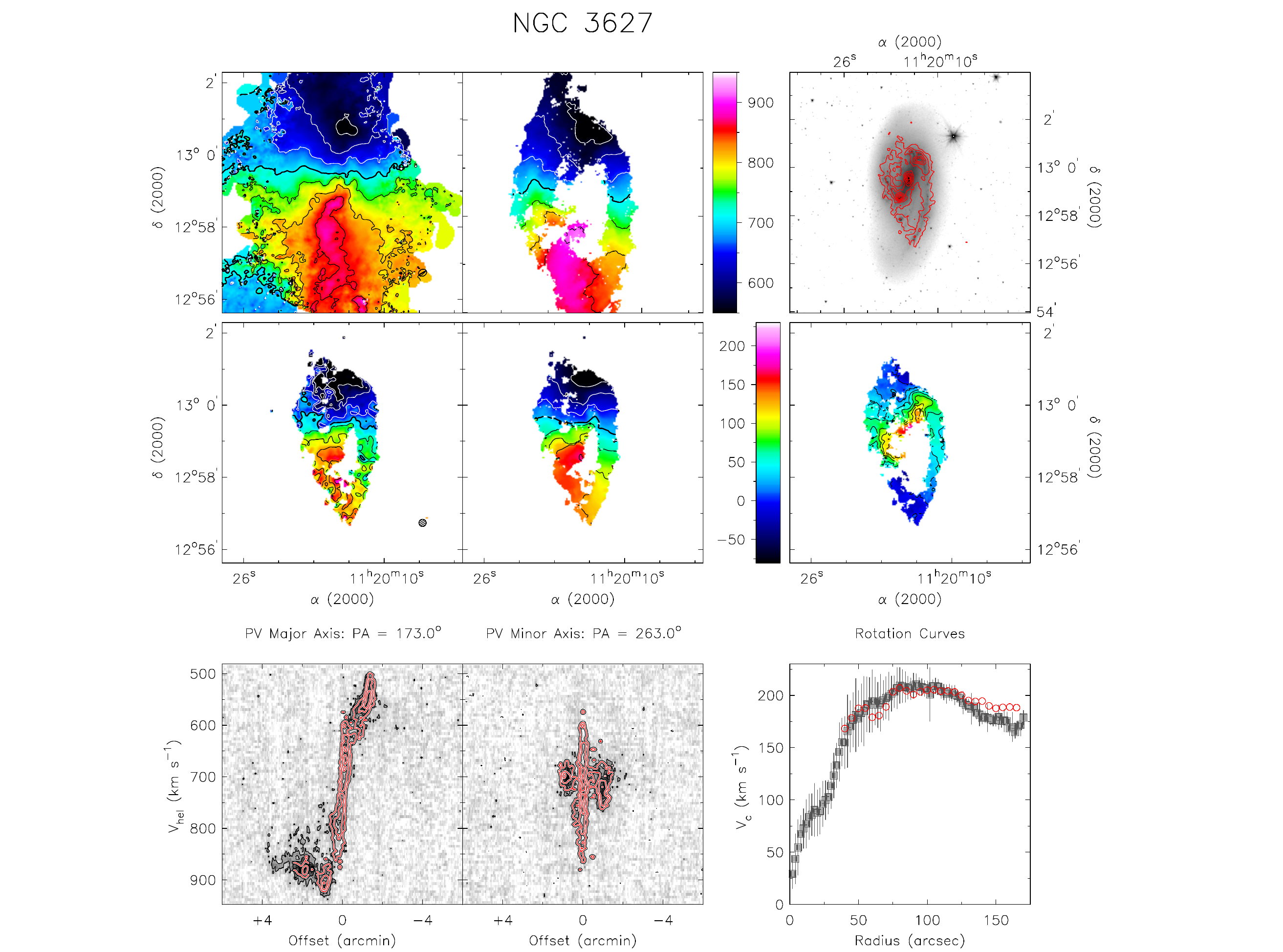}}
	\caption{\label{fig:3627-atlas}Multi-panel plots for NGC 3627. See
	the Appendix for a description of each panel.  $V_{\mathrm{sys}}=708.2\kmss$, $\Delta V=50\kmss$.
	SINGS $3.6\mu$m grayscale image: $0.1\times2^{N}$ MJy/sr,
	$N=0,0.1,...,8.0$; CO Moment-0 contours: $1\,\mathrm{K}\,\kmss$ to $15\,\mathrm{K}\,\kmss$ in
	stpdf of $1.5\,\mathrm{K}\,\kmss$}
\end{figure*}

The comparison of the H\,{\sc i} and CO data is shown in Figure \ref{fig:3627-atlas}. The HERACLES
data shows that the CO distribution is lopsided and although the CO velocity fields shows a general
velocity gradient, the iso-velocity contours and the \textit{pV} diagram show signs of non-circular
motions. The H\,{\sc i}-CO difference velocity field shows large differences in the centre of the
galaxy, where the H\,{\sc i} distribution shows signs of a depression. 

The presence of several different sources of perturbations which could lead to non-circular motions
makes it difficult to determine a tilted-ring model for the CO data. We therefore use the THINGS
parameters to calculate the rotation curve. For the inner part of NGC 3627 we assume the
inclination and position angle of the inner most tilted-ring from the THINGS data. The rotation
curve plotted in Figures \ref{fig:all-tm-rotcurs} and \ref{fig:3627-atlas} show a good agreement between
the H\,{\sc i} and the CO from a radius of approximately $50''$ outwards. Due to the complex
kinematics of the H\,{\sc i} and CO, we do not use the data for NGC 3627 to fit mass models.   

\subsection{NGC 4736}
\begin{figure*}[H]
	\centering
	\resizebox{\hsize}{!}{\includegraphics[origin=lb,width=22cm]{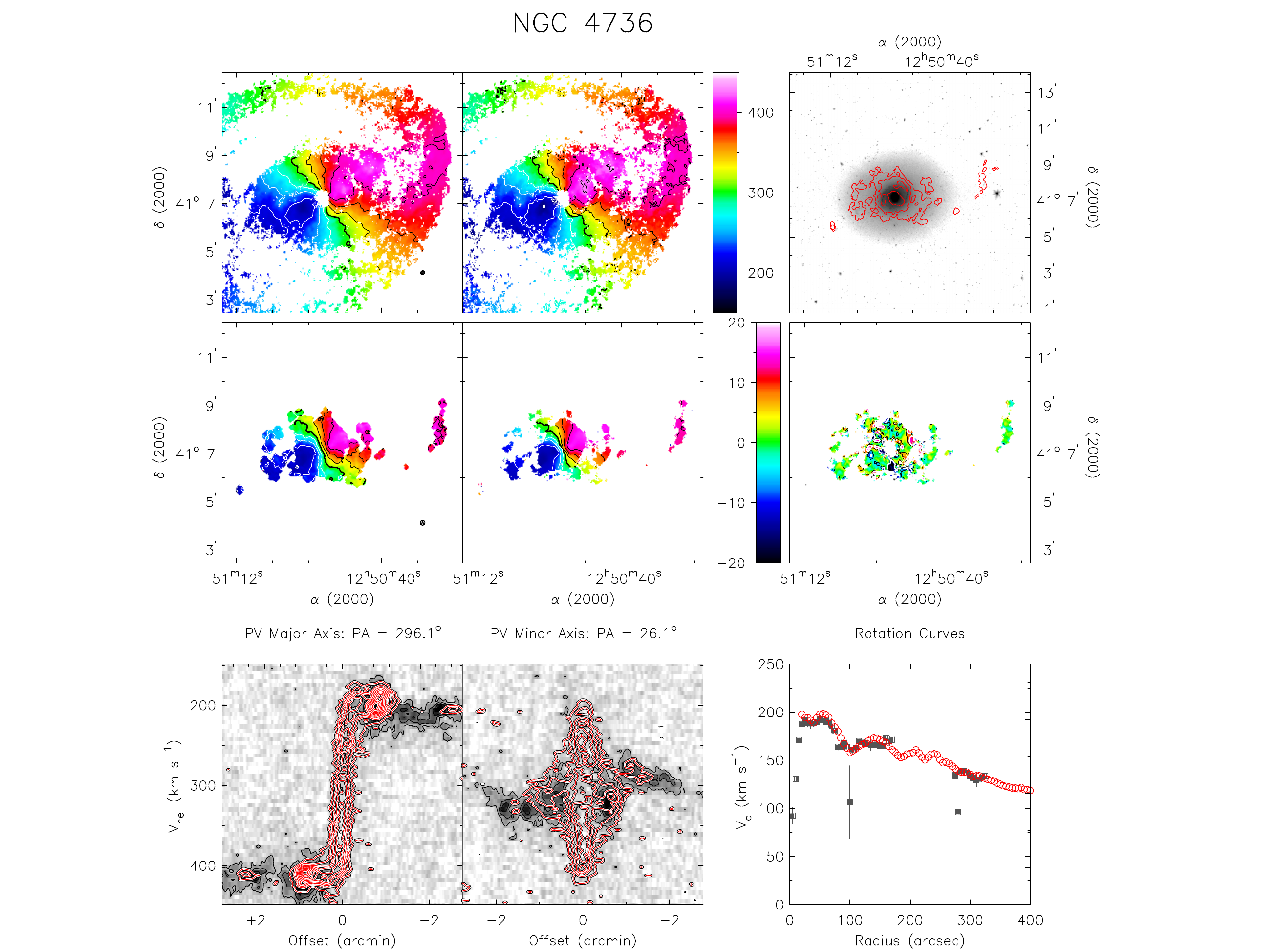}}
	\caption{\label{fig:4736-atlas}Multi-panel plots for NGC 4736. See the Appendix for a
	description of each panel.  $V_{\mathrm{sys}}=306.7\kmss$, $\Delta V=30\kmss$.  SINGS
	$3.6\mu$m grayscale image: $0.1\times2^{N}$ MJy/sr, $N=0,0.1,...,8.0$; CO Moment-0 contours:
	$2^{N}$ K$\,\kmss$, $N=0,2,...,10$.}
\end{figure*}

NGC 4736 is an SAab galaxy. NGC 4736 has a compact molecular
component. This can be see as the excess emission near systemic velocity in
the CO line profile (see Figure
\ref{fig:hico-profiles}) and the minor axis \textit{pV} diagram. 
The velocity fields and the \textit{pV} diagram which are
plotted in Figure \ref{fig:4736-atlas}.  NGC 4736 has also been observed as
part of the survey described in \citet{2007PASJ...59..117K}. We note a good agreement
between the \textit{pV} diagram and velocity field presented therein, and those
presented in this work. There is no systematic difference between the H\,{\sc i} 
and CO velocity fields in the region where they overlap. The tilted-ring
parameters used to compute the CO rotation curve were extrapolated from the
TM by assuming a constant value inwards from the innermost point, 
since there is a central depression of H\,{\sc i}. It was not possible
to compute an independent HM model because of the low inclination. We plot the rotation curve 
in Figure \ref{fig:all-tm-rotcurs}.

\citet{2000APJ...540..771W} did a thorough investigation of the non-circular kinematics in NGC 4736,
solving for inflow and outflow models.  Their rotation curve shows the same features as ours, but
the velocity that they find at the first maximum is higher than ours by approximately $20\kmss$.
\citet{2009APJ...692.1623H} studied NGC 4736 as part of a search for gas radial flows, which was
based on the NUGA survey. They used high-resolution observations of the $J=2\rightarrow1$ and
$J=1\rightarrow0$ transitions of CO to map the central molecular component. Their rotation curve
rises to $200\kmss$ within $0.3\kpc$, which corresponds to a single beam of the HERACLES survey. The
rotation curve that we derive rises steeply (within $0.5\kpc$), but the rise is not as steep as the
rotation curve derived by \citet{2009APJ...692.1623H}, which could be due to differences in the beam
sizes, since the synthesized beam of the Plateau du Bure is approximately $0.5''$.
\subsection{NGC 5055}
NGC 5055 is an SAbc galaxy with extended CO emission across
the optical disk. The H\,{\sc i} and CO profiles are in good agreement with
each other (Figure \ref{fig:hico-profiles}). 

\begin{figure*}[H]
	\centering
	\resizebox{\hsize}{!}{\includegraphics[origin=lb,width=22cm]{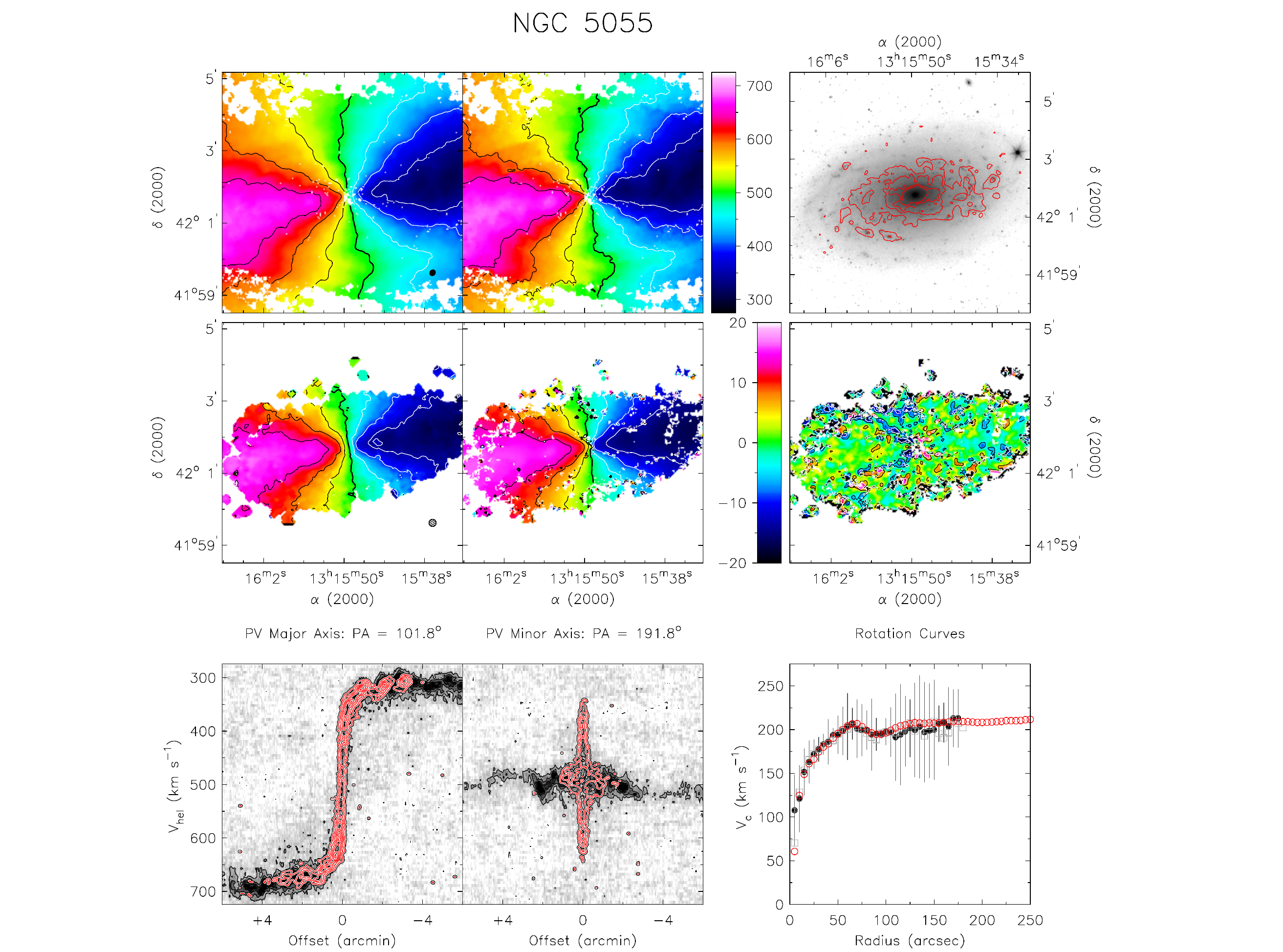}}
	\caption{\label{fig:5055-atlas}Multi-panel plots for NGC 5055. See the Appendix for a
	description of each panel.  $V_{\mathrm{sys}}=496.8\kmss$, $\Delta V=50\kmss$.  SINGS
	$3.6\mu$m grayscale image: $0.1\times2^{N}$ MJy/sr, $N=0,0.1,...,8.0$; CO Moment-0 contours:
	5,10,...,35 K$\,\kmss$.}
\end{figure*}

\citet{0004-637X-484-1-202} studied the morphology and kinematics of NGC
5055 in CO. NGC 5055 has also been observed as part of the
\citet{2007PASJ...59..117K} survey. We note a good agreement between the \textit{pV}
diagram and velocity field presented therein, and those presented in this
work. In Figure \ref{fig:5055-atlas} we plot the H\,{\sc i} and CO velocity
fields and \textit{pV} diagrams for NGC 5055. The velocity fields appear to be in
good agreement, and the difference velocity field shows small
amplitudes. The large difference in the centre of the galaxy is due to a
combination of differential beam smearing (the THINGS and HERACLES beamsizes are
slightly different) and the difference in morphology between the
H\,{\sc i} and the CO.

The major-axis \textit{pV} diagram shows that the H\,{\sc i} and CO distributions are quite
similar - we can thus expect that the H\,{\sc i} and CO rotation curves show
similar features.
 
The CO distribution is sufficient to compute an independent HM rotation
curve. The rotation curves for NGC 5055 are plotted in Figure
\ref{fig:all-hm-rotcurs}.  We also note that \citet{WONG:2004FK} studied CO
and H\,{\sc i} data for NGC 5055; the CO rotation curve derived in their work
(computed from the BIMA-SONG data) goes out to approximately $100''$, and
is sufficient to trace the rising part of the rotation curve. 

The H\,{\sc i} and CO rotation curves are in good agreement with each other, with
the CO rotation velocities being slightly higher (the differences are within the
error-bars and certainly less than about $5\kmss$) than that for the H\,{\sc i} 
within $100''$. 
\subsection{NGC 6946}

\begin{figure*}[H]
	\centering
	\resizebox{\hsize}{!}{\includegraphics[origin=lb,width=22cm]{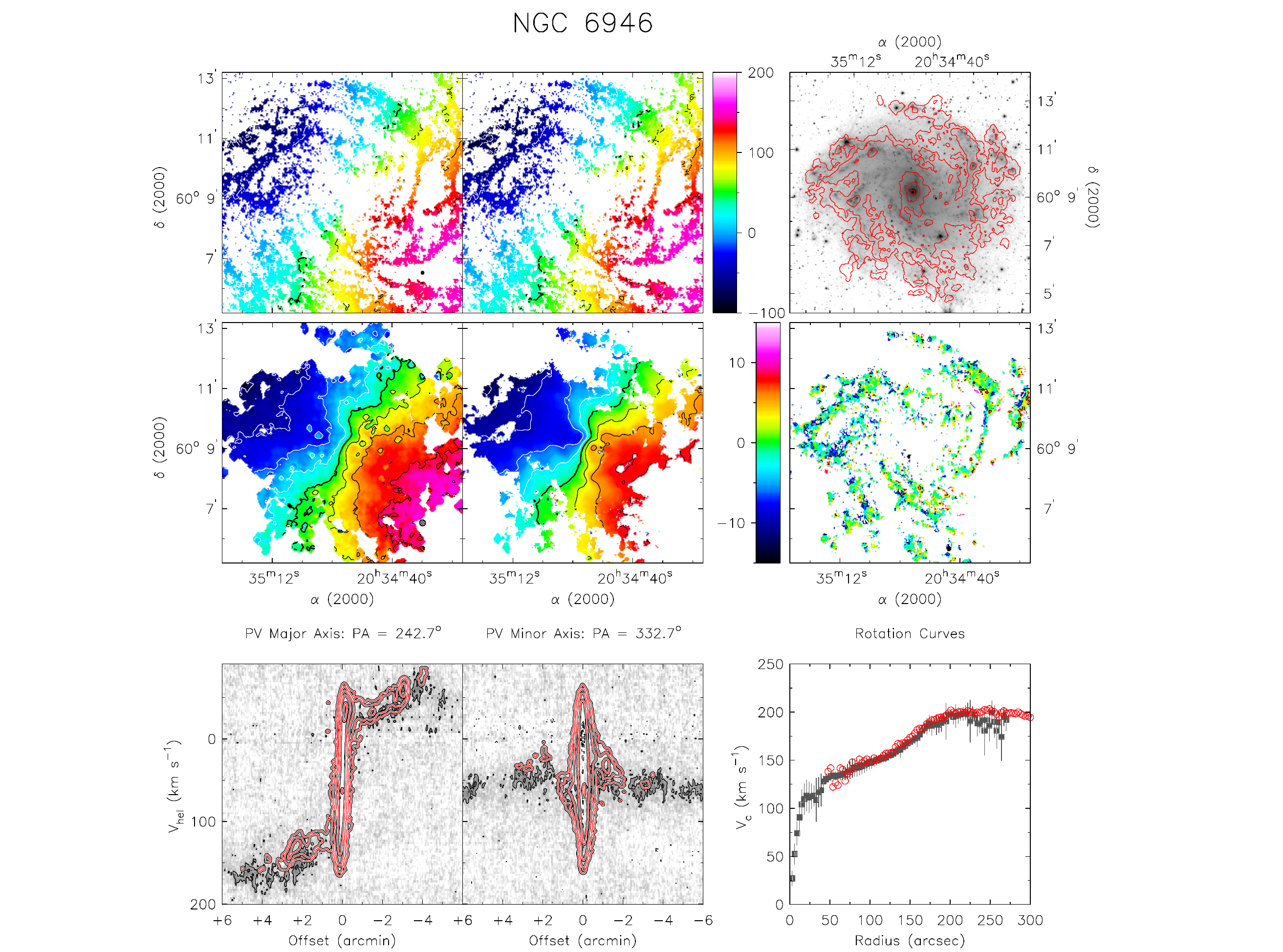}}
	\caption{\label{fig:6946-atlas}Multi-panel plots for NGC 6946. See the Appendix for a
	description of each panel.  $V_{\mathrm{sys}}=43.7\kmss$, $\Delta V=30\kmss$.  SINGS
	$3.6\mu$m grayscale image: $0.1\times2^{N}$ MJy/sr, $N=0,0.1,...,8.0$; CO Moment-0 contours:
	$2^{N}$ K$\,\kmss$, $N=0,2,...,10$.}
\end{figure*}

NGC 6946 is an SABcd galaxy and has historically been a popular target
for CO observations \citep{1982APJ...258..467Y,1985APJ...298L..21B,
1985MNRAS.217..571T,1988PASJ...40..511S,1989APJS...71..455T,1990A&A...233..357C,
2001A&A...371..433I,MEIER_TURNER_NGC6946,0004-637X-649-1-181,2007AJ....134.1827C}. The
CO morphology includes a bright nuclear structure and fainter
spiral arms. The nuclear structure was the object of several
detailed studies
\citep{1985APJ...298L..21B,1990A&A...233..357C,0004-637X-649-1-181}.
There are various studies comparing the distribution and
kinematics of different molecules, e.g., different transitions of
CO, as in \citet{2001A&A...371..433I}.  NGC 6946 has also been
observed as part of the \citet{2007PASJ...59..117K} survey. We note
a good agreement between the \textit{pV} diagram and velocity
field presented therein, and those presented in this work. In
Figure \ref{fig:6946-atlas} we plot the H\,{\sc i} and CO velocity
fields.

The low inclination of the CO emission is at the limit of what can
be modelled using a tilted-ring model. We therefore use the TM as the
estimate of the kinematics, and we compute the rotation curve, which is 
plotted in Figure \ref{fig:all-tm-rotcurs}.

\citet{0004-637X-649-1-181} derived rotation curves from high resolution
Plateau de Bure interferometer observations of CO $J=1\rightarrow0$ and
CO $J=2\rightarrow1$ in a manner very similar to that used in this work,
i.e., velocity field fitting using a tilted-ring model in \texttt{ROTCUR}.
Their rotation curves show a rapid rise to approximately $150\kmss$ within
$10''$, which is well within a single beam of our observations. Even at
$13''$ (the resolution of HERACLES), our rotation curve predicts a
rotational velocity $<100\kmss$, and only reaches $150\kmss$ at a radius of
approximately $100''$. \citet{0004-637X-649-1-181} attribute the steep rise
in the rotation curve to the response of the gas to the inner nuclear bar,
which is unresolved in the HERACLES observations. 
\subsection{NGC 7331}
\begin{figure*}[H]
	\centering
	\resizebox{\hsize}{!}{\includegraphics[origin=lb,width=22cm]{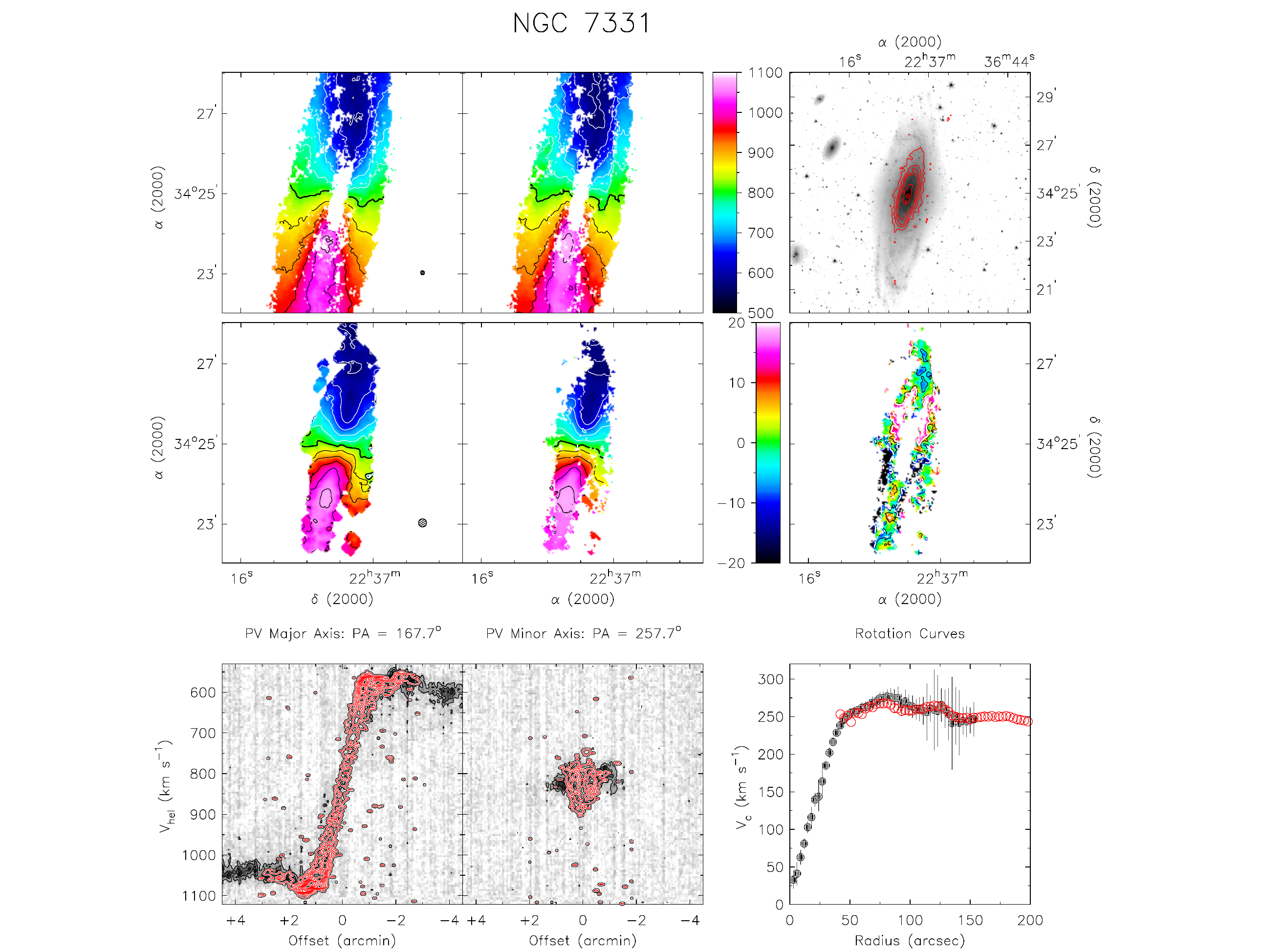}}
	\caption{\label{fig:7331-atlas}Multi-panel plots for NGC 7331. See the Appendix for a
	description of each panel.  $V_{\mathrm{sys}}=818.3\kmss$, $\Delta V=30\kmss$.  SINGS
	$3.6\mu$m grayscale image: $0.1\times2^{N}$ MJy/sr, $N=0,0.1,...,8.0$; CO Moment-0 contours:
	5, 10, 15 and 20 K$\,\kmss$.}
\end{figure*}

NGC 7331 is an SAb 
galaxy. In Figure \ref{fig:7331-atlas} we plot the H\,{\sc i} and CO velocity
fields and \textit{pV} diagrams. The H\,{\sc i} is distributed in a disk with a central
depression. The CO emission is distributed across the extent of the optical
disk and overlaps with the H\,{\sc i} emission. The CO emission is sufficient to
compute an HM rotation curve, which is plotted in Figure
\ref{fig:all-hm-rotcurs}.

\citet{1996A&A...315...52V} performed an in-depth analysis of NGC 7331,
which included simulations of the evolution of the spiral structure of the
galaxy. They derive CO $J=2\rightarrow1$ and $J=1\rightarrow0$ rotation
curves, which are not inclination corrected. The shape of their rotation
curve is consistent with the one derived in this work. They consider the
velocities derived in their rotation curve as being lower than the real
maximum velocity of the galaxy, due to the presence of molecular gas along
the line of sight. The tilted-ring model explicitly accounts for 
the disk geometry, and our maximum velocity ($\sim260\kmss$)
is indeed higher than their associated inclination corrected maximum
velocity $\sim230\kmss$.

\section{Fitting Analytic Expressions to the Rotation Curves}

\begin{figure*}[H]
	\centering
	\includegraphics[origin=lb,width=14cm]{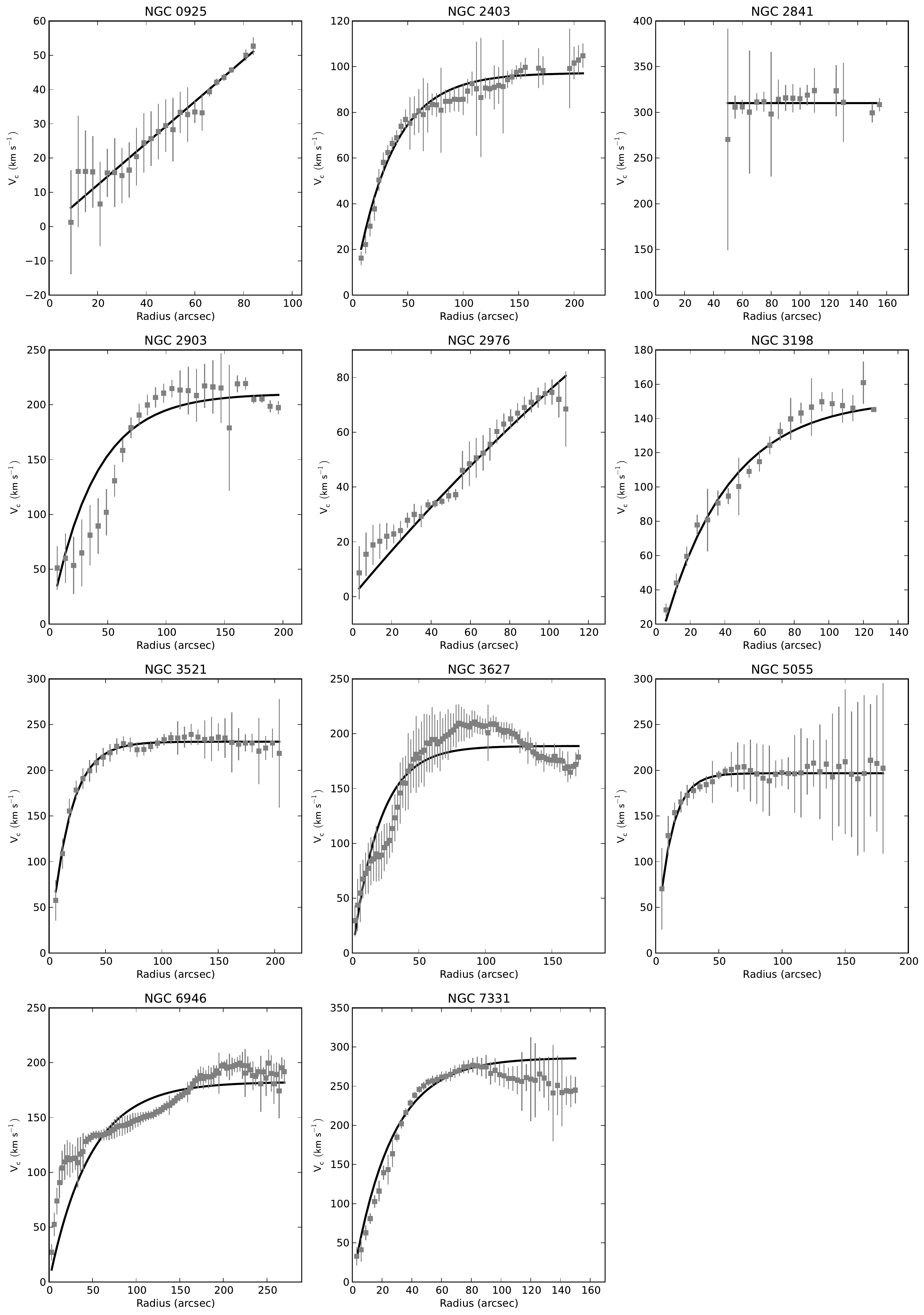}
	\caption{\label{fig:boissier}CO rotation curves (using the THINGS-Model) with the fitted
	analytic rotation curves described in \citet{2003MNRAS.346.1215B}. The CO rotation curves
	are plotted as filled grey squares with error bars; The Boissier fits are plotted as solid
	black lines. The fits for NGC 4736 are not included here, since the
	\citet{2003MNRAS.346.1215B} curves cannot fit a declining rotation curve.}
\end{figure*}

Following
\citet{2008AJ....136.2782L} we fit the analytic expression described in
\citet{2003MNRAS.346.1215B}:

\begin{equation}
	v_{\mathrm{rot}}(r) = v_{\mathrm{flat}}\Bigg[1 - \exp\big(\frac{-r}{l_{\mathrm{flat}}}\big)\Bigg]
\end{equation}

where $v_{\mathrm{rot}}$ is the circular velocity at radius $r$ (the rotation curve) and
$v_{\mathrm{flat}}$ and $l_{\mathrm{flat}}$ are parameters corresponding to the velocity and the
scale length at which the rotation curve approaches flatness, respectively. We use a non-linear
least-square algorithm to fit the analytic rotation curves to the CO rotation curves presented in
this work. The corresponding best fit values for $v_{\mathrm{flat}}$ and $l_{\mathrm{flat}}$  are
presented in Table \ref{tab:vflat-lflat}. In Figure \ref{fig:boissier} we plot the best fit model
rotation curves and the observed CO rotation curves derived using the THINGS Model. We do not
include the results for NGC 4736, since the analytic model cannot fit a declining rotation curve, by
definition.
We find that the $v_{\mathrm{flat}}$ are similar for fits to the CO and H\,{\sc i} rotation curves,
while the values of $l_{\mathrm{flat}}$ are slightly lower when fit to the CO rotation curves. 

\begin{table}
	\caption{\label{tab:vflat-lflat}Fitted parameters of the \citet{2003MNRAS.346.1215B} analytic rotation curve to the CO and
	H\,{\sc i} rotation curves, $v_{\mathrm{flat}}$ and $l_{\mathrm{flat}}$. The parameters for the
	H\,{\sc i} are as in \citet{2008AJ....136.2782L}.}
	\centering
	\begin{tabular}{c l l}
	\hline 
	\hline
	Galaxy & $v_{\mathrm{flat}}$ & $l_{\mathrm{flat}}$ \\
	 & ($\kmss$) & ($\kpc$) \\
	\noalign{\smallskip} \hline 
	NGC 2403 & 97 & 0.5  \\
	NGC 2841 & 310 & 0.1  \\
	NGC 2903 & 210 & 1.6  \\
	NGC 2976 & 305 & 6.1  \\
	NGC 3198 & 152 & 2.5  \\
	NGC 3521 & 231 & 0.9  \\
	NGC 3627 & 88.8 & 0.9 \\
	NGC 5055 & 197 & 0.6  \\
	NGC 6946 & 182 & 1.4  \\
	NGC 7331 & 286 & 0.5 \\
	\hline
	\noalign{\smallskip}
	\multicolumn{3}{c}{ HERACLES Model} \\
	\noalign{\smallskip}
	\hline 
	NGC 2403 & 98 & 0.6\\
	NGC 2976 & 139 & 2.5\\
	NGC 3198 & 159 & 2.7\\
	NGC 3521 & 233 & 0.9\\
	NGC 5055 & 197 & 0.6\\
	NGC 7331 & 286 & 0.5\\
	\hline 
	\end{tabular}
\end{table}

\end{document}